\documentclass{aa}
\usepackage{soul}
\usepackage{graphicx}
\usepackage{txfonts}
\usepackage{amsmath}    
\usepackage{amssymb}    
\usepackage{siunitx}
\usepackage{todonotes}
\usepackage{booktabs}
\usepackage{ulem}
\usepackage{makecell}
\usepackage{pbox}
\usepackage{subcaption}
\usepackage{multirow}
\usepackage{float}

\usetikzlibrary{calc}
\usetikzlibrary{fit}

\usepackage{pifont}

\usepackage{balance}

\usepackage[perpage]{footmisc}

\defcitealias{Eastman1996}{E96}
\defcitealias{Dessart2005a}{D05}
\defcitealias{Hamuy2001}{H01}
\defcitealias{Pastorello2006}{P06}
\defcitealias{Faran2014}{F14}
\defcitealias{Dessart2008}{D08}

\DeclareSIUnit\Mpc{Mpc}
\DeclareSIUnit\pc{pc}
\DeclareSIUnit\year{yr}

\newcommand{\tardis}{\textsc{tardis}}

\newcommand{\myhknot}{\SI[separate-uncertainty=true,multi-part-units = single]{74.9(19)}{\km \per \second \per \Mpc}}

\newcommand{\mysigmaint}{$1.2^{+1.1}_{-0.7}\,$\%\xspace}

\newcommand{\NumZTF}{7}
\newcommand{\NumAtlas}{8}

\usepackage[draft=false]{hyperref}
\usepackage{cleveref}
\crefname{equation}{Eq.}{Eqs.}
\crefname{figure}{Fig.}{Figs.}
\Crefname{figure}{Figure}{Figures}
\crefname{section}{Sect.}{Sects.}
\Crefname{section}{Section}{Sections}
\crefname{table}{Table}{Tables}
\crefname{chapter}{Chapter}{Chapters}
\crefname{appendix}{Appendix}{Appendices}
\bibpunct{(}{)}{,}{a}{}{,} 
\usepackage{color}
\usepackage{epsdice}
\usepackage{pgfplots}
\pgfplotsset{compat=1.16} 

\begin{document}

\title{No rungs attached: A distance-ladder free determination of the Hubble constant through type II supernova spectral modelling}
\titlerunning{A determination of the Hubble constant through type II supernova spectral modelling}

\author{C.~Vogl\inst{1,2,6}
	\and
	S.~Taubenberger\inst{1,2}
	\and
	G.~Cs\"ornyei\inst{5,1}
	\and
 	B.~Leibundgut\inst{5,2}
	\and
         W.~E.~Kerzendorf\inst{7,8}
         \and
         S.~A.~Sim\inst{9}
        \and
	S.~Blondin\inst{4,5}
        \and
        A.~Fl\"ors\inst{10}
	\and
	A.~Holas\inst{3}
	\and
        J.~V.~Shields\inst{7}
        \and
	J.~Spyromilio\inst{5}
	\and
	S.~H.~Suyu\inst{2,1}
	\and
	W.~Hillebrandt\inst{1}
}

\institute{Max-Planck-Institut f\"ur Astrophysik, Karl-Schwarzschild-Str. 1, 85741 Garching, Germany\\
	\email{cvogl@mpa-garching.mpg.de}
	\and
Technische Universit\"at M\"unchen, TUM School of Natural Sciences, Physics Department, James-Franck-Stra\ss{}e 1, 85748 Garching, Germany
	\and
	Heidelberg Institute for Theoretical Studies, Schloss-Wolfsbrunnenweg 35, 69118 Heidelberg, Germany
	\and
	Aix Marseille Univ, CNRS, CNES, LAM, Marseille, France
	\and
	European Southern Observatory, Karl-Schwarzschild-Stra\ss{}e 2, 85748 Garching, Germany
	\and Exzellenzcluster ORIGINS, Boltzmannstr. 2, 85748 Garching, Germany
	   \and
    Department of Physics and Astronomy, Michigan State University, East Lansing, MI 48824, USA
            \and
        Department of Computational Mathematics, Science, and Engineering, Michigan State University, East Lansing, MI 48824, USA
	   \and
	    Astrophysics Research Centre, School of Mathematics and Physics, Queen's University Belfast, Belfast BT7 1NN, UK
     \and
     GSI Helmholtzzentrum f\"ur Schwerionenforschung, Planckstra\ss{}e 1, 64291 Darmstadt, Germany
}

\date{}

\abstract
{The ongoing discrepancy in the Hubble constant ($H_0$) estimates obtained through local distance ladder methods and early universe observations poses a significant challenge to the $\Lambda$CDM model, suggesting potential new physics. Type II supernovae (SNe~II) offer a promising technique for determining $H_0$ in the local universe independently of the traditional distance ladder approach, opening up a complimentary path for testing this discrepancy.}
{We aim to provide the first $H_0$ estimate using the tailored expanding photosphere method (EPM) applied to SNe~II, made possible by recent advancements in spectral modelling that enhance its precision and efficiency.}
{Our tailored EPM measurement utilizes a spectral emulator to interpolate between radiative transfer models calculated with \textsc{tardis}, allowing us to fit supernova spectra efficiently and derive self-consistent values for luminosity-related parameters. We apply the method on public data for ten SNe~II at redshifts between \num{0.01} and \num{0.04}.}
{Our analysis demonstrates that the tailored EPM allows for $H_0$ measurements with precision comparable to the most competitive established techniques, even when applied to literature data not designed for cosmological applications. We find an independent $H_0$ value of $74.9\pm1.9$ (stat) $\mathrm{km}\,\mathrm{s}^{-1}\,\mathrm{Mpc}^{-1}$, which is consistent with most current local measurements. Considering dominant sources of systematic effects, we conclude that our systematic uncertainty is comparable to or less than the current statistical uncertainty.} 
{This proof-of-principle study highlights the potential of the tailored EPM as a robust and precise tool for investigating the Hubble tension independently of the local distance ladder. Observations of SNe~II tailored to $H_0$ estimation can make this an even more powerful tool by improving the precision and by allowing us to better understand and control systematic uncertainties.}


\maketitle
%

\section{Introduction}
The advancements made in the past decade in cosmic distance measurements have brought forward a persistent discrepancy between the Hubble constant ($H_0$) values when estimated through different means: currently, the community faces a $5.8\,\sigma$ tension between the $H_0$ measured directly from redshifts and distances in the local universe ($73.17 \pm 0.86 \textrm{ km s}^{-1}\textrm{ Mpc}^{-1}$), based on the SH0ES analysis \citep{Riess2022, Breuval2024} employing the Cepheid period-luminosity relation and type Ia supernovae (SNe~Ia), and the same parameter estimated through the cosmic microwave background (CMB) assuming a $\Lambda$CDM cosmology ($67.4 \pm 0.5 \textrm{ km s}^{-1}\textrm{ Mpc}^{-1}$, \citealt{Planck2018Overview}). 

The Hubble tension between the local universe and CMB estimates currently constitutes one of the biggest challenges for the successful $\Lambda$CDM model, potentially hinting at new physics. Possible explanations include early dark energy \citep{2019PhRvL.122v1301P,2020PhRvD.101f3523S,2023PhRvD.108d3513H}, new neutrino physics \citep{2020PhRvD.101l3505K,2020PhRvD.102k5008B}, interaction between dark matter and dark energy \citep{2016RPPh...79i6901W, 2023arXiv231007260P}, as reviewed recently by \cite{diValentino21} or in the $H_0$ Olympics \citep{H0Olympics}.

The Hubble tension, which first emerged between the SH0ES (Cepheids and SNe~Ia) and Planck (CMB) observations, has been confirmed by other methods for the local and distant measurements. The lower value from the CMB is supported by independent measurements by  the Wilkinson Microwave Anisotropy Probe \citep[WMAP;][]{2013ApJS..208...19H}, the Atacama Cosmology Telescope \citep[ACT;][]{ACT2020}, and the South Pole Telescope \citep[SPT;][]{SPT2021}. Similar results are obtained through baryonic acoustic oscillations (BAOs), even when calibrated through CMB-independent means such as Big Bang Nucleosynthesis \citep[see, e.g.,][]{2018MNRAS.480.3879A,Schoeneberg2019}. 

In the local universe, mass overdensities induce local disturbances over the pure space expansion \citep[e.g., Cosmicflows-4;][]{2023ApJ...944...94T} and distance indicators reaching into the Hubble flow are required \citep[e.g.,][]{1974ApJ...190..525S}. The local distance ladder has been reduced over the past decades to three steps: geometric distances through parallaxes or detached eclipsing binaries, calibration of intermediate distance indicators (Cepheid stars, Tip of the Red Giant Branch, Mira variables), and the subsequent calibration of luminous objects in the Hubble flow (SNe~Ia, galaxies, etc.). Various recent versions of the local distance ladder agree with the Cepheid-SN~Ia analysis and point to a higher value of $H_0$. Replacing Cepheids by distances to Mira variables for the second rung \citep{Huang2020,Huang2024} or SNe~Ia by surface brightness fluctuations \citep{Blakeslee2021} or Tully-Fischer distances \citep{2020AJ....160...71S,2023ApJ...944...94T} for the third rung yield similar $H_0$ values.

A notable exception to this trend was reported by \citet{CCHP_H0,Freedman2020} and \cite{Freedman2021}, who used the Tip of the Red Giant Branch (TRGB) technique for the second rung and derived a Hubble constant ($H_0 = 69.8 \pm 1.71$ km s$^{-1}$ Mpc$^{-1}$) consistent with both the CMB and SH0ES. Other, independent TRGB analyses, however, have yielded $H_0$ values closer to SH0ES (e.g., \citealt{Anand2022,2023ApJ...954L..31S,2023arXiv230801875U}).

\cite{Freedman2024} presented an analysis based on JWST observations, which also found a lower $H_0$ value ($69.96 \pm 1.53$ km s$^{-1}$ Mpc$^{-1}$) by combining estimates from Cepheids, the J-band Asymptotic Giant Branch method (JAGB, \citealt{Madore2020}), and the TRGB, all applied along the same distance ladder with identical SN~Ia calibrator hosts. Shortly afterward, the SH0ES team presented an estimate based on an extended set of calibrator hosts observed by JWST, finding $H_0 = 72.6 \pm 2.0$ km s$^{-1}$ Mpc$^{-1}$ \citep{Riess2024}, and suggesting that selection effects in calibrator host galaxies explain the discrepancy.

These inconsistencies highlight the importance of assessing systematic effects in the distance ladder. For example, the effect of varying metallicity on the Cepheid \citep{2022ApJ...939...89B} and TRGB distances \citep{2023NatAs...7..590H,2024ApJ...961..166M} is frequently revisited in the literature. Furthermore, it was shown that the treatment of reddening can also cause systematic offsets in the calibration of the Cepheid period-luminosity relation \citep[e.g.,][]{2022ApJ...933..212M,2022ApJ...935...58M}, just as stellar variability may cause variations in the TRGB absolute brightness \citep{Anderson2024, 2024arXiv240619375K}. 

To resolve the ongoing Hubble tension, it is essential to address any systematic effects or find methods with independent systematics. 
A distance indicator bypassing the calibration through a distance ladder and based on known physics is ideal for a cross-check with independent systematics.
Three promising methods to measure individual distances in the Hubble flow have recently been discussed: megamasers around galaxy nuclei, which provide a geometric distance (e.g., \citealt{2019ApJ...886L..27R}), time delay cosmography of lensed quasars or SNe (e.g., \citealt{2017MNRAS.468.2590S}), and the expanding photosphere method with type II supernovae (SNe~II). All methods currently suffer from small number statistics and have individual challenges. 

\cite{Pesce2020} analyzed a sample of five masers. Apart from the currently small sample size, megamasers have not been observed far into the Hubble flow and hence suffer from high uncertainties due to peculiar velocities \citep{Pesce2020}. 

Time-delay cosmography has been performed with eight lensed quasars \citep{2019MNRAS.490.1743C,Wong2020,2020MNRAS.494.6072S,2020A&A...639A.101M,TDCOSMOH0,2024A&A...689A.168W}. Individual distances can be determined to about \SI{4}{\percent} in the best cases.
\cite{2023Sci...380.1322K}, \cite{Grillo2024} and \citet{2024arXiv240318902P} estimated $H_0$ based on SNe lensed by galaxy clusters, while efforts are ongoing to perform similar measurements using SNe behind galaxy-scale lens systems \citep[see, e.g.][]{HOLISMOKES_I}. Time delay lensing measures distances well beyond the local universe (source and lens redshifts are typically at $z \gtrsim 0.5$), and depends to some degree on the assumed cosmological model \citep[e.g.,][]{2019A&A...628L...7T}. The $H_0$ uncertainties depend significantly on the assumed lens density profile, which is required to break the mass-sheet degeneracy. 
Without such assumptions, the $H_0$ uncertainties increase to $\sim$\SI{8}{\percent} for a sample of seven lenses \citep{TDCOSMOH0}, unless spatially-resolved stellar kinematics of the lens galaxies are available \citep[e.g.,][]{2023A&A...673A...9S,2023A&A...675A..21Y}.

Type II supernovae, resulting from the collapse of massive ($\geq 8$ $M_{\odot}$) hydrogen-rich stars, have a long history as distance indicators. Two main techniques are currently used, as reviewed by \cite{deJaegerGalbanyReview}.
The standardizable candle method \citep[SCM;][]{SCMFirstPaper} correlates the luminosity during the light curve plateau phase with the expansion velocity. The method needs to be calibrated with other distance indicators (e.g., Cepheids) and is part of the distance ladder \citep[e.g.,][]{2020MNRAS.496.3402D,deJaeger2022}.

The expanding photosphere method \citep[EPM;][]{KirsherKwan74} relies on a physical model for the photospheric emission and the ejecta’s expansion to relate observed flux to intrinsic luminosity. The EPM provides direct distances to SNe~II without the need for calibration.

Hubble constant determinations with the EPM have been attempted over several decades  \citep[see e.g.][]{1992ApJ...395..366S, Schmidt1994,1996ApJ...466..911E, Gall2016, 2024ApJ...962...60D}.
The first EPM $H_0$ estimate was based on two SNe \citep{KirsherKwan74} and assumed that the SN radiates as a blackbody -- a natural first approximation. However, \citet{1981ApJ...250L..65W} later demonstrated that this assumption did not capture the full complexity of the emission process.
The radiation continuum is formed far below the photosphere due to the dominance of scattering opacity, leading to a diluted continuum. The dilution factor $\xi$ was introduced to incorporate the deviations from blackbody radiation. Non-LTE models of \cite{Eastman1996} and \cite{Dessart2005} have shown this parameter to depend primarily on the temperature and provided the necessary tables of dilution factors, improving the distance estimation accuracy.
\cite{Schmidt1994} were the first to estimate $H_0$ with the improved EPM. \citet{Gall2016} and \citet{Gall2018} demonstrated that this method can be used at redshifts up to $z = 0.3$. 

Tabulated dilution factors present a clear limitation for the EPM. Dilution factors can vary considerably as the SN luminosity depends not only on temperature but, for example, also on the ejecta density profile. This limitation can lead to systematic biases in individual distance estimates if the models used for dilution factors systematically differ from the actual properties of the observed SNe. 
In fact, differences in the ejecta density structure of the underlying SN models are one of the reasons for the significant discrepancies between the dilution factors computed by \citet{Eastman1996} and \citet{Dessart2005} \citep[see][]{Vogl2019}.

\citet{Dessart2006} emphasized the need to find radiative transfer models that reproduce the SN spectrum at each epoch. By matching models to observations, one ensures that the properties influencing the luminosity, including the density profile, are accurately captured. This approach is known as the tailored EPM \citep{Dessart2006} or the spectral-fitting expanding atmosphere method (SEAM; \citealt{Baron2004}). 
While this method improves the accuracy of distance measurements, it is highly time-consuming. Finding suitable models requires calculating many complex radiative transfer simulations, each demanding significant computational resources.
Only three SN~II spectral time series have thus been modeled with the goal of measuring distances \citep{Baron2004, Dessart2006, Dessart2008}. To remedy this situation, we introduced a spectral emulator approach showcased in \cite{Vogl2020}, capable of interpolating radiative transfer models within a pre-defined parameter space in a fraction of the time needed to run a single radiative transfer model. These interpolated models can then be used as an input for distance measurements with the tailored EPM.

Another improvement for the EPM was demonstrated by \cite{Vogl2020}: with a precise estimate of the time of explosion, the EPM can be applied using even a single spectral epoch to constrain the distance with $10-20\,\%$ precision (or better), as each individual epoch can be fitted for luminosity. 

The power of the tailored EPM was tested by applying it on four SN~II siblings -- SNe~II which exploded in the same galaxy -- yielding encouraging consistency \citep{GezaSiblings}. Tailored EPM distance estimates are as precise as state-of-the-art techniques, such as Cepheids and TRGB, and yield similar results \citep{Vogl2020, GezaM51}. Using the tailored EPM, SNe~II are capable of estimating distances in the Hubble flow and provide an independent $H_0$ value with competitive precision.

We present the first $H_0$ estimate using the tailored EPM for SNe~II, building on the improvements of \cite{Dessart2006} and \cite{Vogl2020}. 
We utilize literature datasets not specifically designed for the tailored EPM and demonstrate that even they can provide the basis for a precise $H_0$ estimation.
We perform this analysis as a proof-of-principle to lay the foundation for future studies with dedicated datasets optimized for distance measurements. 

This paper is structured as follows: \Cref{sec:basics} introduces the theoretical framework for the tailored EPM and the application to measure $H_0$. \Cref{sec:obs_data} describes the spectroscopic and photometric data. In
\cref{sec:analysis_steps}, we describe the key analysis steps, including time-of-explosion determination, light curve interpolation, spectral calibration, and fitting of radiative transfer models. \Cref{sec:theta/v} connects these steps to derive the ratios of the photospheric angular diameter and photospheric velocities, as well as the extinction, which is crucial for calculating distances.  We determine $H_0$ through a joint fit of these ratios in \cref{sec:bayes_ensemble_epm}. The results are then discussed in \cref{sec:discussion}, comparing them with other $H_0$ measurements, exploring systematic uncertainties, and assessing the implications for the ongoing $H_0$ tension. Finally, \cref{sec:summary} provides a summary and suggests directions for future research.

\section{Basic principle} \label{sec:basics}
\subsection{Expanding photosphere method} \label{sec:epm_basics}
Like other methods for determining luminosity distances, the EPM relates the observed specific flux to the object's intrinsic specific luminosity.
The relevant equation in the absence of extinction is \citep[e.g,][Eq. 23]{1999astro.ph..5116H}:
\begin{equation} \label{eq:f_obs}
	f_{\lambda,\mathrm{obs}} = \frac{L_{\lambda_\mathrm{em}}}{4 \pi D_\mathrm{L}^2} \frac{1}{1 + z}.
\end{equation}
Due to the cosmic redshift $z$, the observed specific flux $f_{\lambda,\mathrm{obs}}$ at observed wavelength $\lambda$ relates to the luminosity $L_{\lambda_\mathrm{em}}$ at the emitted wavelength $\lambda_\mathrm{em} = \lambda / (1 + z)$. We can solve \cref{eq:f_obs} for the luminosity distance $D_\mathrm{L}$ if we can constrain the SN luminosity.

Assuming spherical symmetry, we express the luminosity in terms of the radius of the emitting region -- the photospheric radius $R_\mathrm{ph}$ -- and the specific flux at that radius $f_{\lambda_\mathrm{em}, \mathrm{ph}}$:
\begin{equation} \label{eq:L_lambda}
	L_{\lambda_\mathrm{em}} = 4 \pi R_\mathrm{ph}^2 f_{\lambda_\mathrm{em}, \mathrm{ph}}.
\end{equation}
We can determine $f_{\lambda_\mathrm{em}, \mathrm{ph}}$
by modelling the SN emission based on spectroscopic observations. The photospheric radius can be calculated from the photospheric velocity $v_\mathrm{ph}$ using the assumption of homologous expansion:
\begin{equation} \label{eq:r_ph}
	R_\mathrm{ph}(t) \approx v_\mathrm{ph}(t) \cdot \frac{t - t_0}{1 + z}.
\end{equation}
The photospheric radius at time $t$ in the observer frame is thus the photospheric velocity multiplied by the time since the explosion in the SN frame, $(t - t_0) / (1 + z)$ \citep[e.g.,][]{1996ApJ...466L..21L,2008ApJ...682..724B}. It is assumed that, after the prompt acceleration in the explosion, the ejected material moves at a constant velocity for each mass coordinate; the initial radius of the ejected material is considered to be negligible.

The photospheric velocity, finally, can be inferred from Doppler-broadened P-Cygni features. If we knew the time of explosion $t_0$, we could now directly solve \cref{eq:f_obs,eq:L_lambda} for the luminosity distance. In practice, however, the EPM involves an additional step since historically the time of explosion was not well-constrained for most objects.

We circumvent this lack of knowledge by calculating the ratio of $R_\mathrm{ph}$ and $D_\mathrm{L}$ as
\begin{equation} \label{eq:theta}
	\theta = \frac{R_\mathrm{ph}}{D_\mathrm{L}} = \sqrt{\frac{f_{\lambda,\mathrm{obs}}}{f_{\lambda_\mathrm{em}, \mathrm{ph}} / (1 + z)}},
\end{equation}
which is commonly called the photospheric angular diameter $\theta$. This name is misleading, however, because the true angular
diameter is calculated using the angular diameter distance
$D_\mathrm{A}$, not the luminosity distance $D_\mathrm{L}$.\footnote{Nevertheless, most studies applying the EPM at non-negligible redshifts use the definition of the photospheric angular diameter $\theta$ given in \cref{eq:theta} \citep[e.g.][]{Schmidt1994,Jones2009}. One exception is \citet{Gall2016} who introduce $\theta^\dag$ as a new name for the quantity to distinguish it from the true photospheric angular diameter.}

\begin{figure}
	\begin{center}
		\includegraphics[]{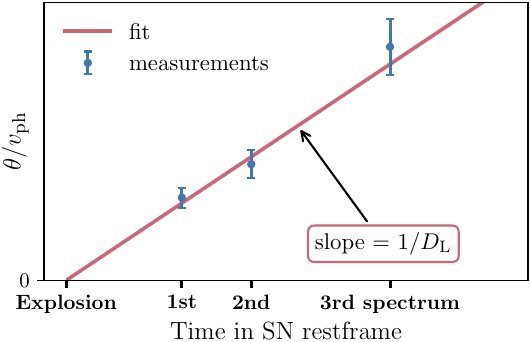}
	\end{center}
	\caption[]{EPM regression principle. The EPM uses multiple spectroscopic observations to measure the ratio of the photospheric angular diameter $\theta$ (Eq. \ref{eq:theta}) and the photospheric velocity $v_\mathrm{ph}$ for different times. These measurements (shown in blue) fall on a straight line (red). We can determine the SN luminosity distance $D_\mathrm{L}$ from the inverse of the slope of this line and the time of explosion from the intercept (see Eq. \ref{eq:theta/v}).}
	\label{fig:cartoon_EPM}
\end{figure}

Finally, we divide the photospheric angular diameter by the photospheric velocity. Using \cref{eq:r_ph} and the definition of $\theta$ we see that this ratio grows linearly with time starting from zero at the time of explosion. The rate of growth is inversely proportional to the distance $D_\mathrm{L}$:
\begin{equation} \label{eq:theta/v}
	\frac{\theta}{v_\mathrm{ph}} = \frac{1}{D_\mathrm{L}} \cdot \frac{t - t_0}{1 + z}.
\end{equation}
If we have measurements of $\theta / v_\mathrm{ph}$ from multiple spectroscopic observations, we can thus fit a straight line to the temporal evolution; the time of explosion is then the intercept of the line and the distance is the inverse of the slope. \Cref{fig:cartoon_EPM} illustrates this basic principle of the EPM.

While the general idea is simple, the details of the implementation are complex. We need measurements of the observed specific flux $f_{\lambda,\mathrm{obs}}$, the photospheric velocities $v_\mathrm{ph}$, and models for the specific flux at the photosphere $f_{\lambda_\mathrm{em}, \mathrm{ph}}$ to calculate $\theta / v_\mathrm{ph}$ values.

Out of the three, only the observed specific flux is straightforward to determine:
we simply interpolate the available photometry to the epochs of spectroscopic observations (see \cref{sec:lc_interpolation}). The other two ($v_\mathrm{ph}$ and $f_{\lambda_\mathrm{em}, \mathrm{ph}}$), for the tailored EPM, come from sophisticated radiative transfer models that are optimized to match the spectroscopic observations (\cref{sec:spectral_fits}).

\subsection{From distances to the Hubble constant} \label{sec:h0_basics}
We use the kinematic expansion of the luminosity distance $D_\mathrm{L}$  \citep[see, e.g.,][]{Riess2004} to relate our EPM distances to the Hubble constant $H_0$:
\begin{equation} \label{eq:DL}
	D_\mathrm{L} = \frac{c z}{H_0} \left[1 + \frac{1}{2} (1 - q_0) z  -  \frac{1}{6} (1 - q_0 - 3q_0^2 + j_0) z^2 + O(z^3)\right].
\end{equation}
We adopt a deceleration parameter $q_0 = -0.55$ and jerk $j_0 = 1$ as in \citet{Riess2016,Riess2022}, corresponding to a flat $\Lambda$CDM cosmology with $\Omega_\mathrm{M} = 0.3$. These values can be measured from high-redshift SNe Ia without requiring an absolute calibration of the SNe through a distance ladder \citep[e.g.,][]{2014A&A...568A..22B}.

The choice of $q_0$ and $j_0$, however, is not critical because we work at relatively low redshifts ($\lesssim 0.04$), where the nonlinear terms in \cref{eq:DL} are $\lesssim \SI{3}{\percent}$. For comparison, in their SN~Ia sample with a higher mean redshift of \num{0.07}, \citet{Riess2016} find that the uncertainty in $ q_0 $ only introduces a \SI{0.1}{\percent} uncertainty in $ H_0 $.

The final challenge is that the redshift $z$ in \cref{eq:DL} is the true cosmological redshift of the SN host galaxy $z_\mathrm{cosmo}$, not the measurable heliocentric redshift. They differ due to peculiar motions of our galaxy and the SN host galaxy \citep[e.g.,][]{2014MNRAS.442.1117D}. We can correct for our own motions by transforming the redshift to the CMB rest frame ($z_\mathrm{CMB}$) using the well-measured CMB dipole parameters \citep[e.g.,][]{Planck2018Overview}.
Assuming that we know the host galaxy peculiar velocity $v_\mathrm{pec}$, $z_\mathrm{cosmo}$ is then given by \citep[see, e.g.,][]{2014MNRAS.442.1117D}:
\begin{equation} \label{eq:z_cosmo}
	z_\mathrm{cosmo} = \frac{1 + z_\mathrm{CMB}}{1 + v_\mathrm{pec} / c} - 1.
\end{equation}
The challenge lies in determining $v_\mathrm{pec}$, which is much harder to quantify than our own peculiar motion.
However, observations of big samples of objects can be used to model the large scale flows in the universe. These models then provide approximate peculiar velocities for individual galaxies (see \cref{sec:flow_corrections}).

\section{Observational data} \label{sec:obs_data}
\begin{table*} 
\begin{center}
\caption[Supernova sample for $H_0$ measurement]{Supernova sample for $H_0$ measurement.}
\setlength\tabcolsep{4.5pt}
\small
\begin{tabular}{lccccccc}
\toprule
Name & Host galaxy &  Nondetection & First detection & $z_\mathrm{helio}$ \tablefootmark{a}&  $z_\mathrm{CMB}$ & $v_\mathrm{pec}$\tablefootmark{b} & $\mathrm{E}(B-V)$ \\
&&UT&UT&&&[km/s]&(MW)\tablefootmark{c}
\\
\midrule
\midrule
\multicolumn{8}{c}{\textbf{Literature}} \\
\midrule
\href{https://www.wis-tns.org/object/2003bn}{SN 2003bn} & \href{https://ned.ipac.caltech.edu/byname?objname=2MASX%20J10023529-2110531&hconst=67.8&omegam=0.308&omegav=0.692&wmap=4&corr_z=1}{2MASX J10023529-2110531} & Feb 21.00 & Feb 22.33 & 0.01277 &0.01393& $-$90 & \href{https://irsa.ipac.caltech.edu/cgi-bin/DUST/nph-dust?locstr=SN+2003bn}{0.0562}\\ 

\href{https://www.wis-tns.org/object/2006it}{SN 2006it} & 
\href{https://ned.ipac.caltech.edu/byname?objname=NGC%206956&hconst=67.8&omegam=0.308&omegav=0.692&wmap=4&corr_z=1}{NGC 6956} &Sep 26.19&Oct 1.17& 0.01550& 0.01449 & $-$127 & \href{https://irsa.ipac.caltech.edu/cgi-bin/DUST/nph-dust?locstr=SN+2006it}{0.0850}\\

\href{https://www.wis-tns.org/object/2010id}{SN 2010id} & \href{https://ned.ipac.caltech.edu/byname?objname=NGC%207483&hconst=67.8&omegam=0.308&omegav=0.692&wmap=4&corr_z=1}{NGC 7483} &Sep 11.34 &Sep 15.24& 0.01648 &0.01523& $-$180 &\href{https://irsa.ipac.caltech.edu/cgi-bin/DUST/nph-dust?locstr=SN+2010id}{0.0521}\\ 

\href{https://www.wis-tns.org/object/2013fs}{SN 2013fs} & \href{https://ned.ipac.caltech.edu/byname?objname=NGC%207610&hconst=67.8&omegam=0.308&omegav=0.692&wmap=4&corr_z=1}{NGC 7610} &Oct 5.34 &  Oct 6.25& 0.01186 &0.01061& $-$68 & \href{https://irsa.ipac.caltech.edu/cgi-bin/DUST/nph-dust?locstr=SN+2013fs}{0.0346}\\ 

\midrule
\midrule
\multicolumn{8}{c}{\textbf{ePESSTO+}} \\
\midrule

\href{https://www.wis-tns.org/object/2019luo}{SN 2019luo} & \href{https://ned.ipac.caltech.edu/byname?objname=MCG+-01-01-040}{MCG -01-01-040} & Jul 13.46  &  Jul 16.48 & 0.04028 & 0.03908 & $-$144 & \href{https://irsa.ipac.caltech.edu/cgi-bin/DUST/nph-dust?locstr=SN+2019luo}{0.0277}\\

\href{https://www.wis-tns.org/object/2020bad}{SN 2020bad} & \href{https://ned.ipac.caltech.edu/byname?objname=WISEA+J092426.48%2B203802.9}{WISEA J092426.48+203802.9} & Jan 21.48  & Jan 23.40 & 0.02711 & 0.02811 & +62 & \href{https://irsa.ipac.caltech.edu/cgi-bin/DUST/nph-dust?locstr=SN+2020bad}{0.0376}\\ 

\href{https://www.wis-tns.org/object/2020cvy}{SN 2020cvy} & \href{https://ned.ipac.caltech.edu/byname?objname=SDSS+J080049.73%2B273001.5}{IC 2217} & Feb 16.20 & Feb 19.19 & 0.01734 & 0.01801 & $-$202 &\href{https://irsa.ipac.caltech.edu/cgi-bin/DUST/nph-dust?locstr=SN+2020cvy}{0.0350}\\ 

\href{https://www.wis-tns.org/object/2021gvv}{SN 2021gvv} & \href{https://ned.ipac.caltech.edu/byname?objname=WISEA+J170435.95%2B140450.4}{WISEA J170435.95+140450.4} & Mar 20.58 & Mar 22.48 & 0.03789 & \phantom{\tablefootmark{d}}0.03789 \tablefootmark{d}& $-$134 &\href{https://irsa.ipac.caltech.edu/cgi-bin/DUST/nph-dust?locstr=SN+2021gvv}{0.0987}\\ 

\href{https://www.wis-tns.org/object/2021hkf}{SN 2021hkf} & \href{https://ned.ipac.caltech.edu/byname?objname=CGCG+040-036}{CGCG 040-036} & Mar 25.40 & Mar 29.26 & 0.01957 & 0.02077 & +256 &\href{https://irsa.ipac.caltech.edu/cgi-bin/DUST/nph-dust?locstr=SN+2021hkf}{0.0203} \\

\href{https://www.wis-tns.org/object/2021acvc}{SN 2021acvc} & \href{https://ned.ipac.caltech.edu/byname?objname=WISEA+J053038.62-225410.2}{WISEA J053038.62-225410.2} & Oct 26.52 & Oct 28.52 & 0.03142 & 0.03157 & +178 &\href{https://irsa.ipac.caltech.edu/cgi-bin/DUST/nph-dust?locstr=82.6588041758+-22.900337229+equ+j2000}{0.0329}\\

\bottomrule
\label{tab:H0_measurement_sample}
\end{tabular}    
\tablefoot{
\tablefoottext{a}{From NED with the exception of SN~2021gvv, where we have measured the redshift from narrow host emission lines. For SN~2021hkf, we use a redshift of \num{0.0208} as measured from narrow lines in the SN spectrum for the heliocentric redshift correction, while the galaxy redshift listed in the table is used in the determination of the Hubble constant. This is necessary because the redshift measured at the SN's position deviates significantly from that of the galaxy center due to the SN being located far from the center in a rotating spiral galaxy. It is the only case in our sample where such a significant mismatch occurs.}
\tablefoottext{b}{The peculiar velocities of the SN host galaxies are calculated as described in \cref{sec:flow_corrections}.}
\tablefoottext{c}{Milky Way (MW) color excess taken from \citet{Schlafly2011}.}
\tablefoottext{d}{The sky position of SN~2021gvv is almost exactly orthogonal to the direction of the CMB dipole.}

}
\end{center}
\end{table*}
\subsection{Sample selection}
The data for a tailored EPM measurement of $H_0$ have to meet very specific requirements.
First, the SNe must have significant redshifts to reduce the impact of peculiar velocities, which remain important even after flow corrections. We only consider SNe at $z_\mathrm{CMB} > 0.01$, where peculiar velocities contribute at most a \SI{10}{\percent} uncertainty in the $H_0$ measurement from a single SN.\footnote{This assumes a residual cosmological redshift uncertainty of \SI{250}{\km\per\second} after flow corrections, similar to the Pantheon$+$ analysis \citep{Pantheon+Cosmo,PetersonPantheonPecV}.}

Second, the SNe also need tight constraints on the time of explosion $t_0$ (see \cref{sec:time_explosion} for details on this constraint). By excluding SNe with $t_0$ uncertainties over two days, we limit the contribution to the distance uncertainties to about \SI{10}{\percent} at a representative phase of \SI{20}{\day}.
The tight $t_0$ constraints have the added benefit that we can
even use objects with only one suitable spectrum.

Third, for a spectrum to be suitable for our purposes, it must meet several criteria. It must be taken within \SI{35}{\day} post-explosion, have accurate relative flux calibration, minimal host contamination, and be sufficiently normal to be modeled with a standard SN~II atmosphere without, for example, accounting for circumstellar material (CSM) interaction.
The \SI{35}{\day} limit ensures that neglecting time dependence in the excitation and ionization balances remains a sound approximation, as this assumption becomes less accurate over time \citep[see discussion in][]{Vogl2019}.
The other criteria help ensure that the spectral modelling results are physical and less likely to be influenced by limitations in the data or the modelling.

Finally, we require photometry in at least two bands to recalibrate the spectra against the photometry (see \cref{sec:flux_calibration}).

While many objects with data published in the literature meet some requirements, very few pass all. Our search of WISeREP \citep{Yaron2012}\footnote{\url{https://www.wiserep.org}} and the Open Supernova Catalog \citep[OSC;][]{2017ApJ...835...64G}\footnote{\url{https://github.com/astrocatalogs/supernovae}} yields only four SNe: SN~2003bn, SN~2006it, SN~2010id, and SN~2013fs.\footnote{We cannot exclude that we missed some objects because the relevant information was not available on WISeREP and the OSC. This applies, in particular, to the early light curve data needed for the time of explosion constraints.} Out of these, SN~2013fs is a borderline case:
its early evolution is marked by CSM interaction \citep{Yaron2017} and it becomes spectroscopically normal (according to our earlier description) only around three weeks post-explosion. We include it as a tentative test whether objects with early CSM interaction can be reliable distance indicators with our approach.

We extend our sample by including SNe without published follow-up data, but with publicly available classification spectra from the advanced Public ESO Spectroscopic Survey for Transient Objects (ePESSTO+) -- the successor of the PESSTO program \citep{2015A&A...579A..40S}. The project classifies transients with apparent magnitudes up to \num{19.5}, which includes normally bright SNe~II up to redshifts of around \num{0.04}. The classified objects are usually young with well-constrained explosion epochs, making them suitable for the EPM despite having only one spectrum.

\cref{tab:H0_measurement_sample} summarizes the properties of our combined sample of literature objects and ePESSTO+ classification targets including redshifts, peculiar velocities (see \cref{sec:flow_corrections}), and Milky Way extinction values.

\subsection{Spectroscopy} \label{sec:spectroscopy}
We retrieved the spectra for the literature sample from WISeREP and the Open Supernova Catalog. \cref{tab:interp_phot} lists their original sources.

We have re-reduced the publicly available ePESSTO+ raw data to ensure uniform data quality from this source. This included flat-fielding, cosmic ray rejection with L.A.Cosmic \citep{2001PASP..113.1420V}, and an optimal, variance-weighted extraction with IRAF’s \citep{1986SPIE..627..733T,1993ASPC...52..173T} \textit{apall} task. The wavelength calibration was performed using arc lamps and verified against night sky lines. 

Given the blue colors of early-phase SNe~II, second order contamination in  EFOSC2 \citep{1984Msngr..38....9B} was a concern. We thus corrected the ePESSTO+ spectra for it using an adapted version of the method of \citet{Stanishev2007}.

For the flux calibration, we constructed sensitivity curves from multiple standard star observations taken close in time to the SN data. We calibrated the spectra using the mean sensitivity curve.
Afterward, we applied a telluric absorption correction using a mean correction function derived from standard star spectra. This was done with the IRAF task \textit{telluric}, which allowed us to fine-tune the correction by adjusting the wavelength shift and scaling the absorption strengths of the $\text{O}_2$ and $\text{H}_2\text{O}$ bands independently.\footnote{The re-reduced ePESSTO+ classification spectra will be made available on the author’s GitHub upon paper acceptance.}

Our full set of spectra is summarized in \cref{tab:interp_phot}, with each row corresponding to one spectrum at the specified time (MJD). Before fitting the spectra with radiative transfer models in \cref{sec:spectral_fits}, we also convert the observed wavelengths from air to vacuum using the method described by \citet{1996ApOpt..35.1566C}.

\begin{sidewaystable*}
\centering
\caption{
        Log of modeled spectra and corresponding interpolated photometry.
    }
\label{tab:interp_phot} 
\begin{tabular}{@{\hskip4pt}l@{\hskip0.75pt}clcccccccccc@{\hskip3pt}}
\toprule
       \textbf{SN}   & &  \phantom{a} MJD & $B$ &   $V$ & $R$ &  $I$ & $g$ & $r$ & $i$ & Photo. & Ref & Ref \\
&&&&&&&&&&System&Photo.&Spec.\\ \midrule
\textbf{2003bn} & \multirow{13}{*}{{}} &  52710.30\tablefootmark{a} &  17.52\,$\pm$\,0.05 &  17.19\,$\pm$\,0.02 &  \textemdash &  16.92\,$\pm$\,0.01 &  \textemdash &  \textemdash &  \textemdash & Bessell & 1,2 &   7 \\
\cline{1-13}
\multirow{2}{*}{\textbf{2006it}} & \multirow{14}{*}{{}} & 54018.04 & 18.19\,$\pm$\,0.03 &  17.91\,$\pm$\,0.02 &  \textemdash &  \textemdash &  \textemdash &  17.85\,$\pm$\,0.01 &  17.91\,$\pm$\,0.02 &  CfA3$_\mathrm{KEP}$ &  3 &  7  \\
          & \multirow{13}{*}{{}} &  54020.99 & 18.28\,$\pm$\,0.02 &  17.92\,$\pm$\,0.01 &  \textemdash &  \textemdash &  \textemdash &  17.85\,$\pm$\,0.01 &  17.89\,$\pm$\,0.02 && & 7  \\
\cline{1-13}
\multirow{2}{*}{\textbf{2010id}} & \multirow{12}{*}{{}} &  55468.29 & 19.15\,$\pm$\,0.04 &  18.89\,$\pm$\,0.02 &  18.64\,$\pm$\,0.02 &  18.52\,$\pm$\,0.03 &  \textemdash &  \textemdash &  \textemdash &  KAIT4 &  4  & 8\\
          & \multirow{11}{*}{{}} &  55470.40 & 19.21\,$\pm$\,0.03 &  18.89\,$\pm$\,0.02 &  18.63\,$\pm$\,0.02 &  18.49\,$\pm$\,0.03 &  \textemdash &  \textemdash &  \textemdash & & & 4 \\
\cline{1-13}
\multirow{5}{*}{\textbf{2013fs}} & \multirow{8}{*}{{}} &  56591.82 & 16.65\,$\pm$\,0.01 &  16.41\,$\pm$\,0.02 &  16.17\,$\pm$\,0.01 &  15.97\,$\pm$\,0.01& 16.57\,$\pm$\,0.01 &  16.31\,$\pm$\,0.01 &  16.40\,$\pm$\,0.01 & Bessell, & 5 &   9 \\
          & \multirow{7}{*}{{}} &  56594.02 & 16.75\,$\pm$\,0.01 &  16.47\,$\pm$\,0.02 &  16.20\,$\pm$\,0.01 &  16.00\,$\pm$\,0.01 & 16.64\,$\pm$\,0.01 &  16.35\,$\pm$\,0.01 &  16.44\,$\pm$\,0.01 & Sloan && 9 \\
          & \multirow{6}{*}{{}} &   56598.72 & 16.95\,$\pm$\,0.01 &  16.56\,$\pm$\,0.02 &  16.24\,$\pm$\,0.01 &  16.05\,$\pm$\,0.01 & 16.78\,$\pm$\,0.01 &  16.40\,$\pm$\,0.01 &  16.48\,$\pm$\,0.01 &&& 9 \\
          & \multirow{5}{*}{{}} &   56600.16 &  17.02\,$\pm$\,0.01 &  16.58\,$\pm$\,0.02 &  16.24\,$\pm$\,0.01 &  16.06\,$\pm$\,0.01 & 16.82\,$\pm$\,0.01 &  16.42\,$\pm$\,0.01 &  16.49\,$\pm$\,0.02 & && 10 \\
          & \multirow{4}{*}{{}} &   56603.22 &  17.14\,$\pm$\,0.01 &  16.61\,$\pm$\,0.02 &  16.26\,$\pm$\,0.01 &  16.07\,$\pm$\,0.01 & 16.90\,$\pm$\,0.01 &  16.44\,$\pm$\,0.02 &  16.49\,$\pm$\,0.02 && &9 \\
\cline{1-13}
\textbf{2019luo} & \multirow{15}{*}{{}} & 58693.28 &  \textemdash &  \textemdash &  \textemdash &  \textemdash &  19.23\,$\pm$\,0.02 &  19.12\,$\pm$\,0.03 &  \textemdash & ZTF & 6 &  11  \\
\cline{1-13}
\textbf{2020bad} & \multirow{15}{*}{{}} & 58883.22 &  \textemdash &  \textemdash &  \textemdash &  \textemdash &  19.06\,$\pm$\,0.05 &  18.98\,$\pm$\,0.03 &  \textemdash & ZTF & 6 &   11 \\
\cline{1-13}
\textbf{2020cvy} & \multirow{15}{*}{{}} & 58905.10 &  \textemdash &  \textemdash &  \textemdash &  \textemdash &  18.71\,$\pm$\,0.01 &  18.54\,$\pm$\,0.04 &  \textemdash & ZTF & 6 &  11  \\
\cline{1-13}
\textbf{2021gvv} & \multirow{15}{*}{{}} & 59315.41 &  \textemdash &  \textemdash &  \textemdash &  \textemdash &  19.27\,$\pm$\,0.02 &  19.01\,$\pm$\,0.02 &  19.06\,$\pm$\,0.04 & ZTF & 6 &  11  \\
\cline{1-13}
\textbf{2021hkf} & \multirow{15}{*}{{}} & 59309.23 & \textemdash &  \textemdash &  \textemdash &  \textemdash &  18.50\,$\pm$\,0.01 &  18.44\,$\pm$\,0.01 &  18.63\,$\pm$\,0.03 & ZTF & 6 &  11  \\
\cline{1-13}
\textbf{2021acvc} & \multirow{15}{*}{{}} & 59524.24  & \textemdash &  \textemdash &  \textemdash &  \textemdash &  18.94\,$\pm$\,0.01 &  18.93\,$\pm$\,0.04 &  \textemdash & ZTF & 6 &  11  \\
\bottomrule
\end{tabular}
\tablefoot{
\tablefoottext{a}{Adopted from \citet{Jones2009}.}
}
\tablebib{
(1) \citet{Anderson2014}; (2) \citet{Galbany2016}; (3) \citet{Hicken2017}; (4) \citet{deJaeger2019}; (5) \citet{Valenti2016}; (6) \citet{ZTF,ztfforcedphot}; (7) \citet{Gutierrez2017b}; (8) \cite{Gal-Yam-2010id}; (9) \citet{Yaron2017}; (10) \citet{2015A&A...579A..40S}; (11) ePESSTO+ (re-reduced cf. \cref{sec:spectroscopy}).
}
\end{sidewaystable*}

\subsection{Photometry}\label{sec:photometry}
The EPM relies on photometric measurements of the observed specific flux $f_{\lambda, \mathrm{obs}}$ to calculate the photospheric angular diameter $\theta$ (Eq. \ref{eq:theta}) and recalibrate the spectra against the photometry (\cref{sec:flux_calibration}).
\cref{tab:interp_phot} lists the sources and photometric systems of our photometry.

We use the natural instrument system of the respective surveys whenever possible to avoid problems associated with first-order color term corrections.
The appropriate passbands are from \citet{2010ApJS..190..418G} for the KAIT4 and \citet{Hicken2009}\footnote{\url{https://lweb.cfa.harvard.edu/supernova/CfA3/keplercam.passbands.txt}} for the CfA3$_\mathrm{KEP}$ systems. We adopt the products of the CCD quantum efficiency (QE), the filter transmissions from \citet{Bellm2019}, and a standard atmospheric transmission at \num{1.3} airmass \citep{2010AJ....139.1628D} as our Zwicky Transient Facility (ZTF) passbands.\footnote{The filter transmissions and QE can be downloaded from \url{https://github.com/ZwickyTransientFacility/ztf_information}.}
We use the QE for the single-layer anti-reflective coating, which covers half of the detector with the other half being covered with a dual-layer coating. The coating choice, however, has only a minor impact on the constructed passbands.
\citet{Bessell2012} and \citet{2002AJ....123.2121S} finally provide the transmission curves for our standard system Bessell and Sloan photometry.

We retrieve the ZTF photometry from the ZTF forced-photometry service
\citep{ztfforcedphot} and process it according to the manual, including quality filtering, baseline correction, and validation of flux-uncertainty estimates.

\section{Analysis steps} \label{sec:analysis_steps}
With the basic principle and input data introduced, we now detail the steps to extract the EPM quantities from the data. This involves determining the time of explosion (\cref{sec:time_explosion}). Next, we interpolate light curves to match photometric data with spectral observation epochs (\cref{sec:lc_interpolation}), perform spectral flux calibration (\cref{sec:flux_calibration}), and fit the calibrated spectra to estimate photospheric velocities and specific fluxes (\cref{sec:spectral_fits}). Finally, in \cref{sec:flow_corrections} we apply flow corrections for peculiar velocities of the SN host galaxies to prepare for the $H_0$ measurement in \cref{sec:bayes_ensemble_epm}.
\subsection{Time-of-explosion fit} \label{sec:time_explosion}

The EPM does not require prior information about the time of explosion, but using it can significantly reduce uncertainties \citep[see, e.g.,][]{Gall2018}.
Including prior information on $t_0$ is also crucial for objects with only one spectrum, like our ePESSTO+ SNe, which would otherwise be unusable.
This information can be extracted from the early light curve, which encompasses the first few days to weeks following the explosion.

The most commonly used method calculates the time of explosion as the midpoint between the last non-detection and the first detection, with an uncertainty equal to the difference between the two \citep[e.g.][]{Gutierrez2017a}. While straightforward, this method ignores much of the information in the early light curve, such as the depth of the non-detection relative to the first detection and any additional points on the light curve rise.

To utilize all available constraints, we perform a Bayesian parametric fit to the early light curve, using an inverse exponential model as first introduced by \citet{Ofek2014} and \citet{Rubin2016} for SN~II light curves and later adopted by \citet{GezaM51,GezaSiblings} for EPM applications. 

The flux $f$ at time $t$ in our model is given by
\begin{equation} \label{eq:inv_exp_model}
	f(t) =
	\begin{cases}
		f_\mathrm{m} \left( 1 - \exp{\left(-\frac{t - t_0}{t_\mathrm{e}} \right)} \right) & \mathrm{if}\,\, t \geq t_0 \\
		0                                                                                 & \mathrm{if}\,\, t<t_0      \\
	\end{cases}
	,
\end{equation}
where $f_\mathrm{m}$ represents the maximum flux and $ t_\mathrm{e}$ denotes the timescale for the exponential rise.
This model typically describes the rise very well but
cannot fully capture the complete light curve evolution. Thus, we only fit the data in the rising part of the light
curve and the immediate few days to weeks after the rise.

Our fitting process involves four parameters: the three parameters of the light curve model ($f_\mathrm{m}$, $ t_\mathrm{e}$, and $t_\mathrm{0}$) and an additional parameter $\sigma_\mathrm{add}$, which is added in quadrature to the measured flux errors to account for underestimated uncertainties and imperfections in the light curve model.\footnote{The process of inferring an unknown uncertainty from the data themselves is motivated, for example, in \citet{2010arXiv1008.4686H}.} When photometry is available in multiple bands, we fit them jointly using a shared value for $t_0$ but different values for $f_\mathrm{m}$, $t_\mathrm{e}$, and $\sigma_\mathrm{add}$.

Furthermore, we incorporate that the rise time increases with the effective wavelength of the passband, as demonstrated by, for example, \citet{GonzalezGaitan2015}, consistent with expectations for shock-heated cooling. For instance, when fitting $g$ and $r$ band data, we impose $t_{\mathrm{e}, g} < t_{\mathrm{e}, r}$.

We apply a uniform prior for $t_0$, a uniform positive prior for $f_\mathrm{m}$, and a standard log-uniform prior for $\sigma_\mathrm{add}$. Because the exponential time scale $t_\mathrm{e}$ is not an intuitive parameter, we express our prior in terms of the rise time $t_\mathrm{rise}$.\footnote{As in \citet{Gall2015_RiseTime} and \citet{GonzalezGaitan2015}, we define the rise time as the time it takes for the magnitude increase to fall below 0.01 mag/day.} We adopt a flat prior for $t_\mathrm{rise}$ with a minimum value of \SI{2}{\day}, corresponding to the fastest rising SNe from \citet{GonzalezGaitan2015}.

We determine the time of explosion along with the other parameters by sampling the posterior with \texttt{nestle}\footnote{\url{https://github.com/kbarbary/nestle}} -- a Python implementation of the nested sampling algorithm using a Gaussian likelihood.

\begin{figure}
	\begin{center}
		\includegraphics[]{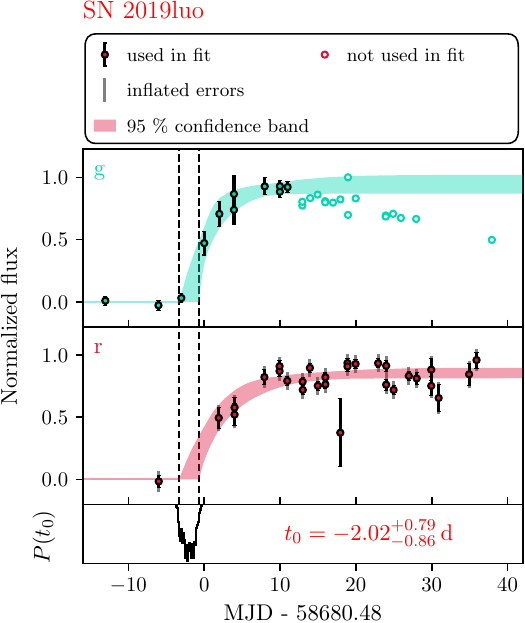}
	\end{center}
	\vspace*{-0.5cm}
	\caption[]{Example of the time-of-explosion determination. We fit an inverse exponential model (Eq. \ref{eq:inv_exp_model}) to the observed flux curves (black error bars), modelling the available bands (here $g$ and $r$) jointly with a shared explosion time $t_0$. The colored bands (cyan for $g$, red for $r$) show the \SI{95}{\percent} confidence regions of the fits. The bottom panel displays the $t_0$ posterior along with the inferred value relative to the first detection and its uncertainty. Finally, the inflated flux errors, including the additional fitted uncertainty $\sigma_\mathrm{add}$, are shown in grey.
		Here, the reported $g$-band uncertainties accurately capture the scatter around the model, resulting in minimal error inflation. In contrast, the $r$-band errors show noticeable inflation, with $\sigma_\mathrm{add}$ around \SI{0.06}{mag}.}
	\label{fig:lc_fit_t0}
\end{figure}

\cref{fig:lc_fit_t0} shows an example of the resulting fits, with plots for the remaining objects provided in \cref{app:t0}. We summarize the derived $t_\mathrm{0}$ values in \cref{tab:lc_fit_t0}, including information on the sources of the photometric data used.
We obtain very tight constraints on the time of explosion, narrowing it down to less than one day for all SNe except SN~2006it. In most cases, we find that $\sigma_\mathrm{add}$ is close to zero. The largest value is observed for SN~2019luo, where $\sigma_\mathrm{add}$ reaches around \SI{0.06}{mag}.

\begin{table}
\begin{center}
\caption{Time-of-explosion fit data and results.}
\addtolength{\tabcolsep}{-3pt} 
\begin{tabular}{@{}lcccc@{}}
\toprule
{} SN &                $t_0\,$[MJD] &  Surveys & Bands & Ref \\
\midrule
\textbf{2003bn} & $52692.21^{+0.08}_{-0.10}$  & NEAT, CATS & clear + $V$\tablefootmark{a} & 1,2 \\
\textbf{2006it} & $54005.48^{+1.59}_{-1.15}$ & LOSS &  clear,$r$ & 3,4 \\
\textbf{2010id}  & $55452.81^{+0.49}_{-0.72}$ & PTF & $r$ & 5    \\
\textbf{2013fs} & $56571.05^{+0.04}_{-0.04}$ & iPTF & $r$  & 6  \\
\textbf{2019luo} & $58678.46^{+0.79}_{-0.86}$ & ZTF & $g,r$ & \NumZTF \\
\textbf{2020bad} & $58870.99^{+0.14}_{-0.16}$ & ZTF & $g,r$ & \NumZTF  \\
\textbf{2020cvy} & $58896.37^{+0.18}_{-0.11}$ & ZTF & $g,r$ & \NumZTF \\
\textbf{2021gvv} & $59294.14^{+0.34}_{-0.54}$ & ATLAS, ZTF & $g,r,o$ & \NumZTF,\NumAtlas\\
\textbf{2021hkf} & $59299.23^{+0.88}_{-0.65}$ & ZTF &  $g,r$ & \NumZTF \\
\textbf{2021acvc} & $59513.87^{+0.51}_{-0.39}$& ATLAS, ZTF &  $r$,$o$ & \NumZTF,\NumAtlas \\
\bottomrule
\label{tab:lc_fit_t0}
\end{tabular}
\addtolength{\tabcolsep}{3pt}
\tablefoot{
\tablefoottext{a}{We combine the deep non-detection (\SI{21.0}{mag}) and detection (\SI{20.2}{mag}) on unfiltered images \citep{Wood-Vasey2003}
with filtered photometry \citep{Anderson2014}. The very blue spectral energy distribution at early times suggests the effective wavelength of the unfiltered images corresponds to a blue band
leading us to adopt the $V$-band. Whether we combine the unfiltered data with $V$ or $B$ band photometry is inconsequential, however, given the significant magnitude difference between the first detection and the peak ($\sim$\SI{3.5}{mag}). Because the first detection does not have a reported uncertainty, we assume a significant error of \SI{0.5}{mag}.}
}
\tablebib{
(1) \citet{Anderson2014}; (2) \citet{Wood-Vasey2003}; (3) \citet{Lee2006_CBET06it}; (4) \citet{Hicken2017}; (5) \citet{Gal-Yam-2010id}; (6) \citet{Yaron2017}; (\NumZTF) \citet{ZTF,ztfforcedphot}; (\NumAtlas) \citet{Tonry2018,Smith2020,ATLASForcedPhot}.
}
\end{center}
\end{table}

Our methodology closely follows \citet{GezaM51,GezaSiblings} but with two key differences. We reformulate the prior for the exponential time scale $t_\mathrm{e}$ in terms of the rise time as discussed earlier and handle non-detections differently. While \citet{GezaM51,GezaSiblings} treated all non-detections as upper limits, with the model flux constrained to not exceed these limits, we use the actual flux measurements with their associated uncertainties, just like detections for the ePESSTO+ sample.
This approach improves constraints on $t_0$ by utilizing all available information whereas the upper limits ignore the actual measured flux.
For the literature sample, we continue using the old method for non-detections because we do not have the necessary flux values available.

\subsection{Light curve interpolation} \label{sec:lc_interpolation}
We use Gaussian processes \citep[GPs; e.g.,][]{Rasmussen06gaussianprocesses} to interpolate the photometry to the epochs of spectral observations, similar to \citet{Inserra2018,Yao2020, Kangas2022,GezaM51,GezaSiblings}.

We adopt a squared-exponential kernel to describe the smooth SN light curve and a white kernel to allow for additional uncertainties in the photometry compared to the reported values.
We model the mean of our GPs with a constant function.

The GP regression is done with the Python package \texttt{george}\footnote{\url{https://github.com/dfm/george}} \citep{2015ITPAM..38..252A}. To avoid overfitting and improve uncertainty estimates, we marginalize over the values of the GP hyperparameters instead of optimizing them with the more commonly used maximum likelihood
method.

We sample the hyperparameters with \texttt{nestle}, which we also used in the time-of-explosion fit (see \cref{sec:time_explosion}). Our prior choice is guided by the Stan User’s Guide\footnote{\url{https://mc-stan.org/docs/2_32/stan-users-guide-2_32.pdf}}:
we use an inverse-gamma distribution for the length scale and a half-normal distribution for the standard deviation of the squared-exponential kernel.

The inverse-gamma prior suppresses small and large length scales where the marginal likelihood function becomes flat.
We set its parameters so that only \SI{1}{\percent} of the prior probability is assigned to length scales below \num{10} or above \num{100} days.

The half-normal prior for the marginal standard deviation is centered at zero with a scale parameter $\sigma=\SI{1.5}{mag}$.
This assigns sufficient probability to small values such that the GP contribution can go to zero, for example, for a nearly flat $R$-band plateau light curve. At the same time it extends to large enough values to describe even the most steeply declining $B$-band light curves. 

A standard log-uniform distribution for the white-kernel standard deviation completes our hyperparameter priors. The final parameter is the value of our constant mean function, which has a uniform prior.

We exclude the rise and fall from the plateau from the fit whenever possible\footnote{For some objects, the spectral observations were taken during the light curve's rise, requiring us to use data points from this period.} because they bias the GP length scales to small values, making the fits too flexible during the plateau phase. 

\begin{figure*}
	\begin{center}
		\includegraphics[]{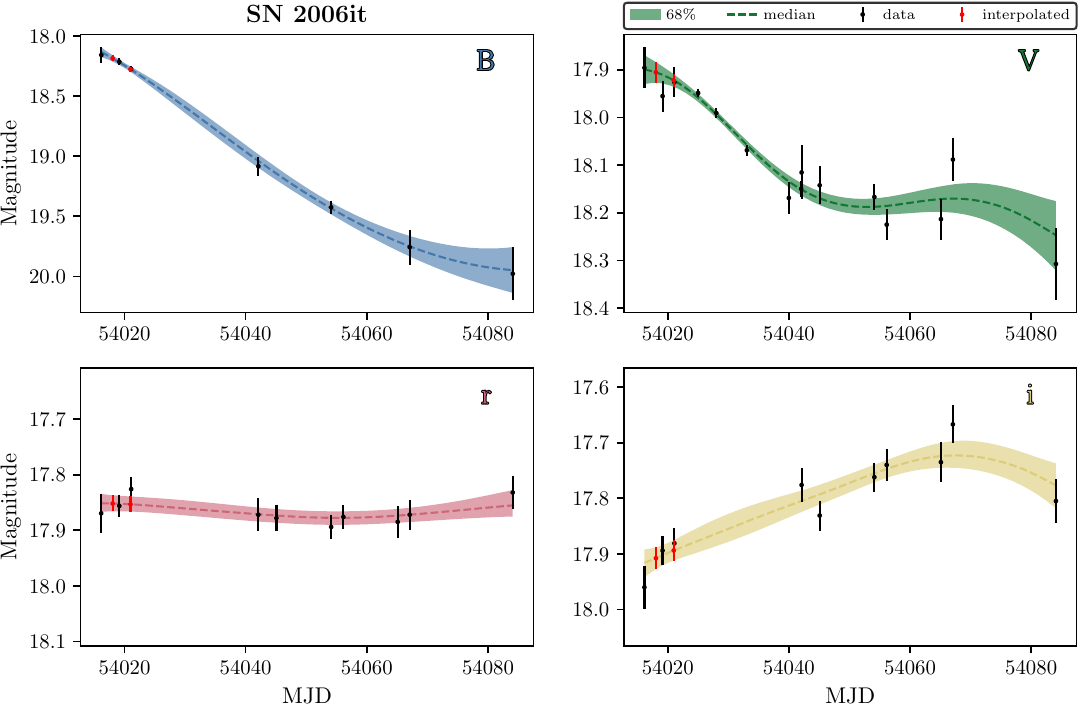}
	\end{center}
	\caption[]{Example of the GP interpolation of the photometry. We plot the observed magnitudes and their uncertainties in black; the interpolated values at the spectral epochs are shown in red. The colored bands, finally, indicate the \SI{68}{\percent} confidence interval of the interpolated light curve and the dashed line the median.}
	\label{fig:lc_interpolation}
\end{figure*}

Finally, we draw a large number of samples ($\sim$ \num[]{100000}) from the marginalized predictive probability distribution to obtain the photometry at the spectral epochs. \Cref{fig:lc_interpolation} shows an example of the interpolated light curves. Plots for the remaining SNe are in \cref{app:lc_interp}. 

The interpolated photometry is not Gaussian: it is a Gaussian mixture. However, since deviations from Gaussianity are small in our sample, we approximate the distributions as Gaussian to simplify the subsequent analysis. 

We list the mean and standard deviation of the interpolated magnitudes in \cref{tab:interp_phot}. The listed uncertainties are typically much smaller than those of the individual data points, despite the error inflation from the white kernel. This is because the GP assumes a smooth light curve, as encoded by the length scale, allowing us to average across multiple data points.

\subsection{Spectral flux calibration} \label{sec:flux_calibration}
Our spectroscopic observations are not spectrophotometric due to the limitations imposed by the slit width and seeing conditions. However, for our radiative transfer models to yield accurate parameter values -- particularly extinction and temperature -- the spectra must have a reliable relative flux calibration. Therefore, we correct our spectra using the photometric measurements interpolated to the epoch of observation. 

We begin by performing synthetic photometry on the observed spectra using the passbands described in \cref{sec:photometry}.
We then use the interpolated magnitudes from the previous section to calculate the ratios of the photometric and spectroscopic flux in the different passbands. Plotting the ratios against the effective wavelength reveals any wavelength-dependent trends in the flux calibration. 
 
We fit these trends with a parametric model, which we later apply to the observed spectrum to correct them.
However, the small number of passbands cannot constrain complex wavelength dependencies, so we use a linear model.

As for the GP hyperparameter estimation, we sample the parameters with \texttt{nestle}. In addition to the slope and intercept of the linear model, we include a parameter that allows for additional uncertainties in the flux ratios.
These uncertainties can arise from mixing different photometric systems, non-linear trends in the flux calibration, or underestimated uncertainties in the interpolated magnitudes.
The additional uncertainty is added in quadrature to the existing values. 

Because this additional uncertainty is often not well constrained by the data -- due to the small number of passbands -- we use a more informative prior than the standard log-uniform one used in the time-of-explosion fit (\cref{sec:time_explosion}) and light curve interpolation (\cref{sec:lc_interpolation}).
We adopt a half-normal distribution with mean zero and a $\sigma$ parameter\footnote{Here $\sigma$ is the standard deviation of the corresponding full normal distribution.} of \SI{0.03}{mag}.
While the log-uniform prior is scale-invariant, the new prior is not, allowing it to better reflect the expected size of the neglected uncertainties, such as photometric system mismatches.

\begin{figure}
	\begin{center}
     	\includegraphics{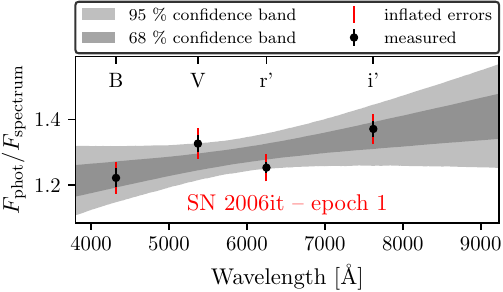}
      \end{center}
	\caption{Example of the linear flux calibration procedure. The plot shows the measured ratios of the photometric and spectroscopic flux $F_\mathrm{phot} / F_\mathrm{spectrum}$ for the first epoch of SN~2006it as a function of wavelength in black. We perform a Bayesian fit of the ratios (see \cref{sec:flux_calibration}) to identify all possible linear flux calibration corrections. Of the possible curves \SI{68}{\percent} fall within the dark grey contour and \SI{95}{\percent} within the light grey contour. Although the underlying curves are straight, the contours naturally exhibit curvature. 
 An important part of the fit is inflating the measured errors if necessary, accounting for additional unquantified uncertainties, such as the mixing of different photometric systems or underestimated uncertainties in the interpolated magnitudes. In the plot, the inflated errors are highlighted in red. In this specific example, the errors are substantially inflated because the data points deviate significantly from a linear trend within the measurement uncertainties.
}
	\label{fig:flux_calibration_example}
\end{figure}

\cref{fig:flux_calibration_example}
shows an example of our linear fit to the flux ratios.\footnote{Plots for the remaining objects can be found in \cref{app:flux_cal}.}
The plot illustrates that there is a wide range of possible flux calibration corrections we can apply to the observations. We store \num{100} randomly-selected corrections, which allows us to propagate the flux calibration uncertainties into the $\theta/v_\mathrm{ph}$ values in \cref{sec:theta/v}.

\subsection{Spectral fits} \label{sec:spectral_fits}
After flux calibration, we fit the calibrated spectra with radiative transfer models to estimate the photospheric velocities and specific fluxes at the photosphere $f_{\lambda_\mathrm{em}, \mathrm{ph}}$ needed for calculating $\theta / v_\mathrm{ph}$ (see Eq. \ref{eq:theta}). The radiative transfer models at the heart of this process are calculated with a custom version of the Monte Carlo code \tardis{} \citep{Kerzendorf2014,kerzendorf_wolfgang_2023_8244935}
that has been modified for use in SNe~II as described in \citet[][]{Vogl2019}. The additional functionality of this version is currently being implemented in the main branch of \tardis{}.

In the transport, we treat the ejecta as spherically symmetric and homologously expanding as in the EPM. We use simple input models that are described by only a handful of parameters so that we can explore the parameter space and find the best-fitting model. The key simplifications are a power-law density profile and uniform composition. Both choices greatly reduce the number of parameters and are well motivated for modelling SNe~II in the photospheric phase \citep[see the discussion in][]{Dessart2006,Dessart2008,Vogl2020}.

Exploring the parameter space automatically remains challenging, however, due to the computational cost of the radiative transfer simulations, which take around a day ($\sim$\SI{e5}{\second}) per spectrum.
To address this, we replace our radiative transfer code with an emulator, drastically reducing the time to generate a spectrum to about \SI{e-2}{\second} -- a difference of seven orders of magnitude.
The emulator, which is trained on a large grid of \tardis{} simulations \citep[see][]{Vogl2020}, predicts the output of \tardis{} with typically sub-percent accuracy. The models used for the training cover a wide range of photospheric temperatures $T_\mathrm{ph}$, velocities $v_\mathrm{ph}$, metallicities $Z$, power-law indices for the density profile $n$, and values for the time since explosion $t_\mathrm{exp}$. \citet{GezaSiblings} describe the most up-to-date models and the parameter space they cover (see their Table 2).

The strategy for finding the best-fitting model is similar to \citet{Vogl2020,Vasylyev2022,Vasylyev2023}, and \citet{GezaM51,GezaSiblings}: we perform maximum likelihood estimation with a Gaussian likelihood. To do this, we
use a simple diagonal covariance matrix
that gives all parts of the spectrum equal weight. This choice provides a reasonable guess for the best fit but it does not yield realistic parameter uncertainties in a full Bayesian analysis. For the latter, we need a covariance matrix that summarizes all sources of uncorrelated and correlated uncertainties. This includes, most importantly, the effect of approximations in our \tardis{} models, which result in complicated correlated fit residuals even for noise-free data. If the covariance matrix does not account for these residuals, the parameter uncertainties will be significantly underestimated \citep[see][]{Czekala2015}.

Given the unresolved challenge of incorporating these residuals into SN fitting, we limit our analysis to the described maximum likelihood estimations.
We will obtain a rough estimate of the impact of SN parameter uncertainties (and other unaccounted-for sources of uncertainty) on distances from the dispersion in $H_0$ between epochs and between objects in \cref{sec:bayes_ensemble_epm}.

Since the host extinction is unknown, we fit for the SN parameters on a fine grid of host color excess values $E(B-V)_\mathrm{host}$. This will allow us in \cref{sec:theta/v} to determine a consistent $E(B-V)_\mathrm{host}$ from all spectra of an object and estimate the contribution of the extinction to uncertainties in $\theta/v_\mathrm{ph}$. In the fits, we correct the observed spectra for Milky Way extinction (\cref{tab:H0_measurement_sample}) and redden the emulated spectra with the host extinction using a \citet{Fitzpatrick2019} extinction law with a total-to-selective extinction ratio $R_V=3.1$.\footnote{We perform the correction using the \texttt{dust\_extinction} Python package \citep{2022zndo...4658887G}.}

We calculate a maximum-likelihood $\theta/v_\mathrm{ph}$ estimate for each $E(B-V)_\mathrm{host}$ value, using the respective best-fit results for $v_\mathrm{ph}$ and the reddened emulated flux at the photosphere. For this, we recast \cref{eq:theta} in terms of magnitudes \citep[as described in][]{Vogl2019} and
use the interpolated magnitudes from \cref{sec:lc_interpolation} corrected for Milky Way extinction as the observed flux. We use all available passbands to constrain $\theta$.

A close examination of \cref{eq:theta} shows K-corrections \citep[e.g.,][]{HoggKcorr} are unnecessary for the observed flux because the relevant transformations are instead applied to the model flux at the photosphere.

\subsection{Flow corrections} \label{sec:flow_corrections}
To accurately measure $H_0$, we must correct for the peculiar velocities of SN host galaxies using a cosmic flow model, as discussed in \cref{sec:h0_basics}. Peculiar velocities can reach \SI{300}{\km\per\second} or more \citep[see, e.g.,][]{2018A&A...615A.162L}, which can easily introduce a \SI{10}{\percent} error in the cosmological redshift and subsequently $H_0$ for our lowest redshift objects (with $cz \sim \SI{3000}{\km\per\second}$).
Due to this high error risk, these corrections are important, and their success in reducing Hubble diagram residuals \citep[see, e.g.,][]{PetersonPantheonPecV} has led to their widespread adoption in recent $H_0$ measurements \citep[e.g.,][]{Pesce2020,deJaeger2022,Riess2022}.

We follow the approach used in the Pantheon$+$ analysis by adopting the 2M++/SDSS model from \citet{CarrPantheonPecV}.
This model is based on the velocity field of \citet{Carrick2015} with updated values for the velocity scale parameter $\beta$ and the external velocity $v_\mathrm{ext}$ from \citet{Said2020}.
We evaluate the model using the publicly available code\footnote{\url{https://github.com/KSaid-1/pvhub}}. The resulting peculiar velocities are listed in \cref{tab:H0_measurement_sample}.

The Pantheon$+$ analysis suggests that we could further minimize the impact of peculiar velocities by assigning galaxies to groups and performing flow corrections on the groups \citep[see][]{PetersonPantheonPecV}. We investigate this using their preferred group catalog from \citet{Tully2015}. However, only three of our SNe (SN~2006it, SN~2010id, and SN~2020cvy) are in groups from this catalog. These groups are small, with a maximum of three members, implying that the small-scale virial motions that could be averaged out are minimal. In fact, the group redshifts deviate from the host galaxy redshifts by less than \SI{70}{\km\per\second}. Given this small difference, we refrain from using group assignments for simplicity.

\section{Extinction and $\theta/v_\mathrm{ph}$ determination} \label{sec:theta/v}
In \cref{sec:spectral_fits}, we described how to determine $\theta/v_\mathrm{ph}$ for a given flux-calibrated spectrum and extinction.
However, in practice, both flux calibration and extinction are subject to uncertainties, which must be accounted for when constraining $\theta/v_\mathrm{ph}$.
The two uncertainties are deeply interconnected because the process of spectral fitting derives extinction from the observed slope of the spectrum. 

While the SN parameters, including the photospheric temperature, can be inferred from spectral features without considering the slope, the difference between the predicted and observed slopes constrains the extinction. Since the observed slope depends on the flux calibration, uncertainties in the calibration directly affect the inferred extinction. 

For example, a \SI{20}{\percent} change in flux calibration from the blue to the red end of the spectrum (\SI{3800}{\AA} to \SI{9000}{\AA}) impacts the slope similarly to a differential extinction with $E(B-V) = 0.05$. Our flux calibration uncertainties can reach this magnitude, making them an important source of uncertainty in the extinction determination.\footnote{This is especially the case for the ePESSTO+ SNe with photometry only in $g$ and $r$, which requires extrapolating the correction significantly to longer and shorter wavelengths.}

The spectral fits are the second key source of uncertainty.
Even with a perfectly flux-calibrated, noise-free spectrum, a range of $E(B-V)$ values can yield reasonable fits due to imperfections in the spectral models.
Determining the exact size of this range is challenging because defining a ``reasonable fit" statistically is difficult (see \cref{sec:spectral_fits}).
 
Based on their expertise as spectroscopists,
\citet{Dessart2006} estimated that $E(B-V)$ can vary by up to $\pm 0.05$ before the fits become unreasonable. This range of acceptable values is comparable to the uncertainties from flux calibration.

To determine $\theta/v_\mathrm{ph}$ given uncertain extinction, we must account for both sources of uncertainty. By assigning reasonable uncertainties to the $E(B-V)$ estimates from individual epochs, we can then combine them into a more precise joint constraint. Prior knowledge about the distribution of extinction values can further refine this estimate.

We explain the prior knowledge used in \cref{sec:ext_prior} and the combination of the constraints and determination of $\theta/v_\mathrm{ph}$ in \cref{sec:theta_v_steps}.

\subsection{Extinction prior} \label{sec:ext_prior}
Since $E(B-V)$ cannot be negative, we must impose a non-negative prior. A simple approach is to use a positive flat prior, but this can bias the extinction toward higher values \citep[see, e.g.,][]{2007ApJ...659..122J}. This bias occurs because measurements that scatter downward into negative $E(B-V)$ values are pushed up to zero by the prior, while upward scatter remains unaffected, leading to an asymmetry.

If the distribution of SN host extinction were truly uniform from zero to infinity, values near zero would be extremely unlikely, making the bias negligible. However, in reality, extinction values close to zero are much more common than large ones. Therefore, using a prior that reflects this reality helps reduce bias by exerting a compensating pull on values that scatter upward.

An example of such a prior is provided by \citet{Hatano98}, who used a simple model of the dust and SN distribution in randomly oriented galaxies to derive the extinction distribution for core-collapse SNe, which is widely used in the literature \citep[e.g.,][]{2009A&A...499..653B, Goldstein19, Vincenzi23}.
Following \citet{2014ApJ...792..135T}, we approximate this distribution as exponential with a mean $A_V$ of \num{0.5}.

\subsection{Method} \label{sec:theta_v_steps}
\begin{figure*}[htbp]
	\begin{center}
		\includegraphics[scale=1.1]{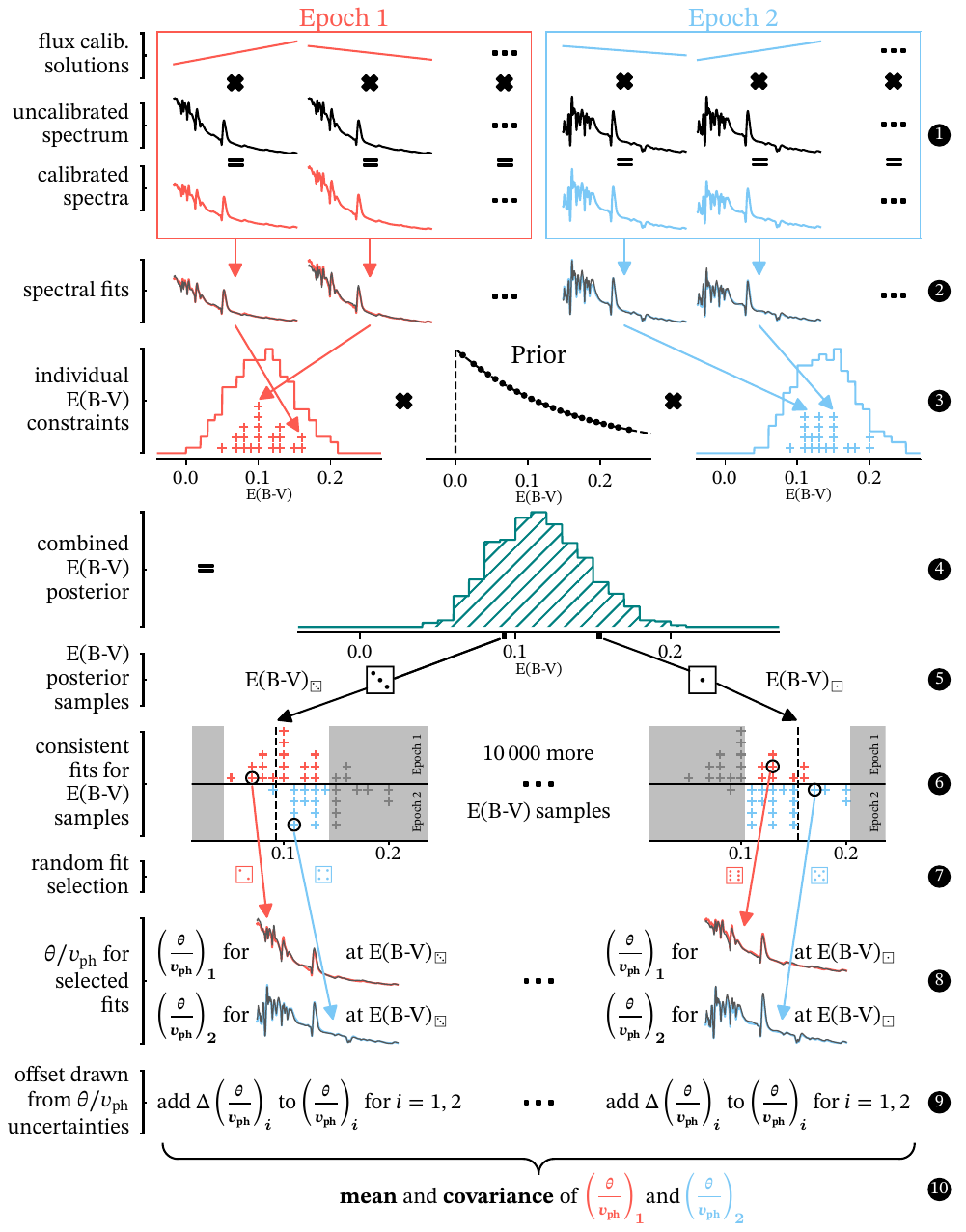}
	\end{center}
	\caption{Procedure to determine $\theta/v_\mathrm{ph}$ and $E(B-V)$ as described in \cref{sec:theta/v}.}
	\label{fig:theta_v_flowchart}
\end{figure*}

To combine our prior knowledge of $E(B-V)$ with the constraints from the individual epochs and determine $\theta/v_\mathrm{ph}$, we propose the approach visualized in \cref{fig:theta_v_flowchart}, which we will explain step by step.
The steps are numbered consistently in the text and the figure for easy reference.

The figure is based on a hypothetical supernova observed at two epochs, but the principle generalizes to any number of epochs. All data and results shown are purely illustrative. We use color coding to distinguish between the epochs: red for epoch 1 and blue for epoch 2.

We split the method into two parts: first, the determination of $E(B-V)$, and second, the determination of $\theta/v_\mathrm{ph}$.
\subsubsection{Determination of $E(B-V)$}
\begin{enumerate}
	\item \textbf{Generating viable flux-calibrated spectra:} To represent the uncertainty in the flux calibration, we generate sets of possible flux-calibrated spectra for each epoch. We randomly select \num{100} linear flux calibration solutions from the posterior of the photometric to spectroscopic flux ratio fits (\cref{sec:flux_calibration}). These flux solutions are multiplied with the original SN spectrum to obtain \num{100} calibrated spectra per epoch.
	\item \textbf{Performing spectral fits:} We perform maximum likelihood spectral fits on the created flux-calibrated spectra for a fine grid of host $E(B-V)$ values (see \cref{sec:spectral_fits}).
	      For each spectrum and each point of the $E(B-V)$ grid, we obtain best-fit values for the SN parameters, plus $\theta/v_\mathrm{ph}$.
    
    To account for flux calibration uncertainties that can make spectra appear artificially bluer, we extend the $E(B-V)$ grid to include negative values, even though negative extinction is unphysical. This allows us to fit these spectra and, by applying the extinction prior to exclude negative values, effectively filter out flux calibration solutions that are too blue (see step 4).
       
     For simplicity, \cref{fig:theta_v_flowchart} displays only a single fit per calibrated spectrum (shown in grey), rather than the full grid of fits for different $E(B-V)$ values.
	\item \textbf{Deriving individual $E(B-V)$ constraints:} We derive constraints on $E(B-V)$ for the individual epochs from the spectral fits. 
    We start by identifying the best-fitting $E(B-V)$ in the grid for each of the calibrated spectra as measured by $\chi^2$, which yields a distribution of possible $E(B-V)$ values for each epoch. This distribution captures the uncertainties in the extinction arising from the flux calibration.
    
    To account for uncertainties from the spectral fits, we consider $E(B-V)$ values within $\pm 0.05$ of the best-fit value, as suggested by \citet{Dessart2006}. We assume all fits within this range are equally likely. For each possible flux-calibrated spectrum, we represent the $E(B-V)$ estimates with a uniform distribution over this range. By summing these uniform distributions from all flux calibrations, we obtain the final $E(B-V)$ estimate for each epoch.

    This method is essentially a kernel density estimation (KDE) using a tophat kernel with a bandwidth of \num{0.05} \citep[see, e.g.,][]{Bishop2007}. We discuss the impact of the bandwidth choice in \cref{sec:theta_v_results}. In the figure, plus symbols indicate the distribution of best-fitting $E(B-V)$ values for each possible flux calibration; the resulting KDEs are shown by solid lines, with epoch 1 in red and epoch 2 in blue. The extinction prior (\cref{sec:ext_prior}) is shown in black, positioned between our estimates derived from the spectral fits.
	\item \textbf{Deriving a joint $E(B-V)$ constraint:}
          To derive the combined $E(B-V)$ constraint, we treat the estimates from the individual epochs as independent and multiply them with the prior. However, directly multiplying the top-hat KDEs
          with the continuous exponential prior creates an artificial sawtooth pattern.
          To prevent this, we approximate the prior as piecewise constant within the regions where the product of the two top-hat KDEs is constant. This ensures a well-behaved joint posterior, which is shown in teal in the figure.
    \end{enumerate}
    \subsubsection{Determination of $\theta/v_\mathrm{ph}$}
    Having arrived at a final $E(B-V)$ distribution, we can now begin to quantify how this translates to the
	      $\theta/v_\mathrm{ph}$ values we are ultimately interested in:
    \begin{enumerate}
    \setcounter{enumi}{4}
	\item \textbf{Drawing $E(B-V)$ posterior samples:}
 The first step is to sample a large number of
	      $E(B-V)$ values (\SI{10000}{}) from the posterior to approximate the distribution.
	      In the diagram, we represent this random process with dice rolls, yielding two example values: $E(B-V)_{\epsdice{3}}$ and \mbox{$E(B-V)_{\epsdice{1}}$}.
	\item \textbf{Identifying consistent flux calibrations for \mbox{$E(B-V)$} samples:}
	      Next, we connect the $E(B-V)$ samples back to the spectral fits, which provide the $\theta/v_\mathrm{ph}$ constraints. We start by identifying the flux calibrations that are consistent with each $E(B-V)$ sample. As established earlier, for each flux calibration we treat all $E(B-V)$ values and associated spectral fits within $\pm 0.05$ of the best fit as valid and equally likely. Reversing this logic, all flux calibrations within this range of an $E(B-V)$ sample are considered consistent with it.
	      
       We illustrate this process in the figure using our two exemplary samples $E(B-V)_{\epsdice{3}}$ and $E(B-V)_{\epsdice{1}}$. For each sample (indicated by the dashed line), we show the distribution of best-fitting $E(B-V)$ values for the different flux calibrations from step \num{3}. The consistent calibrations fall within $\pm 0.05$ of the sample and are shown in color; the inconsistent calibrations are greyed out.
	\item \textbf{Selecting consistent flux calibrations for $E(B-V)$ samples:} We randomly select one consistent flux calibration for each $E(B-V)$ sample and epoch for the determination of $\theta/v_\mathrm{ph}.$ This random selection is represented in \cref{fig:theta_v_flowchart} by two dice rolls for each example extinction value, $E(B-V)_{\epsdice{3}}$ and $E(B-V)_{\epsdice{1}}$.

	\item \textbf{Determining $\theta/v_\mathrm{ph}$ for the selected flux calibrations and extinction:} For each $E(B-V)$ sample (e.g., $E(B-V)_{\epsdice{3}}$), we compute the corresponding $\theta/v_\mathrm{ph}$ values using its selected flux calibration. Each flux calibration has its own grid of maximum-likelihood spectral fits, providing $\theta/v_\mathrm{ph}$ values for a grid of $E(B-V)$ values. 

We interpolate these $\theta/v_\mathrm{ph}$ values to match each $E(B-V)$ sample. By repeating this process for all \SI{10000}{} samples, we generate a distribution of $\theta/v_\mathrm{ph}$ that incorporates uncertainties from both extinction and flux calibration. This approach also captures correlations between epochs, as for each $\theta/v_\mathrm{ph}$ pair, the epochs share the same extinction, pushing them on average in a similar direction with respect to the mean.\footnote{The impact of extinction on $\theta/v_\mathrm{ph}$ is not exactly the same for all epochs due to its interplay with the SN parameters, particularly the temperature.}
	\item \textbf{Including additional $\theta/v_\mathrm{ph}$ uncertainties:} Besides flux calibration, other observational uncertainties affect $\theta/v_\mathrm{ph}$. For this proof-of-principle paper, we focus on only two major sources: uncertainties in the observed flux\footnote{We exclude the impact of weak gravitational lensing on the flux uncertainties, which is commonly considered in SN cosmological analyses \citep[e.g.,][]{2014A&A...568A..22B,2020MNRAS.495.4860D,Pantheon+Cosmo}.
 This effect is negligible for the highest redshifts in our sample, affecting measurements by only about \SI{0.002}{mag} at
		      $z=0.04$ \citep{2010MNRAS.405..535J}.} affecting the $\theta$ calculation (see Eq. \ref{eq:theta}) and errors in the heliocentric redshift correction due to host galaxy rotation. 
        
        We quantify the flux uncertainties through linear error propagation based on the interpolated magnitude uncertainties from \cref{tab:interp_phot}. Given that the magnitude uncertainties are small and multiple bandpasses are combined, the uncertainties in $\theta$ are generally less than $\SI{\sim 1.5}{\percent}$. 
        
        For redshift correction errors, we assume an average uncertainty of \SI{150}{\km \per \second}, as the correction typically uses the redshift of the galaxy center, while the SN is located in a rotating spiral arm, following \citet{2020MNRAS.495.4860D}. This translates to an uncertainty in photospheric velocities of a similar size.\footnote{Changes in photospheric velocity are not entirely equivalent to red- or blueshifts of the spectrum since they also affect line widths and shapes. However, the effects are similar for small velocity offsets.} For our sample, spanning photospheric velocities between \SI{5000}{\km \per \second} and \SI{10000}{\km \per \second}, this contributes $\SI{\sim 1.5}{\percent}$ to \SI{3}{\percent} to the $\theta/v_\mathrm{ph}$ uncertainties. 
        
        To propagate these uncertainties into our samples, we apply random offsets to the $\theta$ values based on the observational flux uncertainties and to $v_\mathrm{ph}$ based on the heliocentric redshift uncertainty.
	\item \textbf{Determining mean and covariance of the $\theta/v_\mathrm{ph}$ samples:}
	      We now determine the mean and covariance $C_\mathrm{SN X, meas}$ of the $\theta/v_\mathrm{ph}$ samples of the different epochs, which we will use in the EPM regressions to constrain the SN distances.
\end{enumerate}

\subsection{Results} \label{sec:theta_v_results}
\begin{figure*}
	\begin{center}
		\includegraphics[scale=0.99]{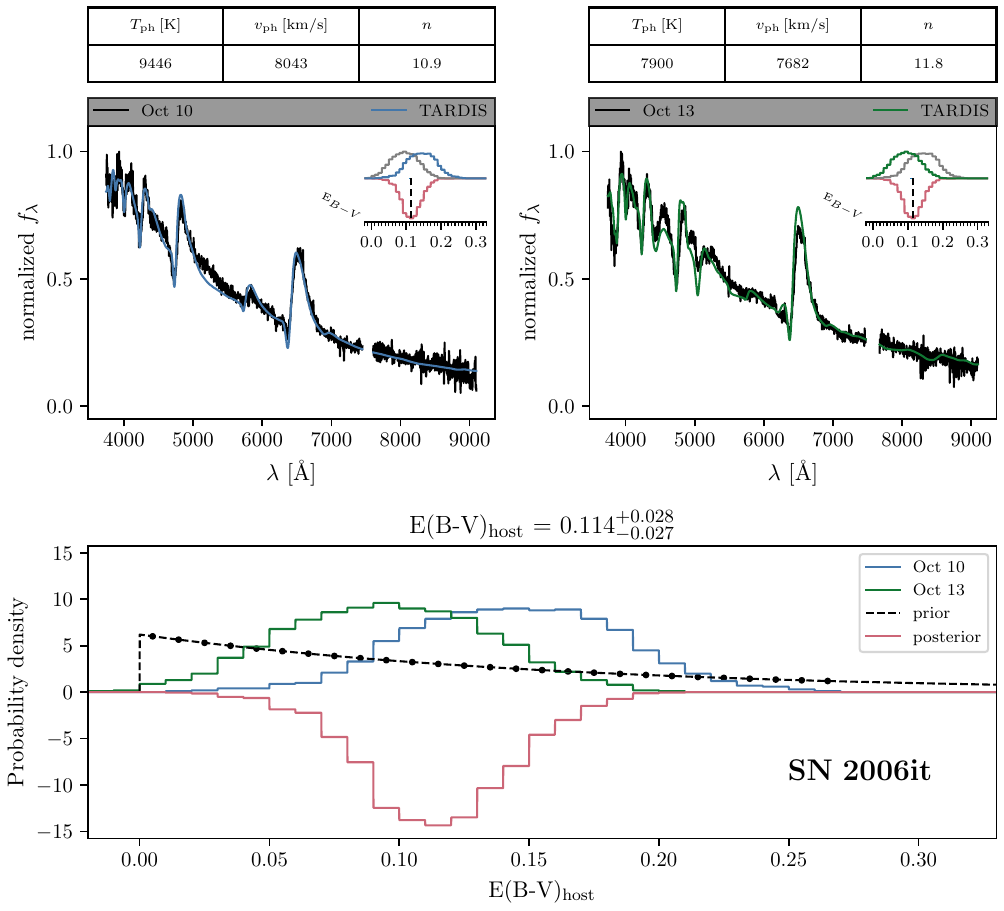}
	\end{center}
	\caption[]{Example spectral fits. We show the the spectral fits for SN~2006it at two epochs (Oct 10 and Oct 13), analogous to the hypothetical SN used in \cref{fig:theta_v_flowchart}.
		The bottom panel displays the $E(B-V)$ constraints from spectral fits of possible flux calibration solutions, with epoch 1 in blue and epoch 2 in green. The constraints from both epochs align well within the uncertainties. The dashed black line represents the exponential approximation of the \citet{Hatano98} extinction prior, with dots indicating evaluation points for multiplication with the individual epoch constraints. We plot the final $E(B-V)$ posterior (red) in the negative probability density direction for visual separation.
  
		The upper part of the figure visualizes the spectral fits contributing to the $E(B-V)$ and subsequently $\theta/v_\mathrm{ph}$ determinations. While fits of many different flux-calibrated spectra and $E(B-V)$ values contribute, we show only one representative fit per epoch for simplicity. To select this fit, we follow a two-step process. First, we choose one of the flux calibration solutions whose best-fit $E(B-V)$ is closest to the median of the posterior. Next, within this flux calibration solution, we analogously identify the fit from the $E(B-V)$ grid that is nearest to the median of the posterior.
		The normalized observed specific flux $f_\lambda$ of this choice appears in black, with the corresponding maximum-likelihood fits in blue (epoch 1) and green (epoch 2). The key physical parameters for the selected fits are listed on the top.
		Small insets in each panel mirror the bottom half of the figure, visualizing the $E(B-V)$ constraints with the other epoch greyed out. The dashed line in the inset marks the median $E(B-V)$ value, which is used in the plotted fits.}
	\label{fig:spectral_fit}
\end{figure*}

We show an example of the extinction determination and the involved spectral fits in \cref{fig:spectral_fit}, while plots for the remaining objects can be found in \cref{app:spec_fits}.

The precision of the host $E(B-V)$ determinations varies significantly, ranging from \num{0.015} to \num{0.08}. This variation reflects differences in flux calibration uncertainties and the number of available epochs. Consequently, the quantified uncertainties in $\theta/v_\mathrm{ph}$ also span a wide range, from roughly \SI{3}{\percent} to \SI{13}{\percent}, with a median of around \SI{4}{\percent}.

Our results are somewhat influenced by the bandwidth choice for the KDE in the $E(B-V)$ determination. To assess this, we increase the bandwidth by \SI{50}{\percent}, from \num{0.05} to \num{0.075}.
This change moderately increases $E(B-V)$ uncertainties by a median of \SI{23}{\percent}, but only slightly increases $\theta/v_\mathrm{ph}$ uncertainties by \SI{4}{\percent}. The mean $\theta/v_\mathrm{ph}$ values change only slightly, with a median decrease of \SI{0.4}{\percent}. Thus, while the bandwidth affects the results, the overall sensitivity is low.

While our quantified uncertainties account for significant sources like extinction, interpolated magnitudes, and heliocentric redshift corrections, they are not exhaustive. We do not consider, for example, uncertainties from model imperfections or the total-to-selective extinction ratio on the modelling side. Observationally, non-linear trends in the flux calibration or wavelength calibration issues are also not included. In the next section, we will attempt to statistically constrain the combined uncertainties from these sources together with the Hubble constant.

\section{Hubble constant}
\label{sec:bayes_ensemble_epm}
The measured $\theta/v_\mathrm{ph}$ values constrain the luminosity distances of our SNe, thereby constraining $H_0$ as explained in \cref{sec:basics}. Traditionally, this involves two steps: fitting the time evolution of $\theta/v_\mathrm{ph}$ for each SN individually and analyzing the resulting Hubble diagram for $H_0$. However, we propose a different strategy where we directly fit for $H_0$ using all SNe.

Our motivation for this departure lies in the need to statistically capture the remaining scatter in $\theta/v_\mathrm{ph}$ not explained by the quantified uncertainties, akin to the intrinsic dispersion term $\sigma_\mathrm{int}$ in SN~Ia cosmology \citep[see e.g.][]{Astier2006,Scolnic2018,Dhawan2018}. Since $\sigma_\mathrm{int}$ is a global parameter, we adopt a collective approach by performing all EPM regressions simultaneously.

Termed Bayesian ensemble EPM, our approach accounts for both the constraints from the scatter of $\theta/v_\mathrm{ph}$ within each SN and the scatter between SNe in the Hubble diagram.
We explain the method in \cref{sec:bayes_ensemble_epm_method}, validate it with simulated data in \cref{sec:bayes_ensemble_epm_validation}, and then apply it to our real data in \cref{sec:bayes_ensemble_epm_results}.

\subsection{Bayesian ensemble EPM} \label{sec:bayes_ensemble_epm_method}
In contrast to the classical EPM regression, we do not use independent values for the SN luminosity distances $D_\mathrm{L,SNX}$ in the fit.
Instead, we propose a shared value for $H_0$ and simultaneously propose values for the true SN cosmological redshifts $z_\mathrm{cosmo,SNX}$. Together, these inputs, along with the cosmological model described by \cref{eq:DL}, generate a trial set of interconnected $D_\mathrm{L,SNX}$, whose viability we evaluate simultaneously.

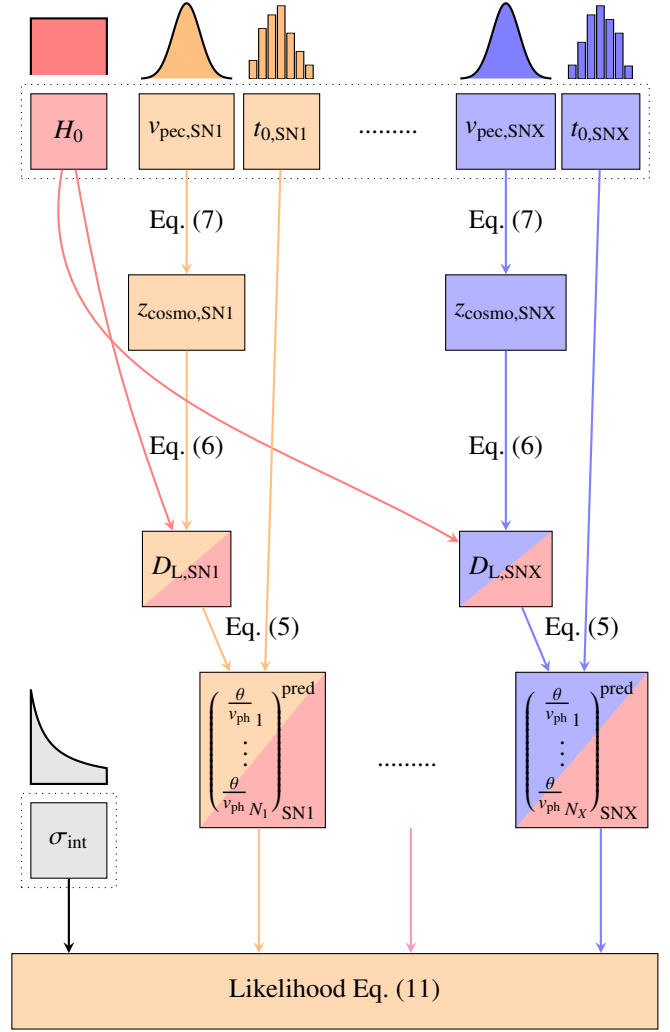
\begin{figure}[htbp]
	\centering
	\tikzstyle{newstyle0}=[fill={rgb,255: red,0; green,128; blue,128}, draw=black, shape=rectangle]

\tikzset{
diagonal fill/.style 2 args={fill=#2, path picture={
\fill[#1, sharp corners] (path picture bounding box.south west) -|
                         (path picture bounding box.north east) -- cycle;}},
reversed diagonal fill/.style 2 args={fill=#2, path picture={
\fill[#1, sharp corners] (path picture bounding box.north west) |- 
                         (path picture bounding box.south east) -- cycle;}}
}

\tikzstyle{sn1} = [rectangle, minimum width=1cm, minimum height=1cm, text centered, draw=black, fill=orange!30]
\tikzstyle{dl1} = [rectangle, minimum width=1cm, minimum height=1cm, text centered, draw=black, diagonal fill={red!30}{orange!30}]
\tikzstyle{dln} = [rectangle, minimum width=1cm, minimum height=1cm, text centered, draw=black, diagonal fill={red!30}{blue!30}]
\tikzstyle{snn} = [rectangle, minimum width=1cm, minimum height=1cm, text centered, draw=black, fill=blue!30]
\tikzstyle{h0} = [rectangle, minimum width=1cm, minimum height=1cm, text centered, draw=black, fill=red!30]
\tikzstyle{likelihood} = [rectangle, minimum width=8.5cm, minimum height=1cm, text centered, draw=black, fill=orange!30]
\tikzstyle{sigmaint} = [rectangle, minimum width=1cm, minimum height=1cm, text centered, draw=black, fill=gray!20]

\tikzstyle{arrow} = [thick,->,>=stealth]
\usetikzlibrary{shapes.geometric, arrows}
\begin{tikzpicture}[node distance=1.4cm]
\node (h0) [h0] {$H_0$};
\node (vpec1) [sn1, right of=h0, xshift=0.15cm] {$v_\mathrm{pec, SN 1}$};
\node (t01) [sn1, right of=vpec1, xshift=-0.15cm] {$t_\mathrm{0, SN 1}$};

\node (dots) [right of=t01]{.........};
\node (vpecn) [snn, right of=dots, xshift=0.15cm] {$v_\mathrm{pec, SN X}$};
\node (t0n) [snn, right of=vpecn, xshift=-0.15cm] {$t_\mathrm{0, SN X}$};

\node (zcosmo) [sn1, below of=vpec1, xshift=0.0cm, yshift=-1cm] {$z_\mathrm{cosmo, SN 1}$};
\node (eq_zcosmo) at ($(vpec1)!0.5!(zcosmo)$) {\cref{eq:z_cosmo}};
\node (zcosmon) [snn, below of=vpecn, xshift=0.0cm, yshift=-1cm] {$z_\mathrm{cosmo, SN X}$};
\node (eq_zcosmon) at ($(vpecn)!0.5!(zcosmon)$) {\cref{eq:z_cosmo}};
\node (dl1) [dl1, below of=zcosmo, yshift=-2cm] {$D_\mathrm{L, SN 1}$};
\node (dln) [dln, below of=zcosmon, yshift=-2cm] {$D_\mathrm{L, SN X}$};
\node (thetav1) [dl1, below of=dl1, xshift=1cm, yshift=-1cm]{$\begin{pmatrix}\frac{\theta}{v_\mathrm{ph}}_1\\ \vdots \\ \frac{\theta}{v_\mathrm{ph}}_{N_1}\end{pmatrix}_\mathrm{SN 1}^\mathrm{pred}$};
\node (thetavn) [dln, below of=dln, xshift=1cm, yshift=-1cm]{$\begin{pmatrix}\frac{\theta}{v_\mathrm{ph}}_1\\ \vdots \\ \frac{\theta}{v_\mathrm{ph}}_{N_X}\end{pmatrix}_\mathrm{SN X}^\mathrm{pred}$};
\node (mdots) [below of=dots, yshift=-7cm, xshift=0.25cm]{.........};
\node (like) [likelihood, below=of $(h0.north)!0.5!(t0n.north)$, yshift=-10cm] {Likelihood \cref{eq:h0_likelihood}};

\node (eq_dl) [above of=dl1, yshift=0.2 cm] {\cref{eq:DL}};
\node (eq_dln) [above of=dln, yshift=0.2 cm] {\cref{eq:DL}};

\node (eq_thetav1) [above of=thetav1, yshift=0.2 cm] {\cref{eq:theta/v}};
\node (eq_thetavn) [above of=thetavn, yshift=0.2 cm] {\cref{eq:theta/v}};
\node (sigmaint) [sigmaint, below of=h0, xshift=0.0cm, yshift=-8.cm] {$\sigma_\mathrm{int}$};

\node[draw,dotted,fit=(h0) (vpec1) (t01) (t0n) ] {};
\node[draw,dotted,fit=(sigmaint) ] {};

\draw [arrow,draw=orange!50] (vpec1) -- (zcosmo);
\draw [arrow,draw=blue!50] (vpecn) -- (zcosmon);
\draw [arrow,draw=orange!50] (zcosmo) -- (dl1);
\draw [arrow,draw=red!50] (h0) to [out=-80,in=110] (dl1);
\draw [arrow,draw=red!50] (h0) to [out=-100,in=-210] (dln);
\draw [arrow,orange!50] (dl1) -- (thetav1);
\draw [arrow,blue!50] (dln) -- (thetavn);
\draw [arrow,draw=blue!50] (zcosmon) -- (dln);
\draw [arrow,draw=orange!50] (t01) -- (thetav1);
\draw [arrow,draw=blue!50] (t0n) -- (thetavn);
\draw [arrow,draw=orange!50] ([xshift=-1. cm]thetav1.south -| like.north) -- ([xshift=-1.0 cm]like.north);
\draw [arrow,draw=blue!50] ([xshift=3.5 cm]thetavn.south -| like.north) -- ([xshift=3.5 cm]like.north);
\draw [arrow,draw=magenta!50] ([xshift=1. cm]thetav1.south -| like.north) -- ([xshift=1.0 cm]like.north);
\draw [arrow] ([xshift=-3.5 cm]sigmaint.south -| like.north) -- ([xshift=-3.5 cm]like.north);

\coordinate (A) at (-0.5,0.75);
\coordinate (B) at (-0.5,1.5);
\coordinate (C) at (0.5,1.5);
\coordinate (D) at (0.5,0.75);

\draw[black, thick, fill=red!50] (A) -- (B) -- (C) -- (D);

  \begin{scope}[shift={(2.3,0.7)}, yscale=0.75] 
    \begin{axis}[
      ybar, 
      bar width=0.0995 cm, 
      xmin=-4, xmax=4, 
      ymin=0, 
      ylabel={},
      xlabel={},
      axis line style={draw=none}, 
      tick style={draw=none}, 
      xtick=\empty, 
      ytick=\empty, 
      symbolic x coords={-4,-3,-2,-1,0,1,2,3,4},
      height=3cm, width=2.57cm 
    ]
      \addplot[fill=orange!50] coordinates {
                (-3,2) (-2,5.5) (-1,7) (0,8) (1,5) (2,3) (3,1.5)
      };
    \end{axis}
  \end{scope}

  \begin{scope}[shift={(6.5,0.7)}, yscale=0.75] 
    \begin{axis}[
      ybar, 
      bar width=0.0995 cm, 
      xmin=-4, xmax=4, 
      ymin=0, 
      ylabel={},
      xlabel={},
      axis line style={draw=none}, 
      tick style={draw=none}, 
      xtick=\empty, 
      ytick=\empty, 
      symbolic x coords={-4,-3,-2,-1,0,1,2,3,4},
      height=3cm, width=2.57cm 
    ]
      \addplot[fill=blue!50] coordinates {
        (-3,1.5) (-2,4) (-1,7) (0,8) (1,6.5) (2,5) (3,1.4)
      };
    \end{axis}
  \end{scope}
  
\draw[thick] (-0.5,-8.65) -- (0.5,-8.65); 
\filldraw[black, thick, fill=gray!20] plot[domain=-0.5:0.5, samples=100] (\x, {-8.65 + 0.75 / (3 * (\x + 0.7))}) -- (0.5,-8.65) -- (-0.5,-8.65) -- cycle;

\draw[black, thick,fill=orange!50] plot[domain=0.95:2.15, samples=100] (\x, {exp(-(\x-1.55)^2 * 16) + 0.7});

\draw[black, thick,fill=blue!50] plot[domain=5.75-0.6:5.75+0.6, samples=100] (\x, {exp(-(\x-5.75)^2 * 16) + 0.7});

\end{tikzpicture}
	\caption{Flowchart of the iterative procedure for parameter estimation in the Bayesian ensemble EPM fit. At each step, we sample a new set of parameters, including the Hubble constant $H_0$, unexplained dispersion $\sigma_\mathrm{int}$, and the peculiar velocity $v_\mathrm{pec}$ and time of explosion $t_0$ for each SN, from the priors. The priors are visually indicated above each parameter. From the parameters, we then calculate true cosmological redshifts $z_\mathrm{cosmo}$ and luminosity distances $D_\mathrm{L}$, ending with the computation of the proposed $\theta/v_\mathrm{ph}^\mathrm{pred}$ vectors using the referenced equations. The process concludes with the comparison of the proposed and measured $\theta/v_\mathrm{ph}$ vectors, evaluating the likelihood for the proposed $\sigma_\mathrm{int}$. This cycle repeats for each new set of parameters, slowly building up the posterior distribution.}
	\label{fig:bayesian_ensemble_epm_flowchart}
\end{figure}

To evaluate this viability, we first calculate the predicted $\theta/v_\mathrm{ph}^\mathrm{pred}$ values from the proposed distances and explosion times $t_\mathrm{0,SNX}$ using \cref{eq:theta/v}.
We then evaluate the likelihood by comparing the predicted and measured $\theta/v_\mathrm{ph}$ values for all SNe. The flowchart in \cref{fig:bayesian_ensemble_epm_flowchart} summarizes the procedure, including the equations used. 

As in the regular EPM regression, we use a Gaussian likelihood $\mathcal{L}$ for each SN:
\begin{equation} \label{eq:h0_likelihood_sn1}
	-2 \ln(\mathcal{L}_\mathrm{SN X}) = \ln{\lvert C_\mathrm{SN X} \rvert} + \Delta \frac{\overrightarrow{\theta}}{v_\mathrm{ph}}^{T}_\mathrm{SN X} \, C^{-1}_\mathrm{SN X}  \, \Delta \frac{\overrightarrow{\theta}}{v_\mathrm{ph}}_\mathrm{SN X}
\end{equation}
Here, $C_\mathrm{SN X}$ is the covariance matrix
and
\begin{equation}
	\Delta \frac{\overrightarrow{\theta}}{v_\mathrm{ph}}_\mathrm{SN X} = \begin{pmatrix}\frac{\theta}{v_\mathrm{ph}}_1\\ \vdots \\ \frac{\theta}{v_\mathrm{ph}}_{N_X}\end{pmatrix}_\mathrm{SN X}^\mathrm{pred} -  \begin{pmatrix}\frac{\theta}{v_\mathrm{ph}}_1\\ \vdots \\ \frac{\theta}{v_\mathrm{ph}}_{N_X}\end{pmatrix}_\mathrm{SN X}^\mathrm{meas}
\end{equation}
are the $\theta/v_\mathrm{ph}$ residuals for the $N_X$ epochs of SN $X$.

The distinguishing feature of the ensemble EPM is the combination of the individual SN likelihoods into a joint likelihood
\begin{equation} \label{eq:h0_likelihood}
	-2 \ln(\mathcal{L}) = \sum_{i=0}^{N_\mathrm{SN}} \left( \ln{\lvert C_\mathrm{SN i} \rvert} + \Delta \frac{\overrightarrow{\theta}}{v_\mathrm{ph}}^{T}_\mathrm{SN i} \, C^{-1}_\mathrm{SN i}  \, \Delta \frac{\overrightarrow{\theta}}{v_\mathrm{ph}}_\mathrm{SN i} \right).
\end{equation}
This approach enables us to quantify the unexplained scatter in $\theta/v_\mathrm{ph}$ by combining constraints from within each SN and between SNe -- our main motivation for the method.

To achieve this, we model the scatter as uncorrelated between epochs and as a consistent fractional uncertainty for $\theta/v_\mathrm{ph}$, which is constant across all SNe. The covariance matrix is thus
\begin{equation} \label{eq:sigma_int}
	C_\mathrm{SN X} = C_\mathrm{SN X, meas} + \left[ \sigma_\mathrm{int}  \, \mathrm{diag }\begin{pmatrix}\frac{\theta}{v_\mathrm{ph}}_1\\ \vdots \\ \frac{\theta}{v_\mathrm{ph}}_{N_X}\end{pmatrix}_\mathrm{SN X}^{\mathrm{pred, H_0 = 70}} \right]^2,
\end{equation}
where $C_\mathrm{SN X, meas}$ is the measured covariance matrix (see \cref{sec:theta/v} and \cref{fig:theta_v_flowchart}) and $\sigma_\mathrm{int}$ the additional fractional uncertainty. To avoid biases, we use the predicted $\theta/v_\mathrm{ph}$ values in the uncertainty calculation rather than the measured ones.\footnote{If the measured quantities are used, those scattered upwards relative to the true value will receive higher uncertainties and thus lower weight in the fit, while those scattered downwards will receive lower uncertainties and higher weight, biasing the results towards smaller values.}

We make one additional adjustment to further prevent biases: we employ a fixed $H_0$ of \SI{70}{\km\per\second\per\Mpc} in computing the fractional uncertainties. This strategy decouples the fitted $H_0$ from the complexity penalty, which penalizes models that can accommodate a wider range of observed data.\footnote{
	Otherwise, higher $H_0$ values lead to larger $\theta/v_\mathrm{ph}$ uncertainties, allowing a wider range of observed data to be accommodated. This increased flexibility results in a greater complexity penalty, thereby reducing the likelihood of higher $H_0$ values.}
We choose \SI{70}{\km\per\second\per\Mpc} for convenience, given its proximity to typical measured values. The specific choice is unimportant since $\sigma_\mathrm{int}$ will be rescaled to match the measured $H_0$ at the end (\cref{sec:bayes_ensemble_epm_results}).

Our assumptions in modelling the unexplained $\theta/v_\mathrm{ph}$ scatter are not entirely accurate. Particularly, the errors resulting from the limitations within the spectral models 
likely vary across different epochs and SNe due to the diverse physical conditions in the spectrum-formation regions. Our current data, however, do not provide sufficient statistical power to detect more complex trends in the uncertainties.

With the likelihood and uncertainty model established, the next step is to define the prior distributions for our parameters:
we adopt a flat prior for $H_0$ and a log-uniform distribution for the unexplained scatter $\sigma_\mathrm{int}$ \citep[e.g.,][]{2015ApJ...813..137R,deJaeger2022}. The priors for explosion times $t_{0,\mathrm{SNX}}$ are based on the posterior distributions from the early light curve fits (see \cref{sec:time_explosion}). Finally, the priors for the peculiar velocities, used to compute the SNe's cosmological redshifts, follow normal distributions with means derived from the flow model (see \cref{sec:flow_corrections}). Their standard deviations
include a peculiar velocity uncertainty of \SI{250}{\km\per\second} similar to the Pantheon$+$ analysis \citep{Pantheon+Cosmo,PetersonPantheonPecV}, and the redshift measurement errors added in quadrature.

The redshift measurement errors are taken from NED except for SN~2021gvv, for which we measured the redshift ourselves from narrow host galaxy emission lines. The median redshift measurement error from NED is \SI{5}{\km\per\second}, and the maximum is about \SI{90}{\km\per\second}, which are negligible compared to the peculiar velocity uncertainty. However, for SN~2021gvv, from the dispersion of the narrow host galaxy emission line measurements, we estimate an error of around \SI{200}{\km\per\second}, which is comparable to the peculiar velocity uncertainties.

\subsection{Validation with simulated data}
\label{sec:bayes_ensemble_epm_validation}
We check that we can retrieve $H_0$ without biases and with meaningful uncertainties by fitting simulated datasets with a known $H_0$ of \SI{70}{\km\per\second\per\Mpc}. The simulated data closely match our real observations; we will not explore whether the method generalizes to new datasets with, for example, a different redshift distribution.

The first step in the mock data generation is to sample true values for the cosmological redshifts and explosion times from the established priors. From these and the selected $H_0$ (\SI{70}{\km\per\second\per\Mpc}), we then calculate true values for $\theta/v_\mathrm{ph}$ at the observed spectral epochs.
The procedure is identical to the computation of the proposed $\theta/v_\mathrm{ph}$ values in the Bayesian ensemble EPM fit.

In the final step, we convert the true $\theta/v_\mathrm{ph}$ values to observed ones by applying random offsets drawn from the covariance matrices $C_\mathrm{SN X}$. Here, the measured covariance matrices $C_\mathrm{SN X, meas}$ are the same as for the real data and we assume two plausible values (\SI{5}{\percent} and \SI{10}{\percent}) for the additional unexplained scatter $\sigma_\mathrm{int}$.

For each $\sigma_\mathrm{int}$ value, we generate \num{100} realizations with different true redshifts, explosion times, and $\theta/v_\mathrm{ph}$, and fit them using our method.
These realizations differ from our actual data only in their mean $\theta/v_\mathrm{ph}$ values.

\Cref{tab:mock_fit_summary} summarizes the fit results. We show the mean inferred values for $H_0$ and $\sigma_\mathrm{int}$ along with the uncertainty on the mean demonstrating that we can retrieve these key parameters without discernible biases.
The accurate inference of the unknown additional uncertainty $\sigma_\mathrm{int}$ is a promising indication that the method provides meaningful $H_0$ uncertainties. To confirm this, we calculate the reduced chi-square of the $H_0$ values around the true value $\chi_{H_0,\,\mathrm{red}}^2$, yielding a result consistent with unity, as expected for accurate uncertainties.
\begin{table} 
\begin{center}
\caption[]{Validation of the Bayesian ensemble EPM with simulated data.}
\small
\addtolength{\tabcolsep}{-3pt}    
\begin{tabular}{@{}ccccc@{}}
\toprule
True $\sigma_\mathrm{int}$ &Fitted $\sigma_\mathrm{int}$ & True $H_0$ & Fitted $H_0$ & $\chi_{H_0,\,\mathrm{red}}^2$ \\
& $\pm$\,error\,of\,mean& &\,$\pm$\,error\,of\,mean\\
\midrule
5\,\% & 5.0\,$\pm$\,0.2\,\% & 70.0 & 70.3\,$\pm$\,0.2 & 0.95\,$\pm$\,0.14 \\
10\,\% & 10.2\,$\pm$\,0.3\% & 70.0 & 69.9\,$\pm$\,0.3 & 1.12\,$\pm$\,0.14 \\

\bottomrule
\label{tab:mock_fit_summary}
\end{tabular}
\addtolength{\tabcolsep}{3pt}    
\tablefoot{Obtained by averaging fits of \num{100} simulated datasets. Consequently, the reported errors of the mean for $H_0$ are approximately ten times smaller (\(\sqrt{100}\)) than the uncertainties of individual fits, which have median values of \SI{2.2}{\km \per \second \per \Mpc} ($\sigma_\mathrm{int} = 5\,\%$) and \SI{2.8}{\km \per \second \per \Mpc} ($\sigma_\mathrm{int} = 10\,\%$).}
\end{center}
\end{table}

\subsection{Results}
\label{sec:bayes_ensemble_epm_results}
\begin{figure}
	\begin{center}
		\includegraphics[]{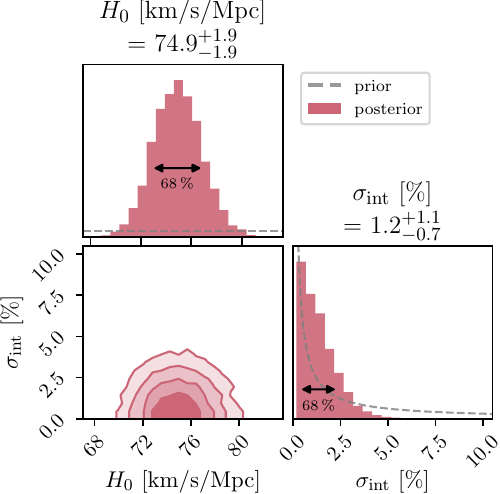}
	\end{center}
	\caption[]{Posterior of the Bayesian ensemble EPM fit (marginalized over the times of explosion and peculiar velocities of the SNe). The plot shows the \SI{68}{\percent} confidence intervals for the one-dimensional distributions. The two dimensional projection displays the \SIlist[list-final-separator = {, }]
		{39.3;67.5;86.5;95.6}{\percent} confidence regions (corresponding to the
		$1\,\sigma$, $1.5\,\sigma$, $2\,\sigma$, $2.5\,\sigma$ levels of a two-dimensional Gaussian distribution, as described in the \texttt{corner} Python package documentation).\footnotemark[1] The prior distributions for the one-dimensional marginalized distributions are shown as grey dashed lines.}
	\label{fig:h0_sigma_int_posterior}
\end{figure}

Building on the successful fits of our simulated data, we apply our method to analyze the actual dataset. Utilizing the \texttt{nestle} algorithm as in \cref{sec:lc_interpolation,sec:flux_calibration,sec:time_explosion}, we sample the parameters to derive a \num{22}-dimensional posterior distribution ($H_0$, $\sigma_\mathrm{int}$, $t_\mathrm{0,SN1}$, \dots, $t_\mathrm{0,SN10}$, $v_\mathrm{pec,SN1}$, \dots, $v_\mathrm{pec,SN10}$).
A two-dimensional projection of the posterior, highlighting the key parameters $H_0$ and $\sigma_\mathrm{int}$, is illustrated in \cref{fig:h0_sigma_int_posterior} using the \texttt{corner} Python package \citep{corner}.
Our fit yields a Hubble constant of \myhknot.

A comprehensive comparison of our $H_0$ with estimates from other methods, alongside a discussion on its precision, is provided in \cref{sec:h0_value_discussion}.
We address potential systematic uncertainties that could impact our findings in \cref{sec:systematic_uncertainties_discussion}.

We determine $\sigma_\mathrm{int}$ to be \mysigmaint, which has been obtained by rescaling the fit result from the fiducial $H_0$ value of \SI{70}{\km\per\second\per\Mpc} to the median measured value of \SI{74.9}{\km\per\second\per\Mpc} through a multiplication by the factor $70\,/\,74.9$ (see Eqs. \ref{eq:theta/v}, \ref{eq:DL}, and \ref{eq:sigma_int}). Our measured $\sigma_\mathrm{int}$ is significantly lower than the median quantified $\theta/v_\mathrm{ph}$ error of approximately \SI{4}{\percent} (see \cref{sec:theta/v}).
As a result, the combined median $\theta/v_\mathrm{ph}$ uncertainty remains essentially the same as the quantified uncertainty, around \SI{4}{\percent}, which is less than half of the heuristic estimate of \SI{10}{\percent} proposed in \citet{Dessart2006} and employed in our previous studies \citep{Vogl2020,GezaM51,GezaSiblings}.

Although the posterior distribution of $\sigma_\mathrm{int}$ includes higher values (up to a few percent), values large enough ($\gtrsim$\SI{9}{\percent}) to push the combined uncertainty beyond the \citet{Dessart2006} threshold are essentially ruled out, comprising less than \SI{0.02}{\percent} of the posterior.
However, this conclusion depends on the accurate quantification of the explosion time and peculiar velocity uncertainties. Should these uncertainties have been overestimated, $\sigma_\mathrm{int}$ might be higher than calculated.

The precision of our $H_0$ measurement, \SI{1.9}{\km \per \second \per \Mpc}, slightly surpasses that achieved in the simulated data fits (see \cref{sec:bayes_ensemble_epm_validation}), where a median precision of \SI{2.2}{\km \per \second \per \Mpc} was obtained with $\sigma_\mathrm{int} = 5\,\%$. This improvement is due to the lower intrinsic scatter, $\sigma_\mathrm{int}=$\mysigmaint, in our real data.

\begin{figure*}
	\begin{center}
		\includegraphics[]{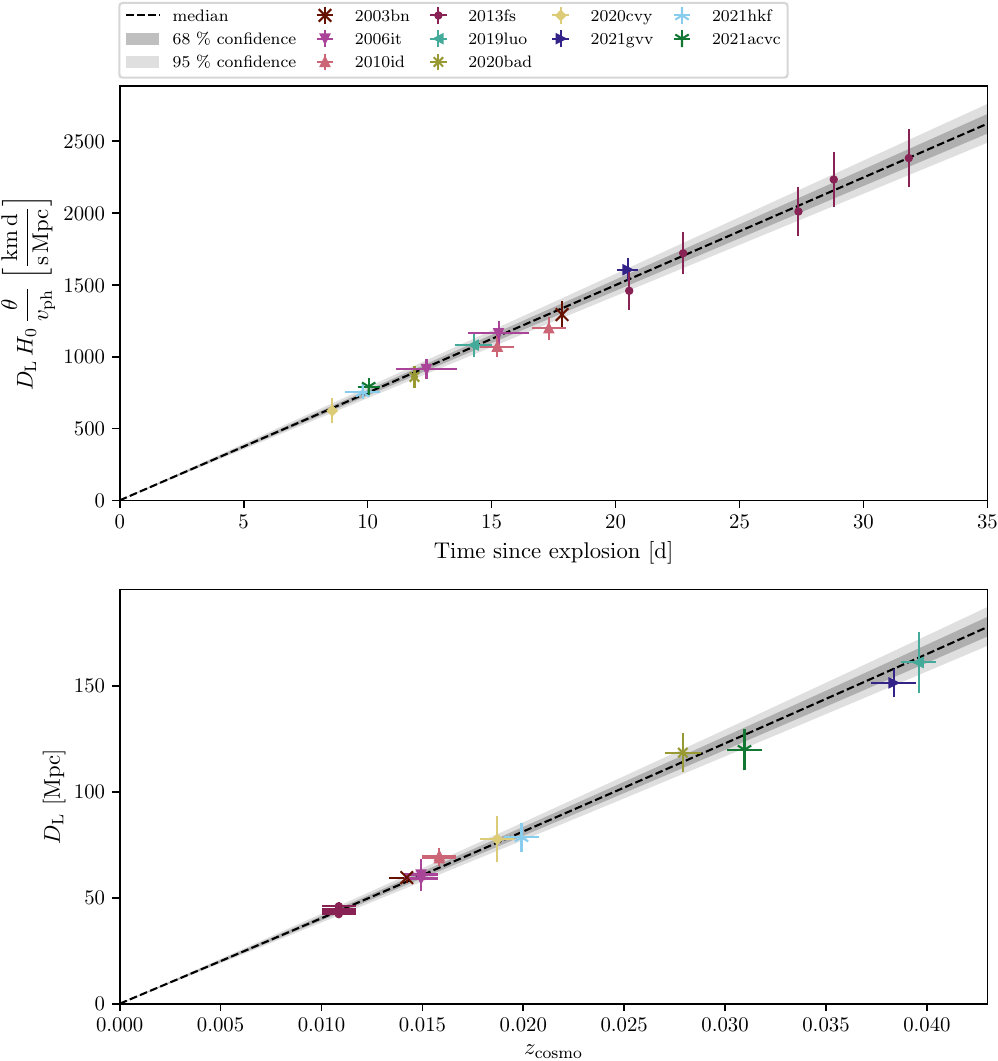}
	\end{center}
	\caption[]{Visualization of the Bayesian ensemble EPM fit. The upper panel presents the combined EPM regressions for all SNe, with $\theta/v_\mathrm{ph}$ values scaled by the Hubble-free luminosity distance $D_\mathrm{L}(z_\mathrm{cosmo}) H_0$, placing them on a uniform absolute scale. Each spectrum contributes one measurement to this plot, with the median of the fit indicated by a dashed line. The \SI{68}{\percent} and \SI{95}{\percent} confidence intervals are shown in light and dark grey, respectively. Each SN is color-coded for clarity. The error bars include the inferred additional uncertainty $\sigma_\mathrm{int}$. In the bottom panel, we complement this visualization with a plot similar to a classical Hubble diagram, where $D_\mathrm{L}$ values, derived by solving \cref{eq:theta/v}, are plotted against their redshifts. Unlike traditional Hubble diagrams, each point represents one epoch of an SN, with all spectra for a given SN appearing at the same redshift.

		Special care is needed when interpreting the error bars. In the EPM-like visualization, the y-errors include both $\theta/v_\mathrm{ph}$ and peculiar velocity uncertainties, whereas in the Hubble-diagram-like visualization, they include both $\theta/v_\mathrm{ph}$ and time of explosion uncertainties. The time of explosion and peculiar velocity uncertainties are completely correlated for all epochs of an SN, introducing correlations in the y-error bars that cannot be captured in this figure. The situation is even worse for the uncertainties in the x-direction, which are solely the time of explosion and peculiar velocity uncertainties, and thus are \SI{100}{\percent} correlated. The data points are thus not independent, making it more difficult to assess aspects like the dispersion around the fit. It is important to remember, however, that we do not use the plotted data points for the fit; instead, we use the $\theta/v_\mathrm{ph}$ values directly.}
	\label{fig:ensemble_epm_visualization}
\end{figure*}

We present a visualization of the fit in \cref{fig:ensemble_epm_visualization}.
To reflect the internal dynamics of the fit, the most precise method would be to display ten separate EPM regression diagrams, each tracking the temporal evolution of $\theta/v_\mathrm{ph}$ for an individual SN (as in the schematic example in \cref{fig:cartoon_EPM}).

However, this approach does not clearly demonstrate the interconnection of SNe through the shared $H_0$ and complicates the assessment of sample properties, such as the dispersion around the fit or suspicious trends that could hint towards systematic errors. To address these issues, we combine all EPM regressions into a single plot, scaling the $\theta/v_\mathrm{ph}$ values by the Hubble free luminosity distance $D_\mathrm{L}(z_\mathrm{cosmo})  H_0$. This places them on a uniform absolute scale, where the slope of the fit corresponds directly to the Hubble constant.

In the bottom panel of \cref{fig:ensemble_epm_visualization}, we complement the EPM regression-like plot with a visualization similar to a classical Hubble diagram. To achieve this, we solve \cref{eq:theta/v} for $D_\mathrm{L}$, using our prior knowledge of $z_\mathrm{cosmo}$ and $t_\mathrm{0}$ for all epochs of all SNe, and plot the resulting $D_\mathrm{L}$ values against their redshifts. Unlike a traditional Hubble diagram, where each point represents a single SN, here each point represents an epoch of an SN.

Both visualizations indicate that the data closely follow the expected trends with minimal dispersion and no obvious systematic errors. For instance, any deviation from a linear trend in the EPM-like visualization after the hydrogen recombination epoch ($\sim$ \SI{20}{\day}) might suggest that neglecting time-dependent ionization introduces errors in the distance calculations at these phases. While the current dataset does not allow for stringent constraints, future datasets will enable these plots to be powerful tools for identifying potential systematics in the method.
\footnotetext[1]{\url{https://corner.readthedocs.io/en/latest/pages/sigmas/}}
We present a visualization of the fit in \cref{fig:ensemble_epm_visualization}.
To reflect the internal dynamics of the fit, the most precise method would be to display ten separate EPM regression diagrams, each tracking the temporal evolution of $\theta/v_\mathrm{ph}$ for an individual SN (as in the schematic example in \cref{fig:cartoon_EPM}).

\section{Discussion} \label{sec:discussion}
In the previous section, we presented our measurement of the Hubble constant $H_0$. While the value of $H_0$ is crucial, the accompanying uncertainty is equally important to understand the significance of the result. To contextualize our measurement, we first address systematic uncertainties not included in the statistical error in \cref{sec:systematic_uncertainties_discussion}. After understanding these uncertainties, we compare our results to other $H_0$ estimates from the distance ladder, early universe measurements, and comparable one-step local methods (see \cref{sec:h0_value_discussion}).

\subsection{Sources of systematic uncertainties}
\label{sec:systematic_uncertainties_discussion}

\subsubsection{Modelling limitations}
Radiative transfer modelling is central to our measurement of $H_0$; however, approximations in both the radiative transfer and parameterized SN models can introduce systematic errors in the inferred distances.

Potential errors stem from several key approximations. For example, in the radiative transfer, the lack of a full non-local thermodynamic equilibrium (NLTE) treatment for all species can affect the results. In the parameterized SN models, the assumption of a power-law density profile may not accurately reflect the true structure of SN ejecta.

These approximations influence our measurements in two ways: they can directly impact the luminosity for a given set of parameters, and they can alter the spectral appearance, leading to incorrect parameter inferences and subsequently biased luminosity estimates.

An example for the first mechanism is the assumption of a power-law density profile. By neglecting the flattening of density profiles toward the center of SNe~II, we might overestimate the density in the continuum-forming regions. This overestimation leads to less flux dilution and, consequently, an overestimate of the luminosity, while causing minimal changes to the spectral appearance.

An example for the second mechanism is the approximate treatment of excitation and ionization for iron-group elements responsible for metal line blanketing. This simplification might cause us to systematically misestimate the photospheric temperature to match the observed strength of metal line blanketing.

Beyond these examples, other approximations could affect our results, such as assuming spherical symmetry, homologous expansion, and steady-state conditions for the radiation field and plasma state. While addressing all these factors in detail is beyond this paper's scope, testing these sources of error is crucial for establishing a solid foundation for our method.

An important approach to assess these limitations is by comparing our models to more comprehensive simulations, either with more sophisticated input models featuring realistic ejecta structures or with more advanced radiative transfer physics -- or both.
The gold standard here are time-dependent NLTE \textsc{cmfgen} models, based on hydrodynamical simulations of exploded stellar evolution models \citep[e.g.,][]{2011MNRAS.410.1739D,2019A&A...631A...8H}. These models are ideal for testing but not suited for spectral fitting because they are computationally intensive and their complex ejecta structures cannot be adequately described by a small set of parameters.

Although we have not conducted extensive tests with these advanced models yet, our code has been applied to some of the same problems as the older, steady-state version of \textsc{cmfgen}, showing good agreement -- specifically in the calculation of dilution factors and distance measurements for SN~1999em and SN~2005cs.

The temperature-binned dilution factors from \citet{Dessart2005} and \citet{Vogl2019} agree within a few percent, with maximum deviations of \SI{8}{\percent} after approximately accounting for differences in average photospheric densities between the model sets.

For SN~1999em, the distance estimates from \citet{Dessart2006} (\SI[separate-uncertainty=true,multi-part-units = single]{11.5(10)}{\Mpc}) and \citet{Vogl2020} (\SI[separate-uncertainty=true,multi-part-units = single]{11.4(10)}{\Mpc}) are in remarkable agreement. Both estimates also align well with the host galaxy's Cepheid distance, \SI[separate-uncertainty=true,multi-part-units = single]{11.3(5)}{\Mpc} \citep{2020MNRAS.496.3402D}. 

For SN~2005cs, our best distance estimate is \SI[separate-uncertainty=true,multi-part-units = single]{7.3(4)}{\Mpc} \citep{GezaM51}, which disagrees by almost \SI{20}{\percent} with the estimate from \citet{Dessart2008} of \SI[separate-uncertainty=true,multi-part-units = single]{8.9(5)}{\Mpc}. This discrepancy can largely be explained by a difference of \SI{1.6}{\day} in the assumed explosion time. Our study incorporates additional constraints from the early light curve, whereas \citet{Dessart2008} rely solely on the EPM regression. This time difference translates to approximately a \SI{16}{\percent} difference in distance at the median phase of the \textsc{cmfgen} models ($\sim$\SI{10}{\day}), explaining most of the \SI{20}{\percent} deviation.

While these agreements are encouraging, they do not eliminate the need for more direct comparisons, especially with more complex time-dependent calculations. 

Nevertheless, we can also assess the impact of modelling limitations empirically, as they do not affect every SN and epoch equally. For example, the approximate NLTE treatment of species other than hydrogen and helium likely becomes significant only after prominent metal lines develop. Similarly, neglecting time dependence in the ionization balance becomes important once hydrogen recombination sets in \citep[see, e.g.,][]{Dessart2008}.

If these approximations cause significant errors, we would expect to see variations in the inferred distances during the phases where these effects are important compared to when they are not -- unless compensated by other errors. However, we do not observe strong evidence for such variations in the objects studied here (see \cref{fig:ensemble_epm_visualization}) or in the sibling SN analysis by \citet{GezaSiblings}. 

Although larger sample studies are necessary for definitive answers, the low dispersion between objects and epochs in our study and in \citet{GezaSiblings} suggests that modelling limitations do not introduce substantial errors. Since most limitations are more significant at certain epochs and for certain objects, they would lead to some dispersion if they were a dominant error source.

In conclusion, while modelling limitations warrant further investigation, the evidence suggests that they are not overwhelmingly detrimental to our distance measurements. Continued efforts to test and refine our models -- through comparisons with more sophisticated simulations and empirical studies -- will enhance the reliability of our method.    

\subsubsection{Circumstellar material}
Narrow emission lines in early-time spectra \citep[e.g.,][]{2016ApJ...818....3K, Bruch2021, Bruch2023} and light curve modelling \citep[e.g.,][]{2017ApJ...838...28M} suggest that confined circumstellar material (CSM) is common around SNe~II, although the mass-loss mechanism responsible remains debated \citep{2017MNRAS.470.1642F,2024OJAp....7E..47F}. 
This topic has gained attention in recent years, sparked by early observations of events like SN~2013fs, which exhibited narrow emission lines for less than two days post-explosion \citep{Yaron2017}, and more extreme cases like SN~2023ixf, where narrow lines persisted for nearly ten days \citep[e.g.,][]{2023ApJ...954L..12T, 2023ApJ...954L..42J}.

The first mechanism through which CSM could affect our measurements is a potential boost in luminosity due to ongoing interaction. If the ejecta-CSM interaction continues beyond the early explosion phases, it can produce additional luminosity currently not accounted for in the models, potentially leading to errors in the inferred luminosity. 

For this effect to significantly impact our measurements, the interaction would need to persist during the timeframes where our models are typically applied -- more than ten days post-explosion. Given the confined nature of the CSM, such ongoing interaction may be unlikely at these phases. For example, \citet{2024ApJ...970..189J} suggested that SNe like SN~2013fs, with short-lived narrow lines, originate from compact CSM with radii smaller than around \SI{5e14}{\cm}. In these cases, the CSM would likely be swept up by the ejecta within a few days, given typical ejecta velocities of around \SI{10000}{\km \per \second}. Additionally, \citet{Bruch2023} found no significant differences in the early light curves of SNe~II with or without early narrow emission lines, suggesting that CSM interaction might not significantly impact the light curve in most cases, except in rare events where narrow lines persist beyond ten days.

The second mechanism by which CSM interaction could affect our measurements is through changes to the ejecta structure and kinematics, influencing the SN’s spectral appearance long after the interaction has ended \citep[e.g.,][]{2019A&A...631A...8H}. Since these effects are not incorporated into our models, they can introduce biases in the inferred parameters and ultimately in the luminosity. 

One example is the deceleration of the fastest-moving ejecta during the interaction, which leads to a deficiency of high-velocity material. To account for this in our power-law density models, we can only mimic the effect by steepening the gradient. However, this adjustment may overestimate the density in the continuum-forming layers, reducing flux dilution and slightly overestimating the luminosity.

While these mechanisms show how CSM interaction could introduce biases, its effect on the tailored EPM has not been extensively studied.
Nevertheless, theoretical and empirical approaches offer promising ways to explore this further. On the theoretical side, fitting simulations that include CSM interaction, such as those by \citet{2019A&A...631A...8H} and \citet{2023A&A...677A.105D}, with our models could help quantify the errors introduced by ignoring this interaction.

Empirically, we could analyze larger SN samples to determine whether objects with CSM interaction show systematically different distance estimates, by studying SNe in the Hubble flow or sibling SNe in the same galaxy. A significant advantage is that CSM interaction often leaves clear spectral imprints, such as an H$\alpha$ profile with a weak or absent absorption component \citep[][]{2019A&A...631A...8H}, which can persist for weeks post-explosion. These features allow us to identify and test objects affected by CSM interaction.

In our current sample, only SN~2013fs shows spectral signs of CSM interaction, and it does not significantly deviate from the rest. Based on this one object, we cannot draw strong conclusions about small systematic errors. However, since only one out of ten SNe in our sample shows evidence of CSM interaction, the overall impact on distance measurements in this study is likely small.

\subsubsection{Total-to-selective extinction ratio}
Systematic errors can also arise from the host galaxy extinction. We can estimate the color excess through our spectral fits, as described in \cref{sec:theta/v}, by aligning the slopes of the model and observed spectra. However, our optical spectra do not strongly constrain the total-to-selective extinction ratio $R_V$ because the slopes of the reddened model spectra are relatively insensitive to $R_V$ for a given $E(B-V)$; the slope in the $B$ and $V$ filters stays constant by definition.
Differences in $R$ and $I$ only become noticeable for large changes in $R_V$.

Consequently, we adopt a fixed $R_V$ of \num{3.1}, typical of Milky Way dust. However, even in the Milky Way, $R_V$ varies slightly \citep[$\sigma(R_V) \approx 0.2$;][]{2016ApJ...821...78S}. Galaxies in the local universe exhibit a wider range of $R_V$ values \citep[e.g.,][]{2018ApJ...859...11S}, and similar diversity is found in SNe~Ia \citep[e.g.,][]{2021ApJ...923..237J,2023MNRAS.525.5187W,2024MNRAS.530.4016T}. While the situation for SNe~II is less studied, they likely also exhibit some variation in $R_V$, introducing at least statistical uncertainty.

A systematic error could arise if the average $R_V$ for SNe~II deviates from \num{3.1}, similar to SNe~Ia, where lower values around \num{2.5} are commonly reported \citep[see, e.g., Table A.1 of][summarizing recent measurements]{2024MNRAS.530.4016T}.
Although SNe~Ia and II occur in different environments and may not share the same dust properties, the lower average $R_V$ in SNe~Ia provides a useful benchmark for assessing the impact of $R_V$.

The effect of $R_V$ is most easily demonstrated in the $V$-band, where the extinction equations are particularly simple. Specifically, the photospheric angular diameter $\theta_V$ depends on the extinction properties as follows:
\begin{equation} \label{eq:theta_ebv_dep}
	\theta_V \propto \sqrt{f_{V,\mathrm{obs}} 10^{0.4\,E(B-V)\,R_V}}
\end{equation}
Assuming that the SN parameters and $E(B-V)$ derived from the spectral fits remain unaffected by $R_V$, which is reasonable given the weak sensitivity of the dereddened model spectrum's slope and shape to $R_V$, we can calculate $\theta$ for any new $R_V$ relative to our reference value of \num{3.1}:
\begin{eqnarray} \label{eq:theta_rv}
	\frac{\theta_{V}(R_V)}{\theta_{V}(R_V=3.1)} = 10^{0.2\,E(B-V)\,(R_V - 3.1)}\\ \approx 1 + 0.46\,\,E(B-V)\,(R_V - 3.1)
\end{eqnarray}
Since the ratio in other bands will be similar, given the weak sensitivity of the optical extinction curve's slope and shape to $R_V$, we estimate that lowering $R_V$ to \num{2.5} results in roughly a \SI{2}{\percent} decrease in $\theta/v_\mathrm{ph}$ for the median $E(B-V)_\mathrm{host}$ of our sample (\num{0.065}). Because $H_0$ is directly proportional to $\theta/v_\mathrm{ph}$, we expect this change to similarly decrease $H_0$.

To verify our back-of-the-envelope calculation, we rescale the $\theta/v_\mathrm{ph}$ values for our spectral fits, including all flux calibration solutions and $E(B-V)$ values. We then obtain updated values for the mean and covariance of the joint $\theta/v_\mathrm{ph}$ distribution through the procedure in \cref{sec:theta/v}, and repeat the $H_0$ fit as in \cref{sec:bayes_ensemble_epm}, confirming the \SI{2}{\percent} decrease in $H_0$.
This \SI{2}{\percent} decrease is comparable to the statistical uncertainty of our measurement, demonstrating that extinction could be an important source of systematic uncertainty.

Given the potential impact, further study of the extinction properties in SNe~II is crucial. This can be achieved either directly, through optical, UV, and infrared data, or indirectly by examining trends in Hubble residuals and distance discrepancies between SN~II siblings as a function of extinction.

However, \citet{2018ApJ...859...11S} found that quiescent galaxies have an average $R_V$ of \num{2.61}, while star-forming galaxies have an average $R_V$ of \num{3.15}.
This suggests that the lower $R_V$ observed in SNe~Ia might result from their presence in both quiescent and star-forming galaxies. In contrast, SNe~II almost exclusively occur in star-forming galaxies, making it likely that their average $R_V$ is closer to the Milky Way value of \num{3.1} used in our study. This would potentially reduce the impact of this systematic uncertainty on our measurements.

\subsubsection{Photometric system} \label{sec:photo_sys_uncerts}
The calibration and characterization of the photometric system represent another significant source of systematic errors, introducing correlated uncertainties across all SNe observed with the same instrument. These uncertainties fall into two main categories:
\begin{enumerate}
    \item Uncertainties in the calibration of photometric zero points (ZPs), directly affecting the observed magnitudes.
    \item Uncertainties in the quantification of passbands, which are crucial for performing synthetic photometry that can be compared to observations.
\end{enumerate}
These uncertainties impact our measurements through two avenues. First, in the direct calculation of $\theta$ (see \cref{eq:theta}), which uses the observed magnitudes ($f_{\lambda,\mathrm{obs}}$) and synthetic model magnitudes ($f_{\lambda_\mathrm{em}, \mathrm{ph}}$). Second, in the spectral recalibration, where observed photometric flux is compared to synthetic flux from observed spectra. Here, photometric uncertainties can systematically alter the color of recalibrated spectra, which then propagates into the parameters inferred from spectral fits.

While ZP and passband uncertainties affect measurements through similar mechanisms, their statistical properties differ: ZP uncertainties influence all SNe (and epochs) observed with the same instrument uniformly. Passband uncertainties depend on the spectral slope and shape, leading to varying effects across SNe and epochs.

Relatively modern surveys, such as CfA3, tend to have small ZP uncertainties ($\sim$\SI{0.01}{mag}), while older surveys can exhibit larger uncertainties, around \SI{0.03}{mag} \citep[see Table 2 of][]{RubinUnion}. Passband uncertainties are more complex, involving the full wavelength-dependent transmission function. To manage this complexity, they are typically approximated by wavelength shifts of the measured passband. \citet{RubinUnion} find that shifts smaller than \SI{100}{\AA} usually suffice to align observed and synthetic color terms, though shifts exceeding \SI{200}{\AA} occasionally occur.

For the ePESSTO+ spectra, passband shifts of \SI{100}{\AA} change synthetic magnitudes on the order of \SI{0.01}{mag}, similar to the estimated ZP uncertainty. However, since passband uncertainties are more statistical, they partially average out. Therefore, we focus primarily on ZP uncertainties when estimating the photometric system's impact.

A typical ZP uncertainty of \SI{0.01}{mag} would cause a \SI{0.5}{\percent} error in $\theta$ (and consequently $H_0$) through avenue one if all SNe were observed in a single band with the same instrument. However, since our measurements span multiple bands and instruments, this small \SI{0.5}{\percent} error becomes even less significant.

In the second avenue (spectral recalibration), the effect is strongest for SNe observed in only two filters, as is the case for much of the ePESSTO+ sample. Here, a systematic error of \SI{0.014}{mag} ($\sqrt{2} \cdot \SI{0.01}{mag}$) in the $g-r$ color could lead to a corresponding $E(B-V)$ error of a similar magnitude.\footnote{The color excesses in $g-r$ and $B-V$ are nearly identical for $R_V = 3.1$.}  This would result in a non-negligible systematic error in $\theta$ and $H_0$ of around \SI{2}{\percent}, calculated using \cref{eq:theta_ebv_dep}, comparable to the statistical uncertainty. The availability of 
$i$-band photometry for some ePESSTO+ objects helps mitigate this effect. Additionally, since the ePESSTO+ sample represents only \SI{60}{\percent} of our dataset, the impact is further reduced.

In summary, photometric system uncertainties, particularly their impact on observed colors, can be significant sources of systematic error when most observations are from the same instrument. However, in our study, these uncertainties are likely subdominant to statistical errors, as we utilize data from multiple instruments and typically observe in more than three filters.

\subsubsection{Conclusion of systematic uncertainties}
We have identified several systematic uncertainties that could impact our measurement of the Hubble constant: modelling limitations, CSM interaction, the total-to-selective extinction ratio, and the photometric system. While we do not find very significant effects from these sources, some uncertainties -- for example, those related to the total-to-selective extinction ratio -- could be comparable to our statistical uncertainties.

Although we have addressed key systematic uncertainties, our discussion is not exhaustive. For example, we have not considered the possibility of a local Hubble bubble \citep[][]{1998ApJ...503..483Z} which could influence measurements at low redshifts and motivates other analyses \citep[e.g.,][]{Riess2016,Riess2022} to use a larger redshift cutoff.

Overall, while some systematic uncertainties could affect our results at the level of our statistical uncertainties, they are unlikely to significantly alter our main conclusions. With this understanding of our errors, we proceed to compare our measurement of the Hubble constant with those derived from other methods.

\subsection{Comparison with other Hubble-constant measurements}
\label{sec:h0_value_discussion}
 In \cref{sec:bayes_ensemble_epm_results}, we determined the Hubble constant to be \myhknot, considering only statistical uncertainties. Systematic uncertainties may be of a similar scale  (see \cref{sec:systematic_uncertainties_discussion}). This is the first $H_0$ determination using the tailored EPM, making it completely independent of existing measurements.

\begin{figure}
	\begin{center}
		\includegraphics[]{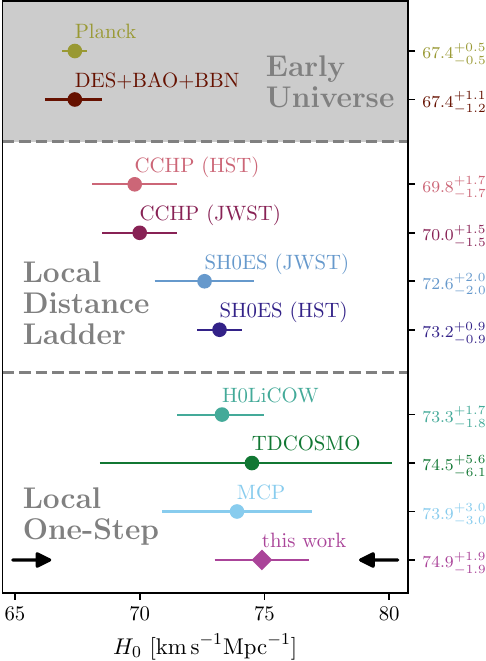}
	\end{center}
	\caption[]{Comparison of our Hubble constant ($H_0$) with other precise, recent measurements. Our result is shown in purple and highlighted with black arrows. Early universe constraints include CMB anisotropies and lensing from Planck \citep{PlanckCosmoParams}, as well as a CMB-independent measurement from BAO and BBN \citep[][]{2018MNRAS.480.3879A}. 
 
 Local distance ladder measurements are shown from the SH0ES project using \textit{HST} data with Cepheids \citep{Breuval2024} and from the CCHP project using \textit{HST} data with the TRGB \citep[]{Freedman2021}. Additionally, for the newer \textit{JWST} data, we include results from both SH0ES \citep[][]{Riess2024} and CCHP \citep{Freedman2024}, where each combines three distance indicators: Cepheids, JAGB stars, and the TRGB. The uncertainties for the distance ladder measurements include systematic uncertainties added in quadrature. For one-step measurements, we show results from time-delay lensing by H0LiCOW \citep{Wong2020} and TDCOSMO \citep{TDCOSMOH0}, megamasers from the MCP \citep[]{Pesce2020}, and our own tailored EPM measurement. Compared to the H0LiCOW analysis, TDCOSMO relaxes assumptions on the lens mass models and therefore has larger uncertainties. 
Among the one-step methods, only the lensing measurements include systematic uncertainties.
 }
	\label{fig:h0_value_comparison} 
\end{figure}

We compare our $H_0$ measurement to other local one-step methods, early universe constraints, and distance ladder results in \cref{fig:h0_value_comparison}. Despite our small sample size, our statistical precision of \SI{2.5}{\percent} is competitive. It matches leading one-step methods such as time-delay lensing by H0LiCOW \citep{Wong2020} and megamasers from the MCP \citep{Pesce2020}. However, it remains less precise than the best distance-ladder estimates, notably those from SH0ES \citep{Breuval2024}.

Our small statistical error results from the low dispersion in $H_0$ estimates across different SNe and epochs, highlighting the promise of the tailored EPM. This required minimal additional uncertainty ($\sigma_\mathrm{int}$) beyond the quantified uncertainties. Expanding the sample size and diversity of SNe~II will be crucial to confirm this consistency.

Our measurement aligns well with the $H_0$ value reported by SH0ES -- the primary local driver of the Hubble tension -- and with other precise one-step methods like time-delay lensing and megamasers.

While our results show a moderate $2\,\sigma$ tension with the CCHP results \citep{Freedman2021,Freedman2024} based on both \textit{HST} and \textit{JWST} data, they indicate a stronger $3.8\,\sigma$ tension with Planck’s $H_0$. However, because systematic errors are not included in our uncertainty budget, this $3.8\,\sigma$ significance is an upper limit. If we assume that systematic uncertainties are comparable to statistical ones, the tension reduces to about $2.7\,\sigma$.

Regardless of the exact significance, the agreement among the three precise, independent one-step measurements -- our tailored EPM, time-delay lensing, and megamasers -- provides compelling evidence that the Hubble tension is not simply a product of measurement systematics. This consistency across independent methods suggests that the tension may reflect a genuine discrepancy between local and early universe measurements.

\section{Summary and conclusion} \label{sec:summary}
We have presented a novel, distance-ladder-independent method to determine the Hubble constant using SNe~II through the tailored EPM. This approach addresses the ongoing discrepancy between local distance-ladder and early universe measurements of $H_0$.
Our tailored EPM technique uses advanced spectral modelling to efficiently fit supernova spectra and derive precise, self-consistent values for luminosity-related parameters. Applying this method to ten SNe~II, comprising both literature objects and ePESSTO+ classification targets, with redshifts from \num{0.01} to \num{0.04}, we find $H_0=\myhknot$.
The quoted uncertainty is statistical. A careful dissection of various dominant sources of systematic effects indicates that systematic errors are likely below or at most comparable to the statistical uncertainties.
Our inferred $H_0$ is consistent with most local measurements but in tension with Planck's early universe constraints.

Our study demonstrates that the tailored EPM achieves a precision comparable to established techniques like the TRGB-SN~Ia distance ladder and strong lensing, highlighting its potential as a robust alternative for $H_0$ estimation. However, improvements are needed, particularly in quantifying systematic uncertainties. Key improvements include a better understanding of the modelling uncertainties, especially through comparisons with time-dependent radiative transfer models like \textsc{cmfgen}, and empirical testing for systematic trends in the distances. 
The latter requires increased statistical power from more SNe and better constraints from individual objects.

Our current sample faces bottlenecks due to data availability or uncertainties from peculiar motions. For ePESSTO+ objects, peculiar motion uncertainties are mostly small, but the precision is limited by having only one spectrum and photometry in two bands. The literature sample has more spectra and photometry in at least three bands but is constrained by low redshifts, leading to significant uncertainties from peculiar velocities. Future work will focus on higher redshift SNe~II with multiple spectroscopic observations and comprehensive multi-band photometry to reduce uncertainties and enhance the method's reliability.

To address these needs, we have assembled two new datasets for future $H_0$ determinations. The ESO VLT large programme adH0cc\footnote{accurate determination of $H_0$ with core-collapse supernovae (\url{https://adh0cc.github.io/})} \citep[][]{adH0cc1st} has collected spectral time series (typically four to five spectra per object) and contemporaneous $BVRI$ photometry for around twenty SNe~II at redshifts between \num{0.03} and \num{0.17}. Additionally, the Nearby Supernova Factory \citep[SNfactory;][]{Aldering2002} has gathered a lower redshift sample ($z$ = $0.01\,$--$\,0.04$) of another twenty objects, with a median of six spectrophotometric observations per object, which will be published in Holas et al. (in preparation). While these objects share similar redshifts with our current sample, the improved data quality and quantity will significantly enhance our analysis.

With a dataset four times larger than our current sample and more data per object, these  observations will substantially reduce statistical uncertainties and allow more powerful tests for systematic trends. Moreover, these datasets, collected specifically for SN~II cosmology and carefully reduced by us, will further minimize observational uncertainties.
This includes the uncertainties related to flux calibration and photometric systems, which are better characterised for these single-instrument data. Future analyses will be fully blinded to ensure unbiased results.

These advancements will make our tailored EPM even more valuable in addressing the Hubble tension, providing an independent and precise tool for measuring $H_0$.

\begin{acknowledgements}
	Based on observations collected at the European
	Organisation for Astronomical Research in the Southern
	Hemisphere, Chile, as part of ePESSTO/ePESSTO+ (the
	extended Public ESO Spectroscopic Survey for Transient
	Objects Survey) under ESO programmes 1103.D-0328, 1106.D-0811, and 1108.D-0740.\\

	Based on observations obtained with the Samuel Oschin 48-inch Telescope at the Palomar Observatory as part of the Zwicky Transient Facility project. ZTF is supported by the National Science Foundation under Grant No. AST-1440341 and a collaboration including Caltech, IPAC, the Weizmann Institute for Science, the Oskar Klein Center at Stockholm University, the University of Maryland, the University of Washington, Deutsches Elektronen-Synchrotron and Humboldt University, Los Alamos National Laboratories, the TANGO Consortium of Taiwan, the University of Wisconsin at Milwaukee, and Lawrence Berkeley National Laboratories. Operations are conducted by COO, IPAC, and UW. The ZTF forced-photometry service was funded under the Heising-Simons Foundation grant \#12540303 (PI: Graham).
	This work has made use of data from the Asteroid Terrestrial-impact Last Alert System (ATLAS) project. The Asteroid Terrestrial-impact Last Alert System (ATLAS) project is primarily funded to search for near earth asteroids through NASA grants NN12AR55G, 80NSSC18K0284, and 80NSSC18K1575; byproducts of the NEO search include images and catalogs from the survey area. This work was partially funded by Kepler/K2 grant J1944/80NSSC19K0112 and HST GO-15889, and STFC grants ST/T000198/1 and ST/S006109/1. The ATLAS science products have been made possible through the contributions of the University of Hawaii Institute for Astronomy, the Queen’s University Belfast, the Space Telescope Science Institute, the South African Astronomical Observatory, and The Millennium Institute of Astrophysics (MAS), Chile.
 The authors would like to acknowledge the use of ChatGPT (version 4.0 and 4.01-preview) developed by OpenAI for assisting with language editing of this manuscript. The model's capabilities were employed to improve clarity, grammar, and coherence in several sections of the text.
	This research has made use of the NASA/IPAC Extragalactic Database (NED),
	which is operated by the Jet Propulsion Laboratory, California Institute of Technology,
	under contract with the National Aeronautics and Space Administration.
	This research made use of \textsc{tardis}, a community-developed software package for spectral
	synthesis in supernovae \citep{Kerzendorf2014, kerzendorf_wolfgang_2023_8244935}. The
	development of \textsc{tardis} received support from GitHub, the Google Summer of Code
	initiative, and from ESA's Summer of Code in Space program. \textsc{tardis} is a fiscally
	sponsored project of NumFOCUS. \textsc{tardis} makes extensive use of Astropy and Pyne. This research is supported in part by the Excellence Cluster ORIGINS, which is funded by the Deutsche Forschungsgemeinschaft (DFG, German Research Foundation) under Germany's Excellence Strategy-EXC-2094-390783311. This work was supported by the ‘Programme National de Physique Stellaire’ (PNPS) of CNRS/INSU cofunded by CEA and CNES. AH acknowledges support by the Klaus Tschira Foundation. AH is a Fellow of the International Max Planck Research School for Astronomy and Cosmic Physics at
the University of Heidelberg (IMPRS-HD). AF acknowledges support by the European Research Council (ERC) under the European Union’s Horizon 2020 research and innovation program (ERC Advanced Grant KILONOVA No. 885281) and the State of Hesse within the Cluster Project ELEMENTS. JVS is supported by the National Science Foundation under Grant No. OAC-2311323 and through the U.S. Department of Energy, Office of Science, Office of Nuclear Physics, award No. DE-SC0017955. SHS thanks the Max Planck Society for support through the Max Planck Fellowship. The authors thank Joe Lyman und Matt Nicholl for taking the calibration data needed for the EFOSC2 second order correction.
\\
Contributor roles according to the CRediT standard:\footnote{\url{https://credit.niso.org/}} \\
Conceptualization: W.~Hillebrandt, B.~Leibundgut, C.~Vogl, S.~Taubenberger\\
Data curation, formal analysis, and visualization: C.~Vogl\\
Funding acquisition and resources: W.~Hillebrandt, S.~H.~Suyu\\
Investigation, methodology, and validation: C.~Vogl, S.~Taubenberger\\
Project administration: C.~Vogl, W.~Hillebrandt, B.~Leibundgut\\
Software: C.~Vogl, W.~E.~Kerzendorf, S.~A.~Sim\\
Supervision: W.~Hillebrandt, S.~H.~Suyu, B.~Leibundgut, W.~E.~Kerzendorf, S.~A.~Sim\\
Writing – original draft: C.~Vogl, G.~Cs\"ornyei\\
Writing – review \& editing:  C.~Vogl, S.~Taubenberger, G.~Cs\"ornyei, B.~Leibundgut, W.~E.~Kerzendorf, S.~A.~Sim, S.~Blondin, A.~Fl\"ors, A.~Holas, J.~V.~Shields, J.~Spyromilio, S.~H.~Suyu, W.~Hillebrandt
\end{acknowledgements}


\bibliographystyle{aa} 
\bibliography{aanda}

\begin{thebibliography}{177}
\expandafter\ifx\csname natexlab\endcsname\relax\def\natexlab#1{#1}\fi

\bibitem[{{Abbott} {et~al.}(2018){Abbott}, {Abdalla}, {Annis}, {Bechtol}, {Blazek}, {Benson}, {Bernstein}, {Bernstein}, {Bertin}, {Brooks}, {Burke}, {Carnero Rosell}, {Carrasco Kind}, {Carretero}, {Castander}, {Chang}, {Crawford}, {Cunha}, {D'Andrea}, {da Costa}, {Davis}, {DeRose}, {Desai}, {Diehl}, {Dietrich}, {Doel}, {Drlica-Wagner}, {Evrard}, {Fernandez}, {Flaugher}, {Fosalba}, {Frieman}, {Garc{\'\i}a-Bellido}, {Gaztanaga}, {Gerdes}, {Giannantonio}, {Gruen}, {Gruendl}, {Gschwend}, {Gutierrez}, {Hartley}, {Henning}, {Honscheid}, {Hoyle}, {Huterer}, {Jain}, {James}, {Jarvis}, {Jeltema}, {Johnson}, {Johnson}, {Krause}, {Kuehn}, {Kuhlmann}, {Kuropatkin}, {Lahav}, {Liddle}, {Lima}, {Lin}, {MacCrann}, {Maia}, {Manzotti}, {March}, {Marshall}, {Miquel}, {Mohr}, {Natoli}, {Nugent}, {Ogando}, {Park}, {Plazas}, {Reichardt}, {Reil}, {Roodman}, {Ross}, {Rozo}, {Rykoff}, {Sanchez}, {Scarpine}, {Schubnell}, {Scolnic}, {Sevilla-Noarbe}, {Sheldon}, {Smith}, {Smith}, {Soares-Santos}, {Sobreira}, {Suchyta}, {Tarle},
  {Thomas}, {Troxel}, {Walker}, {Wechsler}, {Weller}, {Wester}, {Wu}, {Zuntz}, {Dark Energy Survey Collaboration}, \& {South Pole Telescope Collaboration}}]{2018MNRAS.480.3879A}
{Abbott}, T.~M.~C., {Abdalla}, F.~B., {Annis}, J., {et~al.} 2018, \mnras, 480, 3879

\bibitem[{{Aiola} {et~al.}(2020){Aiola}, {Calabrese}, {Maurin}, {Naess}, {Schmitt}, {Abitbol}, {Addison}, {Ade}, {Alonso}, {Amiri}, {Amodeo}, {Angile}, {Austermann}, {Baildon}, {Battaglia}, {Beall}, {Bean}, {Becker}, {Bond}, {Bruno}, {Calafut}, {Campusano}, {Carrero}, {Chesmore}, {Cho}, {Choi}, {Clark}, {Cothard}, {Crichton}, {Crowley}, {Darwish}, {Datta}, {Denison}, {Devlin}, {Duell}, {Duff}, {Duivenvoorden}, {Dunkley}, {D{\"u}nner}, {Essinger-Hileman}, {Fankhanel}, {Ferraro}, {Fox}, {Fuzia}, {Gallardo}, {Gluscevic}, {Golec}, {Grace}, {Gralla}, {Guan}, {Hall}, {Halpern}, {Han}, {Hargrave}, {Hasselfield}, {Helton}, {Henderson}, {Hensley}, {Hill}, {Hilton}, {Hilton}, {Hincks}, {Hlo{\v{z}}ek}, {Ho}, {Hubmayr}, {Huffenberger}, {Hughes}, {Infante}, {Irwin}, {Jackson}, {Klein}, {Knowles}, {Koopman}, {Kosowsky}, {Lakey}, {Li}, {Li}, {Li}, {Lokken}, {Louis}, {Lungu}, {MacInnis}, {Madhavacheril}, {Maldonado}, {Mallaby-Kay}, {Marsden}, {McMahon}, {Menanteau}, {Moodley}, {Morton}, {Namikawa}, {Nati}, {Newburgh},
  {Nibarger}, {Nicola}, {Niemack}, {Nolta}, {Orlowski-Sherer}, {Page}, {Pappas}, {Partridge}, {Phakathi}, {Pisano}, {Prince}, {Puddu}, {Qu}, {Rivera}, {Robertson}, {Rojas}, {Salatino}, {Schaan}, {Schillaci}, {Sehgal}, {Sherwin}, {Sierra}, {Sievers}, {Sifon}, {Sikhosana}, {Simon}, {Spergel}, {Staggs}, {Stevens}, {Storer}, {Sunder}, {Switzer}, {Thorne}, {Thornton}, {Trac}, {Treu}, {Tucker}, {Vale}, {Van Engelen}, {Van Lanen}, {Vavagiakis}, {Wagoner}, {Wang}, {Ward}, {Wollack}, {Xu}, {Zago}, \& {Zhu}}]{ACT2020}
{Aiola}, S., {Calabrese}, E., {Maurin}, L., {et~al.} 2020, \jcap, 2020, 047

\bibitem[{{Aldering} {et~al.}(2002){Aldering}, {Adam}, {Antilogus}, {Astier}, {Bacon}, {Bongard}, {Bonnaud}, {Copin}, {Hardin}, {Henault}, {Howell}, {Lemonnier}, {Levy}, {Loken}, {Nugent}, {Pain}, {Pecontal}, {Pecontal}, {Perlmutter}, {Quimby}, {Schahmaneche}, {Smadja}, \& {Wood-Vasey}}]{Aldering2002}
{Aldering}, G., {Adam}, G., {Antilogus}, P., {et~al.} 2002, Society of Photo-Optical Instrumentation Engineers (SPIE) Conference Series, Vol. 4836, {Overview of the Nearby Supernova Factory}, ed. J.~A. {Tyson} \& S.~{Wolff}, 61--72

\bibitem[{{Ambikasaran} {et~al.}(2015){Ambikasaran}, {Foreman-Mackey}, {Greengard}, {Hogg}, \& {O'Neil}}]{2015ITPAM..38..252A}
{Ambikasaran}, S., {Foreman-Mackey}, D., {Greengard}, L., {Hogg}, D.~W., \& {O'Neil}, M. 2015, IEEE Transactions on Pattern Analysis and Machine Intelligence, 38, 252

\bibitem[{{Anand} {et~al.}(2022){Anand}, {Tully}, {Rizzi}, {Riess}, \& {Yuan}}]{Anand2022}
{Anand}, G.~S., {Tully}, R.~B., {Rizzi}, L., {Riess}, A.~G., \& {Yuan}, W. 2022, \apj, 932, 15

\bibitem[{{Anderson} {et~al.}(2014){Anderson}, {Gonz{\'a}lez-Gait{\'a}n}, {Hamuy}, {Guti{\'e}rrez}, {Stritzinger}, {Olivares E.}, {Phillips}, {Schulze}, {Antezana}, {Bolt}, {Campillay}, {Castell{\'o}n}, {Contreras}, {de Jaeger}, {Folatelli}, {F{\"o}rster}, {Freedman}, {Gonz{\'a}lez}, {Hsiao}, {Krzemi{\'n}ski}, {Krisciunas}, {Maza}, {McCarthy}, {Morrell}, {Persson}, {Roth}, {Salgado}, {Suntzeff}, \& {Thomas-Osip}}]{Anderson2014}
{Anderson}, J.~P., {Gonz{\'a}lez-Gait{\'a}n}, S., {Hamuy}, M., {et~al.} 2014, \apj, 786, 67

\bibitem[{{Anderson} {et~al.}(2024){Anderson}, {Koblischke}, \& {Eyer}}]{Anderson2024}
{Anderson}, R.~I., {Koblischke}, N.~W., \& {Eyer}, L. 2024, \apjl, 963, L43

\bibitem[{{Astier} {et~al.}(2006){Astier}, {Guy}, {Regnault}, {Pain}, {Aubourg}, {Balam}, {Basa}, {Carlberg}, {Fabbro}, {Fouchez}, {Hook}, {Howell}, {Lafoux}, {Neill}, {Palanque-Delabrouille}, {Perrett}, {Pritchet}, {Rich}, {Sullivan}, {Taillet}, {Aldering}, {Antilogus}, {Arsenijevic}, {Balland}, {Baumont}, {Bronder}, {Courtois}, {Ellis}, {Filiol}, {Gon{\c{c}}alves}, {Goobar}, {Guide}, {Hardin}, {Lusset}, {Lidman}, {McMahon}, {Mouchet}, {Mourao}, {Perlmutter}, {Ripoche}, {Tao}, \& {Walton}}]{Astier2006}
{Astier}, P., {Guy}, J., {Regnault}, N., {et~al.} 2006, \aap, 447, 31

\bibitem[{{Baron} {et~al.}(2004){Baron}, {Nugent}, {Branch}, \& {Hauschildt}}]{Baron2004}
{Baron}, E., {Nugent}, P.~E., {Branch}, D., \& {Hauschildt}, P.~H. 2004, \apjl, 616, L91

\bibitem[{{Bazin} {et~al.}(2009){Bazin}, {Palanque-Delabrouille}, {Rich}, {Ruhlmann-Kleider}, {Aubourg}, {Le Guillou}, {Astier}, {Balland}, {Basa}, {Carlberg}, {Conley}, {Fouchez}, {Guy}, {Hardin}, {Hook}, {Howell}, {Pain}, {Perrett}, {Pritchet}, {Regnault}, {Sullivan}, {Antilogus}, {Arsenijevic}, {Baumont}, {Fabbro}, {Le Du}, {Lidman}, {Mouchet}, {Mour{\~a}o}, \& {Walker}}]{2009A&A...499..653B}
{Bazin}, G., {Palanque-Delabrouille}, N., {Rich}, J., {et~al.} 2009, \aap, 499, 653

\bibitem[{{Bellm} {et~al.}(2019){Bellm}, {Kulkarni}, {Graham}, {Dekany}, {Smith}, {Riddle}, {Masci}, {Helou}, {Prince}, {Adams}, {Barbarino}, {Barlow}, {Bauer}, {Beck}, {Belicki}, {Biswas}, {Blagorodnova}, {Bodewits}, {Bolin}, {Brinnel}, {Brooke}, {Bue}, {Bulla}, {Burruss}, {Cenko}, {Chang}, {Connolly}, {Coughlin}, {Cromer}, {Cunningham}, {De}, {Delacroix}, {Desai}, {Duev}, {Eadie}, {Farnham}, {Feeney}, {Feindt}, {Flynn}, {Franckowiak}, {Frederick}, {Fremling}, {Gal-Yam}, {Gezari}, {Giomi}, {Goldstein}, {Golkhou}, {Goobar}, {Groom}, {Hacopians}, {Hale}, {Henning}, {Ho}, {Hover}, {Howell}, {Hung}, {Huppenkothen}, {Imel}, {Ip}, {Ivezi{\'c}}, {Jackson}, {Jones}, {Juric}, {Kasliwal}, {Kaspi}, {Kaye}, {Kelley}, {Kowalski}, {Kramer}, {Kupfer}, {Landry}, {Laher}, {Lee}, {Lin}, {Lin}, {Lunnan}, {Giomi}, {Mahabal}, {Mao}, {Miller}, {Monkewitz}, {Murphy}, {Ngeow}, {Nordin}, {Nugent}, {Ofek}, {Patterson}, {Penprase}, {Porter}, {Rauch}, {Rebbapragada}, {Reiley}, {Rigault}, {Rodriguez}, {van Roestel}, {Rusholme}, {van
  Santen}, {Schulze}, {Shupe}, {Singer}, {Soumagnac}, {Stein}, {Surace}, {Sollerman}, {Szkody}, {Taddia}, {Terek}, {Van Sistine}, {van Velzen}, {Vestrand}, {Walters}, {Ward}, {Ye}, {Yu}, {Yan}, \& {Zolkower}}]{Bellm2019}
{Bellm}, E.~C., {Kulkarni}, S.~R., {Graham}, M.~J., {et~al.} 2019, \pasp, 131, 018002

\bibitem[{{Berbig} {et~al.}(2020){Berbig}, {Jana}, \& {Trautner}}]{2020PhRvD.102k5008B}
{Berbig}, M., {Jana}, S., \& {Trautner}, A. 2020, \prd, 102, 115008

\bibitem[{{Bessell} \& {Murphy}(2012)}]{Bessell2012}
{Bessell}, M. \& {Murphy}, S. 2012, \pasp, 124, 140

\bibitem[{{Betoule} {et~al.}(2014){Betoule}, {Kessler}, {Guy}, {Mosher}, {Hardin}, {Biswas}, {Astier}, {El-Hage}, {Konig}, {Kuhlmann}, {Marriner}, {Pain}, {Regnault}, {Balland}, {Bassett}, {Brown}, {Campbell}, {Carlberg}, {Cellier-Holzem}, {Cinabro}, {Conley}, {D'Andrea}, {DePoy}, {Doi}, {Ellis}, {Fabbro}, {Filippenko}, {Foley}, {Frieman}, {Fouchez}, {Galbany}, {Goobar}, {Gupta}, {Hill}, {Hlozek}, {Hogan}, {Hook}, {Howell}, {Jha}, {Le Guillou}, {Leloudas}, {Lidman}, {Marshall}, {M{\"o}ller}, {Mour{\~a}o}, {Neveu}, {Nichol}, {Olmstead}, {Palanque-Delabrouille}, {Perlmutter}, {Prieto}, {Pritchet}, {Richmond}, {Riess}, {Ruhlmann-Kleider}, {Sako}, {Schahmaneche}, {Schneider}, {Smith}, {Sollerman}, {Sullivan}, {Walton}, \& {Wheeler}}]{2014A&A...568A..22B}
{Betoule}, M., {Kessler}, R., {Guy}, J., {et~al.} 2014, \aap, 568, A22

\bibitem[{{Birrer} {et~al.}(2020){Birrer}, {Shajib}, {Galan}, {Millon}, {Treu}, {Agnello}, {Auger}, {Chen}, {Christensen}, {Collett}, {Courbin}, {Fassnacht}, {Koopmans}, {Marshall}, {Park}, {Rusu}, {Sluse}, {Spiniello}, {Suyu}, {Wagner-Carena}, {Wong}, {Barnab{\`e}}, {Bolton}, {Czoske}, {Ding}, {Frieman}, \& {Van de Vyvere}}]{TDCOSMOH0}
{Birrer}, S., {Shajib}, A.~J., {Galan}, A., {et~al.} 2020, \aap, 643, A165

\bibitem[{{Bishop} \& {Nasrabadi}(2007)}]{Bishop2007}
{Bishop}, C.~M. \& {Nasrabadi}, N.~M. 2007, Journal of Electronic Imaging, 16, 049901

\bibitem[{{Blakeslee} {et~al.}(2021){Blakeslee}, {Jensen}, {Ma}, {Milne}, \& {Greene}}]{Blakeslee2021}
{Blakeslee}, J.~P., {Jensen}, J.~B., {Ma}, C.-P., {Milne}, P.~A., \& {Greene}, J.~E. 2021, \apj, 911, 65

\bibitem[{{Blondin} {et~al.}(2008){Blondin}, {Davis}, {Krisciunas}, {Schmidt}, {Sollerman}, {Wood-Vasey}, {Becker}, {Challis}, {Clocchiatti}, {Damke}, {Filippenko}, {Foley}, {Garnavich}, {Jha}, {Kirshner}, {Leibundgut}, {Li}, {Matheson}, {Miknaitis}, {Narayan}, {Pignata}, {Rest}, {Riess}, {Silverman}, {Smith}, {Spyromilio}, {Stritzinger}, {Stubbs}, {Suntzeff}, {Tonry}, {Tucker}, \& {Zenteno}}]{2008ApJ...682..724B}
{Blondin}, S., {Davis}, T.~M., {Krisciunas}, K., {et~al.} 2008, \apj, 682, 724

\bibitem[{{Breuval} {et~al.}(2024){Breuval}, {Riess}, {Casertano}, {Yuan}, {Macri}, {Romaniello}, {Murakami}, {Scolnic}, {Anand}, \& {Soszy{\'n}ski}}]{Breuval2024}
{Breuval}, L., {Riess}, A.~G., {Casertano}, S., {et~al.} 2024, \apj, 973, 30

\bibitem[{{Breuval} {et~al.}(2022){Breuval}, {Riess}, {Kervella}, {Anderson}, \& {Romaniello}}]{2022ApJ...939...89B}
{Breuval}, L., {Riess}, A.~G., {Kervella}, P., {Anderson}, R.~I., \& {Romaniello}, M. 2022, \apj, 939, 89

\bibitem[{{Brout} {et~al.}(2022){Brout}, {Scolnic}, {Popovic}, {Riess}, {Carr}, {Zuntz}, {Kessler}, {Davis}, {Hinton}, {Jones}, {Kenworthy}, {Peterson}, {Said}, {Taylor}, {Ali}, {Armstrong}, {Charvu}, {Dwomoh}, {Meldorf}, {Palmese}, {Qu}, {Rose}, {Sanchez}, {Stubbs}, {Vincenzi}, {Wood}, {Brown}, {Chen}, {Chambers}, {Coulter}, {Dai}, {Dimitriadis}, {Filippenko}, {Foley}, {Jha}, {Kelsey}, {Kirshner}, {M{\"o}ller}, {Muir}, {Nadathur}, {Pan}, {Rest}, {Rojas-Bravo}, {Sako}, {Siebert}, {Smith}, {Stahl}, \& {Wiseman}}]{Pantheon+Cosmo}
{Brout}, D., {Scolnic}, D., {Popovic}, B., {et~al.} 2022, \apj, 938, 110

\bibitem[{{Bruch} {et~al.}(2021){Bruch}, {Gal-Yam}, {Schulze}, {Yaron}, {Yang}, {Soumagnac}, {Rigault}, {Strotjohann}, {Ofek}, {Sollerman}, {Masci}, {Barbarino}, {Ho}, {Fremling}, {Perley}, {Nordin}, {Cenko}, {Adams}, {Adreoni}, {Bellm}, {Blagorodnova}, {Bulla}, {Burdge}, {De}, {Dhawan}, {Drake}, {Duev}, {Dugas}, {Graham}, {Graham}, {Irani}, {Jencson}, {Karamehmetoglu}, {Kasliwal}, {Kim}, {Kulkarni}, {Kupfer}, {Liang}, {Mahabal}, {Miller}, {Prince}, {Riddle}, {Sharma}, {Smith}, {Taddia}, {Taggart}, {Walters}, \& {Yan}}]{Bruch2021}
{Bruch}, R.~J., {Gal-Yam}, A., {Schulze}, S., {et~al.} 2021, \apj, 912, 46

\bibitem[{{Bruch} {et~al.}(2023){Bruch}, {Gal-Yam}, {Yaron}, {Chen}, {Strotjohann}, {Irani}, {Zimmerman}, {Schulze}, {Yang}, {Kim}, {Bulla}, {Sollerman}, {Rigault}, {Ofek}, {Soumagnac}, {Masci}, {Fremling}, {Perley}, {Nordin}, {Cenko}, {Ho}, {Adams}, {Adreoni}, {Bellm}, {Blagorodnova}, {Burdge}, {De}, {Dekany}, {Dhawan}, {Drake}, {Duev}, {Graham}, {Graham}, {Jencson}, {Karamehmetoglu}, {Kasliwal}, {Kulkarni}, {Miller}, {Neill}, {Prince}, {Riddle}, {Rusholme}, {Sharma}, {Smith}, {Sravan}, {Taggart}, {Walters}, \& {Yan}}]{Bruch2023}
{Bruch}, R.~J., {Gal-Yam}, A., {Yaron}, O., {et~al.} 2023, \apj, 952, 119

\bibitem[{{Buzzoni} {et~al.}(1984){Buzzoni}, {Delabre}, {Dekker}, {Dodorico}, {Enard}, {Focardi}, {Gustafsson}, {Nees}, {Paureau}, \& {Reiss}}]{1984Msngr..38....9B}
{Buzzoni}, B., {Delabre}, B., {Dekker}, H., {et~al.} 1984, The Messenger, 38, 9

\bibitem[{{Carr} {et~al.}(2022){Carr}, {Davis}, {Scolnic}, {Said}, {Brout}, {Peterson}, \& {Kessler}}]{CarrPantheonPecV}
{Carr}, A., {Davis}, T.~M., {Scolnic}, D., {et~al.} 2022, \pasa, 39, e046

\bibitem[{{Carrick} {et~al.}(2015){Carrick}, {Turnbull}, {Lavaux}, \& {Hudson}}]{Carrick2015}
{Carrick}, J., {Turnbull}, S.~J., {Lavaux}, G., \& {Hudson}, M.~J. 2015, \mnras, 450, 317

\bibitem[{{Chen} {et~al.}(2019){Chen}, {Fassnacht}, {Suyu}, {Rusu}, {Chan}, {Wong}, {Auger}, {Hilbert}, {Bonvin}, {Birrer}, {Millon}, {Koopmans}, {Lagattuta}, {McKean}, {Vegetti}, {Courbin}, {Ding}, {Halkola}, {Jee}, {Shajib}, {Sluse}, {Sonnenfeld}, \& {Treu}}]{2019MNRAS.490.1743C}
{Chen}, G. C.~F., {Fassnacht}, C.~D., {Suyu}, S.~H., {et~al.} 2019, \mnras, 490, 1743

\bibitem[{{Ciddor}(1996)}]{1996ApOpt..35.1566C}
{Ciddor}, P.~E. 1996, \ao, 35, 1566

\bibitem[{{Cs{\"o}rnyei} {et~al.}(2023{\natexlab{a}}){Cs{\"o}rnyei}, {Anderson}, {Vogl}, {Taubenberger}, {Blondin}, {Leibundgut}, \& {Hillebrandt}}]{GezaM51}
{Cs{\"o}rnyei}, G., {Anderson}, R.~I., {Vogl}, C., {et~al.} 2023{\natexlab{a}}, \aap, 678, A44

\bibitem[{{Cs{\"o}rnyei} {et~al.}(2023{\natexlab{b}}){Cs{\"o}rnyei}, {Vogl}, {Taubenberger}, {Fl{\"o}rs}, {Blondin}, {Cudmani}, {Holas}, {Kressierer}, {Leibundgut}, \& {Hillebrandt}}]{GezaSiblings}
{Cs{\"o}rnyei}, G., {Vogl}, C., {Taubenberger}, S., {et~al.} 2023{\natexlab{b}}, \aap, 672, A129

\bibitem[{{Czekala} {et~al.}(2015){Czekala}, {Andrews}, {Mandel}, {Hogg}, \& {Green}}]{Czekala2015}
{Czekala}, I., {Andrews}, S.~M., {Mandel}, K.~S., {Hogg}, D.~W., \& {Green}, G.~M. 2015, \apj, 812, 128

\bibitem[{{Davis} \& {Scrimgeour}(2014)}]{2014MNRAS.442.1117D}
{Davis}, T.~M. \& {Scrimgeour}, M.~I. 2014, \mnras, 442, 1117

\bibitem[{{de Jaeger} \& {Galbany}(2023)}]{deJaegerGalbanyReview}
{de Jaeger}, T. \& {Galbany}, L. 2023, arXiv e-prints, arXiv:2305.17243

\bibitem[{{de Jaeger} {et~al.}(2020{\natexlab{a}}){de Jaeger}, {Galbany}, {Gonz{\'a}lez-Gait{\'a}n}, {Kessler}, {Filippenko}, {F{\"o}rster}, {Hamuy}, {Brown}, {Davis}, {Guti{\'e}rrez}, {Inserra}, {Lewis}, {M{\"o}ller}, {Scolnic}, {Smith}, {Brout}, {Carollo}, {Foley}, {Glazebrook}, {Hinton}, {Macaulay}, {Nichol}, {Sako}, {Sommer}, {Tucker}, {Abbott}, {Aguena}, {Allam}, {Annis}, {Avila}, {Bertin}, {Bhargava}, {Brooks}, {Burke}, {Carnero Rosell}, {Carrasco Kind}, {Carretero}, {Costanzi}, {Crocce}, {da Costa}, {De Vicente}, {Desai}, {Diehl}, {Doel}, {Drlica-Wagner}, {Eifler}, {Estrada}, {Everett}, {Flaugher}, {Fosalba}, {Frieman}, {Garc{\'\i}a-Bellido}, {Gaztanaga}, {Gruen}, {Gruendl}, {Gschwend}, {Gutierrez}, {Hartley}, {Hollowood}, {Honscheid}, {James}, {Kuehn}, {Kuropatkin}, {Li}, {Lima}, {Maia}, {Menanteau}, {Miquel}, {Palmese}, {Paz-Chinch{\'o}n}, {Plazas}, {Romer}, {Roodman}, {Sanchez}, {Scarpine}, {Schubnell}, {Serrano}, {Sevilla-Noarbe}, {Soares-Santos}, {Suchyta}, {Swanson}, {Tarle}, {Thomas}, {Tucker},
  {Varga}, {Walker}, {Weller}, {Wilkinson}, \& {DES Collaboration}}]{2020MNRAS.495.4860D}
{de Jaeger}, T., {Galbany}, L., {Gonz{\'a}lez-Gait{\'a}n}, S., {et~al.} 2020{\natexlab{a}}, \mnras, 495, 4860

\bibitem[{{de Jaeger} {et~al.}(2022){de Jaeger}, {Galbany}, {Riess}, {Stahl}, {Shappee}, {Filippenko}, \& {Zheng}}]{deJaeger2022}
{de Jaeger}, T., {Galbany}, L., {Riess}, A.~G., {et~al.} 2022, \mnras, 514, 4620

\bibitem[{{de Jaeger} {et~al.}(2020{\natexlab{b}}){de Jaeger}, {Stahl}, {Zheng}, {Filippenko}, {Riess}, \& {Galbany}}]{2020MNRAS.496.3402D}
{de Jaeger}, T., {Stahl}, B.~E., {Zheng}, W., {et~al.} 2020{\natexlab{b}}, \mnras, 496, 3402

\bibitem[{{de Jaeger} {et~al.}(2019){de Jaeger}, {Zheng}, {Stahl}, {Filippenko}, {Brink}, {Bigley}, {Blanchard}, {Blanchard}, {Bradley}, {Cargill}, {Casper}, {Cenko}, {Channa}, {Choi}, {Clubb}, {Cobb}, {Cohen}, {de Kouchkovsky}, {Ellison}, {Falcon}, {Fox}, {Fuller}, {Ganeshalingam}, {Gould}, {Graham}, {Halevi}, {Hayakawa}, {Hestenes}, {Hyland}, {Jeffers}, {Joubert}, {Kandrashoff}, {Kelly}, {Kim}, {Kim}, {Kumar}, {Leonard}, {Li}, {Lowe}, {Lu}, {Mason}, {McAllister}, {Mauerhan}, {Modjaz}, {Molloy}, {Perley}, {Pina}, {Poznanski}, {Ross}, {Shivvers}, {Silverman}, {Soler}, {Stegman}, {Taylor}, {Tang}, {Wilkins}, {Wang}, {Wang}, {Yuk}, {Yunus}, \& {Zhang}}]{deJaeger2019}
{de Jaeger}, T., {Zheng}, W., {Stahl}, B.~E., {et~al.} 2019, \mnras, 490, 2799

\bibitem[{{Dessart} {et~al.}(2008){Dessart}, {Blondin}, {Brown}, {Hicken}, {Hillier}, {Holland}, {Immler}, {Kirshner}, {Milne}, {Modjaz}, \& {Roming}}]{Dessart2008}
{Dessart}, L., {Blondin}, S., {Brown}, P.~J., {et~al.} 2008, \apj, 675, 644

\bibitem[{{Dessart} \& {Hillier}(2005)}]{Dessart2005}
{Dessart}, L. \& {Hillier}, D.~J. 2005, \aap, 437, 667

\bibitem[{{Dessart} \& {Hillier}(2006)}]{Dessart2006}
{Dessart}, L. \& {Hillier}, D.~J. 2006, \aap, 447, 691

\bibitem[{{Dessart} \& {Hillier}(2011)}]{2011MNRAS.410.1739D}
{Dessart}, L. \& {Hillier}, D.~J. 2011, \mnras, 410, 1739

\bibitem[{{Dessart} \& {Jacobson-Gal{\'a}n}(2023)}]{2023A&A...677A.105D}
{Dessart}, L. \& {Jacobson-Gal{\'a}n}, W.~V. 2023, \aap, 677, A105

\bibitem[{{Dhawan} {et~al.}(2018){Dhawan}, {Jha}, \& {Leibundgut}}]{Dhawan2018}
{Dhawan}, S., {Jha}, S.~W., \& {Leibundgut}, B. 2018, \aap, 609, A72

\bibitem[{{Dhungana} {et~al.}(2024){Dhungana}, {Kehoe}, {Staten}, {Vinko}, {Wheeler}, {Akerlof}, {Doss}, {Ferrante}, {Gibson}, {Lasker}, {Marion}, {Pandey}, {Quimby}, {Rykoff}, {Smith}, {Yuan}, \& {Zheng}}]{2024ApJ...962...60D}
{Dhungana}, G., {Kehoe}, R., {Staten}, R., {et~al.} 2024, \apj, 962, 60

\bibitem[{{Di Valentino} {et~al.}(2021){Di Valentino}, {Mena}, {Pan}, {Visinelli}, {Yang}, {Melchiorri}, {Mota}, {Riess}, \& {Silk}}]{diValentino21}
{Di Valentino}, E., {Mena}, O., {Pan}, S., {et~al.} 2021, Classical and Quantum Gravity, 38, 153001

\bibitem[{{Doi} {et~al.}(2010){Doi}, {Tanaka}, {Fukugita}, {Gunn}, {Yasuda}, {Ivezi{\'c}}, {Brinkmann}, {de Haars}, {Kleinman}, {Krzesinski}, \& {French Leger}}]{2010AJ....139.1628D}
{Doi}, M., {Tanaka}, M., {Fukugita}, M., {et~al.} 2010, \aj, 139, 1628

\bibitem[{{Dutcher} {et~al.}(2021){Dutcher}, {Balkenhol}, {Ade}, {Ahmed}, {Anderes}, {Anderson}, {Archipley}, {Avva}, {Aylor}, {Barry}, {Basu Thakur}, {Benabed}, {Bender}, {Benson}, {Bianchini}, {Bleem}, {Bouchet}, {Bryant}, {Byrum}, {Carlstrom}, {Carter}, {Cecil}, {Chang}, {Chaubal}, {Chen}, {Cho}, {Chou}, {Cliche}, {Crawford}, {Cukierman}, {Daley}, {de Haan}, {Denison}, {Dibert}, {Ding}, {Dobbs}, {Everett}, {Feng}, {Ferguson}, {Foster}, {Fu}, {Galli}, {Gambrel}, {Gardner}, {Goeckner-Wald}, {Gualtieri}, {Guns}, {Gupta}, {Guyser}, {Halverson}, {Harke-Hosemann}, {Harrington}, {Henning}, {Hilton}, {Hivon}, {Holder}, {Holzapfel}, {Hood}, {Howe}, {Huang}, {Irwin}, {Jeong}, {Jonas}, {Jones}, {Khaire}, {Knox}, {Kofman}, {Korman}, {Kubik}, {Kuhlmann}, {Kuo}, {Lee}, {Leitch}, {Lowitz}, {Lu}, {Meyer}, {Michalik}, {Millea}, {Montgomery}, {Nadolski}, {Natoli}, {Nguyen}, {Noble}, {Novosad}, {Omori}, {Padin}, {Pan}, {Paschos}, {Pearson}, {Posada}, {Prabhu}, {Quan}, {Raghunathan}, {Rahlin}, {Reichardt}, {Riebel}, {Riedel},
  {Rouble}, {Ruhl}, {Sayre}, {Schiappucci}, {Shirokoff}, {Smecher}, {Sobrin}, {Stark}, {Stephen}, {Story}, {Suzuki}, {Thompson}, {Thorne}, {Tucker}, {Umilta}, {Vale}, {Vanderlinde}, {Vieira}, {Wang}, {Whitehorn}, {Wu}, {Yefremenko}, {Yoon}, {Young}, \& {SPT-3G Collaboration}}]{SPT2021}
{Dutcher}, D., {Balkenhol}, L., {Ade}, P.~A.~R., {et~al.} 2021, \prd, 104, 022003

\bibitem[{{Eastman} {et~al.}(1996{\natexlab{a}}){Eastman}, {Schmidt}, \& {Kirshner}}]{1996ApJ...466..911E}
{Eastman}, R.~G., {Schmidt}, B.~P., \& {Kirshner}, R. 1996{\natexlab{a}}, \apj, 466, 911

\bibitem[{{Eastman} {et~al.}(1996{\natexlab{b}}){Eastman}, {Schmidt}, \& {Kirshner}}]{Eastman1996}
{Eastman}, R.~G., {Schmidt}, B.~P., \& {Kirshner}, R. 1996{\natexlab{b}}, \apj, 466, 911

\bibitem[{{Fitzpatrick} {et~al.}(2019){Fitzpatrick}, {Massa}, {Gordon}, {Bohlin}, \& {Clayton}}]{Fitzpatrick2019}
{Fitzpatrick}, E.~L., {Massa}, D., {Gordon}, K.~D., {Bohlin}, R., \& {Clayton}, G.~C. 2019, \apj, 886, 108

\bibitem[{Foreman-Mackey(2016)}]{corner}
Foreman-Mackey, D. 2016, The Journal of Open Source Software, 1, 24

\bibitem[{{Freedman}(2021)}]{Freedman2021}
{Freedman}, W.~L. 2021, \apj, 919, 16

\bibitem[{{Freedman} {et~al.}(2019){Freedman}, {Madore}, {Hatt}, {Hoyt}, {Jang}, {Beaton}, {Burns}, {Lee}, {Monson}, {Neeley}, {Phillips}, {Rich}, \& {Seibert}}]{CCHP_H0}
{Freedman}, W.~L., {Madore}, B.~F., {Hatt}, D., {et~al.} 2019, \apj, 882, 34

\bibitem[{{Freedman} {et~al.}(2020){Freedman}, {Madore}, {Hoyt}, {Jang}, {Beaton}, {Lee}, {Monson}, {Neeley}, \& {Rich}}]{Freedman2020}
{Freedman}, W.~L., {Madore}, B.~F., {Hoyt}, T., {et~al.} 2020, \apj, 891, 57

\bibitem[{{Freedman} {et~al.}(2024){Freedman}, {Madore}, {Jang}, {Hoyt}, {Lee}, \& {Owens}}]{Freedman2024}
{Freedman}, W.~L., {Madore}, B.~F., {Jang}, I.~S., {et~al.} 2024, arXiv e-prints, arXiv:2408.06153

\bibitem[{{Fuller}(2017)}]{2017MNRAS.470.1642F}
{Fuller}, J. 2017, \mnras, 470, 1642

\bibitem[{{Fuller} \& {Tsuna}(2024)}]{2024OJAp....7E..47F}
{Fuller}, J. \& {Tsuna}, D. 2024, The Open Journal of Astrophysics, 7, 47

\bibitem[{{Gal-Yam} {et~al.}(2011){Gal-Yam}, {Kasliwal}, {Arcavi}, {Green}, {Yaron}, {Ben-Ami}, {Xu}, {Sternberg}, {Quimby}, {Kulkarni}, {Ofek}, {Walters}, {Nugent}, {Poznanski}, {Bloom}, {Cenko}, {Filippenko}, {Li}, {Silverman}, {Walker}, {Sullivan}, {Maguire}, {Howell}, {Mazzali}, {Frail}, {Bersier}, {James}, {Akerlof}, {Yuan}, {Law}, {Fox}, \& {Gehrels}}]{Gal-Yam-2010id}
{Gal-Yam}, A., {Kasliwal}, M.~M., {Arcavi}, I., {et~al.} 2011, \apj, 736, 159

\bibitem[{{Galbany} {et~al.}(2016){Galbany}, {Hamuy}, {Phillips}, {Suntzeff}, {Maza}, {de Jaeger}, {Moraga}, {Gonz{\'a}lez-Gait{\'a}n}, {Krisciunas}, {Morrell}, {Thomas-Osip}, {Krzeminski}, {Gonz{\'a}lez}, {Antezana}, {Wishnjewski}, {McCarthy}, {Anderson}, {Guti{\'e}rrez}, {Stritzinger}, {Folatelli}, {Anguita}, {Galaz}, {Green}, {Impey}, {Kim}, {Kirhakos}, {Malkan}, {Mulchaey}, {Phillips}, {Pizzella}, {Prosser}, {Schmidt}, {Schommer}, {Sherry}, {Strolger}, {Wells}, \& {Williger}}]{Galbany2016}
{Galbany}, L., {Hamuy}, M., {Phillips}, M.~M., {et~al.} 2016, \aj, 151, 33

\bibitem[{{Gall} {et~al.}(2016){Gall}, {Kotak}, {Leibundgut}, {Taubenberger}, {Hillebrandt}, \& {Kromer}}]{Gall2016}
{Gall}, E.~E.~E., {Kotak}, R., {Leibundgut}, B., {et~al.} 2016, \aap, 592, A129

\bibitem[{{Gall} {et~al.}(2018){Gall}, {Kotak}, {Leibundgut}, {Taubenberger}, {Hillebrandt}, {Kromer}, {Burgett}, {Chambers}, {Flewelling}, {Huber}, {Kaiser}, {Kudritzki}, {Magnier}, {Metcalfe}, {Smith}, {Tonry}, {Wainscoat}, \& {Waters}}]{Gall2018}
{Gall}, E.~E.~E., {Kotak}, R., {Leibundgut}, B., {et~al.} 2018, \aap, 611, A25

\bibitem[{{Gall} {et~al.}(2015){Gall}, {Polshaw}, {Kotak}, {Jerkstrand}, {Leibundgut}, {Rabinowitz}, {Sollerman}, {Sullivan}, {Smartt}, {Anderson}, {Benetti}, {Baltay}, {Feindt}, {Fraser}, {Gonz{\'a}lez-Gait{\'a}n}, {Inserra}, {Maguire}, {McKinnon}, {Valenti}, \& {Young}}]{Gall2015_RiseTime}
{Gall}, E.~E.~E., {Polshaw}, J., {Kotak}, R., {et~al.} 2015, \aap, 582, A3

\bibitem[{{Ganeshalingam} {et~al.}(2010){Ganeshalingam}, {Li}, {Filippenko}, {Anderson}, {Foster}, {Gates}, {Griffith}, {Grigsby}, {Joubert}, {Leja}, {Lowe}, {Macomber}, {Pritchard}, {Thrasher}, \& {Winslow}}]{2010ApJS..190..418G}
{Ganeshalingam}, M., {Li}, W., {Filippenko}, A.~V., {et~al.} 2010, \apjs, 190, 418

\bibitem[{{Goldstein} {et~al.}(2019){Goldstein}, {Nugent}, \& {Goobar}}]{Goldstein19}
{Goldstein}, D.~A., {Nugent}, P.~E., \& {Goobar}, A. 2019, \apjs, 243, 6

\bibitem[{{Gonz{\'a}lez-Gait{\'a}n} {et~al.}(2015){Gonz{\'a}lez-Gait{\'a}n}, {Tominaga}, {Molina}, {Galbany}, {Bufano}, {Anderson}, {Gutierrez}, {F{\"o}rster}, {Pignata}, {Bersten}, {Howell}, {Sullivan}, {Carlberg}, {de Jaeger}, {Hamuy}, {Baklanov}, \& {Blinnikov}}]{GonzalezGaitan2015}
{Gonz{\'a}lez-Gait{\'a}n}, S., {Tominaga}, N., {Molina}, J., {et~al.} 2015, \mnras, 451, 2212

\bibitem[{{Gordon}(2024)}]{2022zndo...4658887G}
{Gordon}, K. 2024, {dust\_extinction}, v1.4.1, \url{https://doi.org/10.5281/zenodo.4658887}

\bibitem[{{Grillo} {et~al.}(2024){Grillo}, {Pagano}, {Rosati}, \& {Suyu}}]{Grillo2024}
{Grillo}, C., {Pagano}, L., {Rosati}, P., \& {Suyu}, S.~H. 2024, \aap, 684, L23

\bibitem[{{Guillochon} {et~al.}(2017){Guillochon}, {Parrent}, {Kelley}, \& {Margutti}}]{2017ApJ...835...64G}
{Guillochon}, J., {Parrent}, J., {Kelley}, L.~Z., \& {Margutti}, R. 2017, \apj, 835, 64

\bibitem[{{Guti{\'e}rrez} {et~al.}(2017{\natexlab{a}}){Guti{\'e}rrez}, {Anderson}, {Hamuy}, {Gonz{\'a}lez-Gaitan}, {Galbany}, {Dessart}, {Stritzinger}, {Phillips}, {Morrell}, \& {Folatelli}}]{Gutierrez2017b}
{Guti{\'e}rrez}, C.~P., {Anderson}, J.~P., {Hamuy}, M., {et~al.} 2017{\natexlab{a}}, \apj, 850, 90

\bibitem[{{Guti{\'e}rrez} {et~al.}(2017{\natexlab{b}}){Guti{\'e}rrez}, {Anderson}, {Hamuy}, {Morrell}, {Gonz{\'a}lez-Gaitan}, {Stritzinger}, {Phillips}, {Galbany}, {Folatelli}, {Dessart}, {Contreras}, {Della Valle}, {Freedman}, {Hsiao}, {Krisciunas}, {Madore}, {Maza}, {Suntzeff}, {Prieto}, {Gonz{\'a}lez}, {Cappellaro}, {Navarrete}, {Pizzella}, {Ruiz}, {Smith}, \& {Turatto}}]{Gutierrez2017a}
{Guti{\'e}rrez}, C.~P., {Anderson}, J.~P., {Hamuy}, M., {et~al.} 2017{\natexlab{b}}, \apj, 850, 89

\bibitem[{{Hamuy} \& {Pinto}(2002)}]{SCMFirstPaper}
{Hamuy}, M. \& {Pinto}, P.~A. 2002, \apjl, 566, L63

\bibitem[{{Hatano} {et~al.}(1998){Hatano}, {Branch}, \& {Deaton}}]{Hatano98}
{Hatano}, K., {Branch}, D., \& {Deaton}, J. 1998, \apj, 502, 177

\bibitem[{{Herold} \& {Ferreira}(2023)}]{2023PhRvD.108d3513H}
{Herold}, L. \& {Ferreira}, E. G.~M. 2023, \prd, 108, 043513

\bibitem[{{Hicken} {et~al.}(2009){Hicken}, {Challis}, {Jha}, {Kirshner}, {Matheson}, {Modjaz}, {Rest}, {Wood-Vasey}, {Bakos}, {Barton}, {Berlind}, {Bragg}, {Brice{\~n}o}, {Brown}, {Caldwell}, {Calkins}, {Cho}, {Ciupik}, {Contreras}, {Dendy}, {Dosaj}, {Durham}, {Eriksen}, {Esquerdo}, {Everett}, {Falco}, {Fernandez}, {Gaba}, {Garnavich}, {Graves}, {Green}, {Groner}, {Hergenrother}, {Holman}, {Hradecky}, {Huchra}, {Hutchison}, {Jerius}, {Jordan}, {Kilgard}, {Krauss}, {Luhman}, {Macri}, {Marrone}, {McDowell}, {McIntosh}, {McNamara}, {Megeath}, {Mochejska}, {Munoz}, {Muzerolle}, {Naranjo}, {Narayan}, {Pahre}, {Peters}, {Peterson}, {Rines}, {Ripman}, {Roussanova}, {Schild}, {Sicilia-Aguilar}, {Sokoloski}, {Smalley}, {Smith}, {Spahr}, {Stanek}, {Barmby}, {Blondin}, {Stubbs}, {Szentgyorgyi}, {Torres}, {Vaz}, {Vikhlinin}, {Wang}, {Westover}, {Woods}, \& {Zhao}}]{Hicken2009}
{Hicken}, M., {Challis}, P., {Jha}, S., {et~al.} 2009, \apj, 700, 331

\bibitem[{{Hicken} {et~al.}(2017){Hicken}, {Friedman}, {Blondin}, {Challis}, {Berlind}, {Calkins}, {Esquerdo}, {Matheson}, {Modjaz}, {Rest}, \& {Kirshner}}]{Hicken2017}
{Hicken}, M., {Friedman}, A.~S., {Blondin}, S., {et~al.} 2017, \apjs, 233, 6

\bibitem[{{Hillier} \& {Dessart}(2019)}]{2019A&A...631A...8H}
{Hillier}, D.~J. \& {Dessart}, L. 2019, \aap, 631, A8

\bibitem[{{Hinshaw} {et~al.}(2013){Hinshaw}, {Larson}, {Komatsu}, {Spergel}, {Bennett}, {Dunkley}, {Nolta}, {Halpern}, {Hill}, {Odegard}, {Page}, {Smith}, {Weiland}, {Gold}, {Jarosik}, {Kogut}, {Limon}, {Meyer}, {Tucker}, {Wollack}, \& {Wright}}]{2013ApJS..208...19H}
{Hinshaw}, G., {Larson}, D., {Komatsu}, E., {et~al.} 2013, \apjs, 208, 19

\bibitem[{{Hogg}(1999)}]{1999astro.ph..5116H}
{Hogg}, D.~W. 1999, arXiv e-prints, astro

\bibitem[{{Hogg} {et~al.}(2002){Hogg}, {Baldry}, {Blanton}, \& {Eisenstein}}]{HoggKcorr}
{Hogg}, D.~W., {Baldry}, I.~K., {Blanton}, M.~R., \& {Eisenstein}, D.~J. 2002, arXiv e-prints, astro

\bibitem[{{Hogg} {et~al.}(2010){Hogg}, {Bovy}, \& {Lang}}]{2010arXiv1008.4686H}
{Hogg}, D.~W., {Bovy}, J., \& {Lang}, D. 2010, arXiv e-prints, arXiv:1008.4686

\bibitem[{{Hoyt}(2023)}]{2023NatAs...7..590H}
{Hoyt}, T.~J. 2023, Nature Astronomy, 7, 590

\bibitem[{{Huang} {et~al.}(2020){Huang}, {Riess}, {Yuan}, {Macri}, {Zakamska}, {Casertano}, {Whitelock}, {Hoffmann}, {Filippenko}, \& {Scolnic}}]{Huang2020}
{Huang}, C.~D., {Riess}, A.~G., {Yuan}, W., {et~al.} 2020, \apj, 889, 5

\bibitem[{{Huang} {et~al.}(2024){Huang}, {Yuan}, {Riess}, {Hack}, {Whitelock}, {Zakamska}, {Casertano}, {Macri}, {Marengo}, {Menzies}, \& {Smith}}]{Huang2024}
{Huang}, C.~D., {Yuan}, W., {Riess}, A.~G., {et~al.} 2024, \apj, 963, 83

\bibitem[{{Inserra} {et~al.}(2018){Inserra}, {Smartt}, {Gall}, {Leloudas}, {Chen}, {Schulze}, {Jerkstrand}, {Nicholl}, {Anderson}, {Arcavi}, {Benetti}, {Cartier}, {Childress}, {Della Valle}, {Flewelling}, {Fraser}, {Gal-Yam}, {Guti{\'e}rrez}, {Hosseinzadeh}, {Howell}, {Huber}, {Kankare}, {Kr{\"u}hler}, {Magnier}, {Maguire}, {McCully}, {Prajs}, {Primak}, {Scalzo}, {Schmidt}, {Smith}, {Smith}, {Tucker}, {Valenti}, {Wilman}, {Young}, \& {Yuan}}]{Inserra2018}
{Inserra}, C., {Smartt}, S.~J., {Gall}, E.~E.~E., {et~al.} 2018, \mnras, 475, 1046

\bibitem[{{Jacobson-Gal{\'a}n} {et~al.}(2024){Jacobson-Gal{\'a}n}, {Dessart}, {Davis}, {Kilpatrick}, {Margutti}, {Foley}, {Chornock}, {Terreran}, {Hiramatsu}, {Newsome}, {Padilla Gonzalez}, {Pellegrino}, {Howell}, {Filippenko}, {Anderson}, {Angus}, {Auchettl}, {Bostroem}, {Brink}, {Cartier}, {Coulter}, {de Boer}, {Drout}, {Earl}, {Ertini}, {Farah}, {Farias}, {Gall}, {Gao}, {Gerlach}, {Guo}, {Haynie}, {Hosseinzadeh}, {Ibik}, {Jha}, {Jones}, {Langeroodi}, {LeBaron}, {Magnier}, {Piro}, {Raimundo}, {Rest}, {Rest}, {Rich}, {Rojas-Bravo}, {Sears}, {Taggart}, {Villar}, {Wainscoat}, {Wang}, {Wasserman}, {Yan}, {Yang}, {Zhang}, \& {Zheng}}]{2024ApJ...970..189J}
{Jacobson-Gal{\'a}n}, W.~V., {Dessart}, L., {Davis}, K.~W., {et~al.} 2024, \apj, 970, 189

\bibitem[{{Jacobson-Gal{\'a}n} {et~al.}(2023){Jacobson-Gal{\'a}n}, {Dessart}, {Margutti}, {Chornock}, {Foley}, {Kilpatrick}, {Jones}, {Taggart}, {Angus}, {Bhattacharjee}, {Braff}, {Brethauer}, {Burgasser}, {Cao}, {Carlile}, {Chambers}, {Coulter}, {Dominguez-Ruiz}, {Dickinson}, {de Boer}, {Gagliano}, {Gall}, {Gao}, {Gates}, {Gomez}, {Guolo}, {Halford}, {Hjorth}, {Huber}, {Johnson}, {Karpoor}, {Laskar}, {LeBaron}, {Li}, {Lin}, {Loch}, {Lynam}, {Magnier}, {Maloney}, {Matthews}, {McDonald}, {Miao}, {Milisavljevic}, {Pan}, {Pradyumna}, {Ransome}, {Rees}, {Rest}, {Rojas-Bravo}, {Sandford}, {Ascencio}, {Sanjaripour}, {Savino}, {Sears}, {Sharei}, {Smartt}, {Softich}, {Theissen}, {Tinyanont}, {Tohfa}, {Villar}, {Wang}, {Wainscoat}, {Westerling}, {Wiston}, {Wozniak}, {Yadavalli}, \& {Zenati}}]{2023ApJ...954L..42J}
{Jacobson-Gal{\'a}n}, W.~V., {Dessart}, L., {Margutti}, R., {et~al.} 2023, \apjl, 954, L42

\bibitem[{{Jha} {et~al.}(2007){Jha}, {Riess}, \& {Kirshner}}]{2007ApJ...659..122J}
{Jha}, S., {Riess}, A.~G., \& {Kirshner}, R.~P. 2007, \apj, 659, 122

\bibitem[{{Johansson} {et~al.}(2021){Johansson}, {Cenko}, {Fox}, {Dhawan}, {Goobar}, {Stanishev}, {Butler}, {Lee}, {Watson}, {Fremling}, {Kasliwal}, {Nugent}, {Petrushevska}, {Sollerman}, {Yan}, {Burke}, {Hosseinzadeh}, {Howell}, {McCully}, \& {Valenti}}]{2021ApJ...923..237J}
{Johansson}, J., {Cenko}, S.~B., {Fox}, O.~D., {et~al.} 2021, \apj, 923, 237

\bibitem[{{Jones} {et~al.}(2009){Jones}, {Hamuy}, {Lira}, {Maza}, {Clocchiatti}, {Phillips}, {Morrell}, {Roth}, {Suntzeff}, {Matheson}, {Filippenko}, {Foley}, \& {Leonard}}]{Jones2009}
{Jones}, M.~I., {Hamuy}, M., {Lira}, P., {et~al.} 2009, \apj, 696, 1176

\bibitem[{{J{\"o}nsson} {et~al.}(2010){J{\"o}nsson}, {Sullivan}, {Hook}, {Basa}, {Carlberg}, {Conley}, {Fouchez}, {Howell}, {Perrett}, \& {Pritchet}}]{2010MNRAS.405..535J}
{J{\"o}nsson}, J., {Sullivan}, M., {Hook}, I., {et~al.} 2010, \mnras, 405, 535

\bibitem[{{Kangas} {et~al.}(2022){Kangas}, {Yan}, {Schulze}, {Fransson}, {Sollerman}, {Lunnan}, {Omand}, {Andreoni}, {Burruss}, {Chen}, {Drake}, {Fremling}, {Gal-Yam}, {Graham}, {Groom}, {Lezmy}, {Mahabal}, {Masci}, {Perley}, {Riddle}, {Tartaglia}, \& {Yao}}]{Kangas2022}
{Kangas}, T., {Yan}, L., {Schulze}, S., {et~al.} 2022, \mnras, 516, 1193

\bibitem[{{Kelly} {et~al.}(2023){Kelly}, {Rodney}, {Treu}, {Oguri}, {Chen}, {Zitrin}, {Birrer}, {Bonvin}, {Dessart}, {Diego}, {Filippenko}, {Foley}, {Gilman}, {Hjorth}, {Jauzac}, {Mandel}, {Millon}, {Pierel}, {Sharon}, {Thorp}, {Williams}, {Broadhurst}, {Dressler}, {Graur}, {Jha}, {McCully}, {Postman}, {Schmidt}, {Tucker}, \& {von der Linden}}]{2023Sci...380.1322K}
{Kelly}, P.~L., {Rodney}, S., {Treu}, T., {et~al.} 2023, Science, 380, abh1322

\bibitem[{Kerzendorf {et~al.}(2023)Kerzendorf, Sim, Vogl, Williamson, Pássaro, Flörs, Camacho, Jančauskas, Harpole, Nöbauer, Lietzau, Mishin, Tsamis, Boyle, Shingles, Gupta, Desai, Klauser, Beaujean, Suban-Loewen, Heringer, Barna, Gautam, Fullard, Smith, Cawley, Singhal, Arya, O'Brien, Barbosa, Sondhi, Yu, Patel, Varanasi, Gillanders, Rathi, Chitchyan, Savel, Reinecke, Eweis, Bylund, Bentil, Black, Shields, Eguren, Alam, Kumar, Bartnik, Magee, Singh, Varma~Buddaraju, Livneh, Kambham, Rajagopalan, Daksh, Mishra, Jain, Reichenbach, Floers, Actions, Holas, Singh, Brar, Chaumal, Bhakar, Selsing, Kowalski, Kumar, Patidar, Talegaonkar, Sofiatti, Venkat, Sharma, Sarafina, Patel, Singh~Rathore, Patra, Lu, Zaheer, Sandler, Truong, Yap, Buchner, Gupta, Prasad, Kolliboyina, Lemoine, Wahi, Aggarwal, Matsumura, Gupta, Volodin, PATIDAR, Martinez, Kharkar, Nayak~U, Dasgupta, \& Kumar}]{kerzendorf_wolfgang_2023_8244935}
Kerzendorf, W., Sim, S., Vogl, C., {et~al.} 2023, tardis-sn/tardis: TARDIS v2023.08.13

\bibitem[{{Kerzendorf} \& {Sim}(2014)}]{Kerzendorf2014}
{Kerzendorf}, W.~E. \& {Sim}, S.~A. 2014, \mnras, 440, 387

\bibitem[{{Khazov} {et~al.}(2016){Khazov}, {Yaron}, {Gal-Yam}, {Manulis}, {Rubin}, {Kulkarni}, {Arcavi}, {Kasliwal}, {Ofek}, {Cao}, {Perley}, {Sollerman}, {Horesh}, {Sullivan}, {Filippenko}, {Nugent}, {Howell}, {Cenko}, {Silverman}, {Ebeling}, {Taddia}, {Johansson}, {Laher}, {Surace}, {Rebbapragada}, {Wozniak}, \& {Matheson}}]{2016ApJ...818....3K}
{Khazov}, D., {Yaron}, O., {Gal-Yam}, A., {et~al.} 2016, \apj, 818, 3

\bibitem[{{Kirshner} \& {Kwan}(1974)}]{KirsherKwan74}
{Kirshner}, R.~P. \& {Kwan}, J. 1974, \apj, 193, 27

\bibitem[{{Koblischke} \& {Anderson}(2024)}]{2024arXiv240619375K}
{Koblischke}, N.~W. \& {Anderson}, R.~I. 2024, \apj, 974, 181

\bibitem[{{Kreisch} {et~al.}(2020){Kreisch}, {Cyr-Racine}, \& {Dor{\'e}}}]{2020PhRvD.101l3505K}
{Kreisch}, C.~D., {Cyr-Racine}, F.-Y., \& {Dor{\'e}}, O. 2020, \prd, 101, 123505

\bibitem[{{Lee} \& {Li}(2006)}]{Lee2006_CBET06it}
{Lee}, N. \& {Li}, W. 2006, Central Bureau Electronic Telegrams, 660, 1

\bibitem[{{L{\'e}get} {et~al.}(2018){L{\'e}get}, {Pruzhinskaya}, {Ciulli}, {Gangler}, {Aldering}, {Antilogus}, {Aragon}, {Bailey}, {Baltay}, {Barbary}, {Bongard}, {Boone}, {Buton}, {Childress}, {Chotard}, {Copin}, {Dixon}, {Fagrelius}, {Feindt}, {Fouchez}, {Gris}, {Hayden}, {Hillebrandt}, {Howell}, {Kim}, {Kowalski}, {Kuesters}, {Lombardo}, {Lin}, {Nordin}, {Pain}, {Pecontal}, {Pereira}, {Perlmutter}, {Rabinowitz}, {Rigault}, {Runge}, {Rubin}, {Saunders}, {Says}, {Smadja}, {Sofiatti}, {Suzuki}, {Taubenberger}, {Tao}, {Thomas}, \& {Nearby Supernova Factory}}]{2018A&A...615A.162L}
{L{\'e}get}, P.~F., {Pruzhinskaya}, M.~V., {Ciulli}, A., {et~al.} 2018, \aap, 615, A162

\bibitem[{{Leibundgut} {et~al.}(1996){Leibundgut}, {Schommer}, {Phillips}, {Riess}, {Schmidt}, {Spyromilio}, {Walsh}, {Suntzeff}, {Hamuy}, {Maza}, {Kirshner}, {Challis}, {Garnavich}, {Smith}, {Dressler}, \& {Ciardullo}}]{1996ApJ...466L..21L}
{Leibundgut}, B., {Schommer}, R., {Phillips}, M., {et~al.} 1996, \apjl, 466, L21

\bibitem[{{Leibundgut} {et~al.}(2019){Leibundgut}, {Spyromilio}, {Vogl}, {Fl{\"o}rs}, {Taubenberger}, {Hillebrandt}, {Suyu}, {Smartt}, {Dobson}, {Kotak}, {Bruch}, \& {Gal-Yam}}]{adH0cc1st}
{Leibundgut}, B., {Spyromilio}, J., {Vogl}, C., {et~al.} 2019, Transient Name Server AstroNote, 103, 1

\bibitem[{{Madore} \& {Freedman}(2020)}]{Madore2020}
{Madore}, B.~F. \& {Freedman}, W.~L. 2020, \apj, 899, 66

\bibitem[{{Madore} \& {Freedman}(2024)}]{2024ApJ...961..166M}
{Madore}, B.~F. \& {Freedman}, W.~L. 2024, \apj, 961, 166

\bibitem[{{Masci} {et~al.}(2023){Masci}, {Laher}, {Rusholme}, {Shupe}, {Paladini}, {Groom}, {Wold}, {Miller}, \& {Drake}}]{ztfforcedphot}
{Masci}, F.~J., {Laher}, R.~R., {Rusholme}, B., {et~al.} 2023, arXiv e-prints, arXiv:2305.16279

\bibitem[{{Masci} {et~al.}(2019){Masci}, {Laher}, {Rusholme}, {Shupe}, {Groom}, {Surace}, {Jackson}, {Monkewitz}, {Beck}, {Flynn}, {Terek}, {Landry}, {Hacopians}, {Desai}, {Howell}, {Brooke}, {Imel}, {Wachter}, {Ye}, {Lin}, {Cenko}, {Cunningham}, {Rebbapragada}, {Bue}, {Miller}, {Mahabal}, {Bellm}, {Patterson}, {Juri{\'c}}, {Golkhou}, {Ofek}, {Walters}, {Graham}, {Kasliwal}, {Dekany}, {Kupfer}, {Burdge}, {Cannella}, {Barlow}, {Van Sistine}, {Giomi}, {Fremling}, {Blagorodnova}, {Levitan}, {Riddle}, {Smith}, {Helou}, {Prince}, \& {Kulkarni}}]{ZTF}
{Masci}, F.~J., {Laher}, R.~R., {Rusholme}, B., {et~al.} 2019, \pasp, 131, 018003

\bibitem[{{Millon} {et~al.}(2020){Millon}, {Galan}, {Courbin}, {Treu}, {Suyu}, {Ding}, {Birrer}, {Chen}, {Shajib}, {Sluse}, {Wong}, {Agnello}, {Auger}, {Buckley-Geer}, {Chan}, {Collett}, {Fassnacht}, {Hilbert}, {Koopmans}, {Motta}, {Mukherjee}, {Rusu}, {Sonnenfeld}, {Spiniello}, \& {Van de Vyvere}}]{2020A&A...639A.101M}
{Millon}, M., {Galan}, A., {Courbin}, F., {et~al.} 2020, \aap, 639, A101

\bibitem[{{Morozova} {et~al.}(2017){Morozova}, {Piro}, \& {Valenti}}]{2017ApJ...838...28M}
{Morozova}, V., {Piro}, A.~L., \& {Valenti}, S. 2017, \apj, 838, 28

\bibitem[{{M{\"o}rtsell} {et~al.}(2022{\natexlab{a}}){M{\"o}rtsell}, {Goobar}, {Johansson}, \& {Dhawan}}]{2022ApJ...933..212M}
{M{\"o}rtsell}, E., {Goobar}, A., {Johansson}, J., \& {Dhawan}, S. 2022{\natexlab{a}}, \apj, 933, 212

\bibitem[{{M{\"o}rtsell} {et~al.}(2022{\natexlab{b}}){M{\"o}rtsell}, {Goobar}, {Johansson}, \& {Dhawan}}]{2022ApJ...935...58M}
{M{\"o}rtsell}, E., {Goobar}, A., {Johansson}, J., \& {Dhawan}, S. 2022{\natexlab{b}}, \apj, 935, 58

\bibitem[{{Ofek} {et~al.}(2014){Ofek}, {Arcavi}, {Tal}, {Sullivan}, {Gal-Yam}, {Kulkarni}, {Nugent}, {Ben-Ami}, {Bersier}, {Cao}, {Cenko}, {De Cia}, {Filippenko}, {Fransson}, {Kasliwal}, {Laher}, {Surace}, {Quimby}, \& {Yaron}}]{Ofek2014}
{Ofek}, E.~O., {Arcavi}, I., {Tal}, D., {et~al.} 2014, \apj, 788, 154

\bibitem[{{Pan} \& {Yang}(2023)}]{2023arXiv231007260P}
{Pan}, S. \& {Yang}, W. 2023, arXiv e-prints, arXiv:2310.07260

\bibitem[{{Pascale} {et~al.}(2024){Pascale}, {Frye}, {Pierel}, {Chen}, {Kelly}, {Cohen}, {Windhorst}, {Riess}, {Kamieneski}, {Diego}, {Meena}, {Cha}, {Oguri}, {Zitrin}, {Jee}, {Foo}, {Leimbach}, {Koekemoer}, {Conselice}, {Dai}, {Goobar}, {Siebert}, {Strolger}, \& {Willner}}]{2024arXiv240318902P}
{Pascale}, M., {Frye}, B.~L., {Pierel}, J. D.~R., {et~al.} 2024, arXiv e-prints, arXiv:2403.18902

\bibitem[{{Pesce} {et~al.}(2020){Pesce}, {Braatz}, {Reid}, {Riess}, {Scolnic}, {Condon}, {Gao}, {Henkel}, {Impellizzeri}, {Kuo}, \& {Lo}}]{Pesce2020}
{Pesce}, D.~W., {Braatz}, J.~A., {Reid}, M.~J., {et~al.} 2020, \apjl, 891, L1

\bibitem[{{Peterson} {et~al.}(2022){Peterson}, {Kenworthy}, {Scolnic}, {Riess}, {Brout}, {Carr}, {Courtois}, {Davis}, {Dwomoh}, {Jones}, {Popovic}, {Rose}, \& {Said}}]{PetersonPantheonPecV}
{Peterson}, E.~R., {Kenworthy}, W.~D., {Scolnic}, D., {et~al.} 2022, \apj, 938, 112

\bibitem[{{Planck Collaboration} {et~al.}(2020{\natexlab{a}}){Planck Collaboration}, {Aghanim}, {Akrami}, {Arroja}, {Ashdown}, {Aumont}, {Baccigalupi}, {Ballardini}, {Banday}, {Barreiro}, {Bartolo}, {Basak}, {Battye}, {Benabed}, {Bernard}, {Bersanelli}, {Bielewicz}, {Bock}, {Bond}, {Borrill}, {Bouchet}, {Boulanger}, {Bucher}, {Burigana}, {Butler}, {Calabrese}, {Cardoso}, {Carron}, {Casaponsa}, {Challinor}, {Chiang}, {Colombo}, {Combet}, {Contreras}, {Crill}, {Cuttaia}, {de Bernardis}, {de Zotti}, {Delabrouille}, {Delouis}, {D{\'e}sert}, {Di Valentino}, {Dickinson}, {Diego}, {Donzelli}, {Dor{\'e}}, {Douspis}, {Ducout}, {Dupac}, {Efstathiou}, {Elsner}, {En{\ss}lin}, {Eriksen}, {Falgarone}, {Fantaye}, {Fergusson}, {Fernandez-Cobos}, {Finelli}, {Forastieri}, {Frailis}, {Franceschi}, {Frolov}, {Galeotta}, {Galli}, {Ganga}, {G{\'e}nova-Santos}, {Gerbino}, {Ghosh}, {Gonz{\'a}lez-Nuevo}, {G{\'o}rski}, {Gratton}, {Gruppuso}, {Gudmundsson}, {Hamann}, {Handley}, {Hansen}, {Helou}, {Herranz}, {Hildebrandt}, {Hivon},
  {Huang}, {Jaffe}, {Jones}, {Karakci}, {Keih{\"a}nen}, {Keskitalo}, {Kiiveri}, {Kim}, {Kisner}, {Knox}, {Krachmalnicoff}, {Kunz}, {Kurki-Suonio}, {Lagache}, {Lamarre}, {Langer}, {Lasenby}, {Lattanzi}, {Lawrence}, {Le Jeune}, {Leahy}, {Lesgourgues}, {Levrier}, {Lewis}, {Liguori}, {Lilje}, {Lilley}, {Lindholm}, {L{\'o}pez-Caniego}, {Lubin}, {Ma}, {Mac{\'\i}as-P{\'e}rez}, {Maggio}, {Maino}, {Mandolesi}, {Mangilli}, {Marcos-Caballero}, {Maris}, {Martin}, {Martinelli}, {Mart{\'\i}nez-Gonz{\'a}lez}, {Matarrese}, {Mauri}, {McEwen}, {Meerburg}, {Meinhold}, {Melchiorri}, {Mennella}, {Migliaccio}, {Millea}, {Mitra}, {Miville-Desch{\^e}nes}, {Molinari}, {Moneti}, {Montier}, {Morgante}, {Moss}, {Mottet}, {M{\"u}nchmeyer}, {Natoli}, {N{\o}rgaard-Nielsen}, {Oxborrow}, {Pagano}, {Paoletti}, {Partridge}, {Patanchon}, {Pearson}, {Peel}, {Peiris}, {Perrotta}, {Pettorino}, {Piacentini}, {Polastri}, {Polenta}, {Puget}, {Rachen}, {Reinecke}, {Remazeilles}, {Renault}, {Renzi}, {Rocha}, {Rosset}, {Roudier},
  {Rubi{\~n}o-Mart{\'\i}n}, {Ruiz-Granados}, {Salvati}, {Sandri}, {Savelainen}, {Scott}, {Shellard}, {Shiraishi}, {Sirignano}, {Sirri}, {Spencer}, {Sunyaev}, {Suur-Uski}, {Tauber}, {Tavagnacco}, {Tenti}, {Terenzi}, {Toffolatti}, {Tomasi}, {Trombetti}, {Valiviita}, {Van Tent}, {Vibert}, {Vielva}, {Villa}, {Vittorio}, {Wandelt}, {Wehus}, {White}, {White}, {Zacchei}, \& {Zonca}}]{Planck2018Overview}
{Planck Collaboration}, {Aghanim}, N., {Akrami}, Y., {et~al.} 2020{\natexlab{a}}, \aap, 641, A1

\bibitem[{{Planck Collaboration} {et~al.}(2020{\natexlab{b}}){Planck Collaboration}, {Aghanim}, {Akrami}, {Ashdown}, {Aumont}, {Baccigalupi}, {Ballardini}, {Banday}, {Barreiro}, {Bartolo}, {Basak}, {Battye}, {Benabed}, {Bernard}, {Bersanelli}, {Bielewicz}, {Bock}, {Bond}, {Borrill}, {Bouchet}, {Boulanger}, {Bucher}, {Burigana}, {Butler}, {Calabrese}, {Cardoso}, {Carron}, {Challinor}, {Chiang}, {Chluba}, {Colombo}, {Combet}, {Contreras}, {Crill}, {Cuttaia}, {de Bernardis}, {de Zotti}, {Delabrouille}, {Delouis}, {Di Valentino}, {Diego}, {Dor{\'e}}, {Douspis}, {Ducout}, {Dupac}, {Dusini}, {Efstathiou}, {Elsner}, {En{\ss}lin}, {Eriksen}, {Fantaye}, {Farhang}, {Fergusson}, {Fernandez-Cobos}, {Finelli}, {Forastieri}, {Frailis}, {Fraisse}, {Franceschi}, {Frolov}, {Galeotta}, {Galli}, {Ganga}, {G{\'e}nova-Santos}, {Gerbino}, {Ghosh}, {Gonz{\'a}lez-Nuevo}, {G{\'o}rski}, {Gratton}, {Gruppuso}, {Gudmundsson}, {Hamann}, {Handley}, {Hansen}, {Herranz}, {Hildebrandt}, {Hivon}, {Huang}, {Jaffe}, {Jones}, {Karakci},
  {Keih{\"a}nen}, {Keskitalo}, {Kiiveri}, {Kim}, {Kisner}, {Knox}, {Krachmalnicoff}, {Kunz}, {Kurki-Suonio}, {Lagache}, {Lamarre}, {Lasenby}, {Lattanzi}, {Lawrence}, {Le Jeune}, {Lemos}, {Lesgourgues}, {Levrier}, {Lewis}, {Liguori}, {Lilje}, {Lilley}, {Lindholm}, {L{\'o}pez-Caniego}, {Lubin}, {Ma}, {Mac{\'\i}as-P{\'e}rez}, {Maggio}, {Maino}, {Mandolesi}, {Mangilli}, {Marcos-Caballero}, {Maris}, {Martin}, {Martinelli}, {Mart{\'\i}nez-Gonz{\'a}lez}, {Matarrese}, {Mauri}, {McEwen}, {Meinhold}, {Melchiorri}, {Mennella}, {Migliaccio}, {Millea}, {Mitra}, {Miville-Desch{\^e}nes}, {Molinari}, {Montier}, {Morgante}, {Moss}, {Natoli}, {N{\o}rgaard-Nielsen}, {Pagano}, {Paoletti}, {Partridge}, {Patanchon}, {Peiris}, {Perrotta}, {Pettorino}, {Piacentini}, {Polastri}, {Polenta}, {Puget}, {Rachen}, {Reinecke}, {Remazeilles}, {Renzi}, {Rocha}, {Rosset}, {Roudier}, {Rubi{\~n}o-Mart{\'\i}n}, {Ruiz-Granados}, {Salvati}, {Sandri}, {Savelainen}, {Scott}, {Shellard}, {Sirignano}, {Sirri}, {Spencer}, {Sunyaev}, {Suur-Uski},
  {Tauber}, {Tavagnacco}, {Tenti}, {Toffolatti}, {Tomasi}, {Trombetti}, {Valenziano}, {Valiviita}, {Van Tent}, {Vibert}, {Vielva}, {Villa}, {Vittorio}, {Wandelt}, {Wehus}, {White}, {White}, {Zacchei}, \& {Zonca}}]{PlanckCosmoParams}
{Planck Collaboration}, {Aghanim}, N., {Akrami}, Y., {et~al.} 2020{\natexlab{b}}, \aap, 641, A6

\bibitem[{{Poulin} {et~al.}(2019){Poulin}, {Smith}, {Karwal}, \& {Kamionkowski}}]{2019PhRvL.122v1301P}
{Poulin}, V., {Smith}, T.~L., {Karwal}, T., \& {Kamionkowski}, M. 2019, \prl, 122, 221301

\bibitem[{Rasmussen \& Williams(2006)}]{Rasmussen06gaussianprocesses}
Rasmussen, C.~E. \& Williams, C. K.~I. 2006, Gaussian Processes for Machine Learning (The MIT Press)

\bibitem[{{Reid} {et~al.}(2019){Reid}, {Pesce}, \& {Riess}}]{2019ApJ...886L..27R}
{Reid}, M.~J., {Pesce}, D.~W., \& {Riess}, A.~G. 2019, \apjl, 886, L27

\bibitem[{{Riess} {et~al.}(2016){Riess}, {Macri}, {Hoffmann}, {Scolnic}, {Casertano}, {Filippenko}, {Tucker}, {Reid}, {Jones}, {Silverman}, {Chornock}, {Challis}, {Yuan}, {Brown}, \& {Foley}}]{Riess2016}
{Riess}, A.~G., {Macri}, L.~M., {Hoffmann}, S.~L., {et~al.} 2016, \apj, 826, 56

\bibitem[{{Riess} {et~al.}(2024){Riess}, {Scolnic}, {Anand}, {Breuval}, {Casertano}, {Macri}, {Li}, {Yuan}, {Huang}, {Jha}, {Murakami}, {Beaton}, {Brout}, {Wu}, {Addison}, {Bennett}, {Anderson}, {Filippenko}, \& {Carr}}]{Riess2024}
{Riess}, A.~G., {Scolnic}, D., {Anand}, G.~S., {et~al.} 2024, arXiv e-prints, arXiv:2408.11770

\bibitem[{{Riess} {et~al.}(2004){Riess}, {Strolger}, {Tonry}, {Casertano}, {Ferguson}, {Mobasher}, {Challis}, {Filippenko}, {Jha}, {Li}, {Chornock}, {Kirshner}, {Leibundgut}, {Dickinson}, {Livio}, {Giavalisco}, {Steidel}, {Ben{\'\i}tez}, \& {Tsvetanov}}]{Riess2004}
{Riess}, A.~G., {Strolger}, L.-G., {Tonry}, J., {et~al.} 2004, \apj, 607, 665

\bibitem[{{Riess} {et~al.}(2022){Riess}, {Yuan}, {Macri}, {Scolnic}, {Brout}, {Casertano}, {Jones}, {Murakami}, {Anand}, {Breuval}, {Brink}, {Filippenko}, {Hoffmann}, {Jha}, {D'arcy Kenworthy}, {Mackenty}, {Stahl}, \& {Zheng}}]{Riess2022}
{Riess}, A.~G., {Yuan}, W., {Macri}, L.~M., {et~al.} 2022, \apjl, 934, L7

\bibitem[{{Rubin} {et~al.}(2016){Rubin}, {Gal-Yam}, {De Cia}, {Horesh}, {Khazov}, {Ofek}, {Kulkarni}, {Arcavi}, {Manulis}, {Yaron}, {Vreeswijk}, {Kasliwal}, {Ben-Ami}, {Perley}, {Cao}, {Cenko}, {Rebbapragada}, {Wo{\'z}niak}, {Filippenko}, {Clubb}, {Nugent}, {Pan}, {Badenes}, {Howell}, {Valenti}, {Sand}, {Sollerman}, {Johansson}, {Leonard}, {Horst}, {Armen}, {Fedrow}, {Quimby}, {Mazzali}, {Pian}, {Sternberg}, {Matheson}, {Sullivan}, {Maguire}, \& {Lazarevic}}]{Rubin2016}
{Rubin}, A., {Gal-Yam}, A., {De Cia}, A., {et~al.} 2016, \apj, 820, 33

\bibitem[{{Rubin} {et~al.}(2015){Rubin}, {Aldering}, {Barbary}, {Boone}, {Chappell}, {Currie}, {Deustua}, {Fagrelius}, {Fruchter}, {Hayden}, {Lidman}, {Nordin}, {Perlmutter}, {Saunders}, {Sofiatti}, \& {Supernova Cosmology Project}}]{2015ApJ...813..137R}
{Rubin}, D., {Aldering}, G., {Barbary}, K., {et~al.} 2015, \apj, 813, 137

\bibitem[{{Rubin} {et~al.}(2023){Rubin}, {Aldering}, {Betoule}, {Fruchter}, {Huang}, {Kim}, {Lidman}, {Linder}, {Perlmutter}, {Ruiz-Lapuente}, \& {Suzuki}}]{RubinUnion}
{Rubin}, D., {Aldering}, G., {Betoule}, M., {et~al.} 2023, arXiv e-prints, arXiv:2311.12098

\bibitem[{{Said} {et~al.}(2020){Said}, {Colless}, {Magoulas}, {Lucey}, \& {Hudson}}]{Said2020}
{Said}, K., {Colless}, M., {Magoulas}, C., {Lucey}, J.~R., \& {Hudson}, M.~J. 2020, \mnras, 497, 1275

\bibitem[{{Salim} {et~al.}(2018){Salim}, {Boquien}, \& {Lee}}]{2018ApJ...859...11S}
{Salim}, S., {Boquien}, M., \& {Lee}, J.~C. 2018, \apj, 859, 11

\bibitem[{{Sandage} \& {Tammann}(1974)}]{1974ApJ...190..525S}
{Sandage}, A. \& {Tammann}, G.~A. 1974, \apj, 190, 525

\bibitem[{{Schlafly} \& {Finkbeiner}(2011)}]{Schlafly2011}
{Schlafly}, E.~F. \& {Finkbeiner}, D.~P. 2011, \apj, 737, 103

\bibitem[{{Schlafly} {et~al.}(2016){Schlafly}, {Meisner}, {Stutz}, {Kainulainen}, {Peek}, {Tchernyshyov}, {Rix}, {Finkbeiner}, {Covey}, {Green}, {Bell}, {Burgett}, {Chambers}, {Draper}, {Flewelling}, {Hodapp}, {Kaiser}, {Magnier}, {Martin}, {Metcalfe}, {Wainscoat}, \& {Waters}}]{2016ApJ...821...78S}
{Schlafly}, E.~F., {Meisner}, A.~M., {Stutz}, A.~M., {et~al.} 2016, \apj, 821, 78

\bibitem[{{Schmidt} {et~al.}(1992){Schmidt}, {Kirshner}, \& {Eastman}}]{1992ApJ...395..366S}
{Schmidt}, B.~P., {Kirshner}, R.~P., \& {Eastman}, R.~G. 1992, \apj, 395, 366

\bibitem[{{Schmidt} {et~al.}(1994){Schmidt}, {Kirshner}, {Eastman}, {Phillips}, {Suntzeff}, {Hamuy}, {Maza}, \& {Aviles}}]{Schmidt1994}
{Schmidt}, B.~P., {Kirshner}, R.~P., {Eastman}, R.~G., {et~al.} 1994, \apj, 432, 42

\bibitem[{{Schombert} {et~al.}(2020){Schombert}, {McGaugh}, \& {Lelli}}]{2020AJ....160...71S}
{Schombert}, J., {McGaugh}, S., \& {Lelli}, F. 2020, \aj, 160, 71

\bibitem[{{Sch{\"o}neberg} {et~al.}(2022){Sch{\"o}neberg}, {Abell{\'a}n}, {S{\'a}nchez}, {Witte}, {Poulin}, \& {Lesgourgues}}]{H0Olympics}
{Sch{\"o}neberg}, N., {Abell{\'a}n}, G.~F., {S{\'a}nchez}, A.~P., {et~al.} 2022, \physrep, 984, 1

\bibitem[{{Sch{\"o}neberg} {et~al.}(2019){Sch{\"o}neberg}, {Lesgourgues}, \& {Hooper}}]{Schoeneberg2019}
{Sch{\"o}neberg}, N., {Lesgourgues}, J., \& {Hooper}, D.~C. 2019, \jcap, 2019, 029

\bibitem[{{Scolnic} {et~al.}(2023){Scolnic}, {Riess}, {Wu}, {Li}, {Anand}, {Beaton}, {Casertano}, {Anderson}, {Dhawan}, \& {Ke}}]{2023ApJ...954L..31S}
{Scolnic}, D., {Riess}, A.~G., {Wu}, J., {et~al.} 2023, \apjl, 954, L31

\bibitem[{{Scolnic} {et~al.}(2018){Scolnic}, {Jones}, {Rest}, {Pan}, {Chornock}, {Foley}, {Huber}, {Kessler}, {Narayan}, {Riess}, {Rodney}, {Berger}, {Brout}, {Challis}, {Drout}, {Finkbeiner}, {Lunnan}, {Kirshner}, {Sanders}, {Schlafly}, {Smartt}, {Stubbs}, {Tonry}, {Wood-Vasey}, {Foley}, {Hand}, {Johnson}, {Burgett}, {Chambers}, {Draper}, {Hodapp}, {Kaiser}, {Kudritzki}, {Magnier}, {Metcalfe}, {Bresolin}, {Gall}, {Kotak}, {McCrum}, \& {Smith}}]{Scolnic2018}
{Scolnic}, D.~M., {Jones}, D.~O., {Rest}, A., {et~al.} 2018, \apj, 859, 101

\bibitem[{{Shajib} {et~al.}(2020){Shajib}, {Birrer}, {Treu}, {Agnello}, {Buckley-Geer}, {Chan}, {Christensen}, {Lemon}, {Lin}, {Millon}, {Poh}, {Rusu}, {Sluse}, {Spiniello}, {Chen}, {Collett}, {Courbin}, {Fassnacht}, {Frieman}, {Galan}, {Gilman}, {More}, {Anguita}, {Auger}, {Bonvin}, {McMahon}, {Meylan}, {Wong}, {Abbott}, {Annis}, {Avila}, {Bechtol}, {Brooks}, {Brout}, {Burke}, {Carnero Rosell}, {Carrasco Kind}, {Carretero}, {Castander}, {Costanzi}, {da Costa}, {De Vicente}, {Desai}, {Dietrich}, {Doel}, {Drlica-Wagner}, {Evrard}, {Finley}, {Flaugher}, {Fosalba}, {Garc{\'\i}a-Bellido}, {Gerdes}, {Gruen}, {Gruendl}, {Gschwend}, {Gutierrez}, {Hollowood}, {Honscheid}, {Huterer}, {James}, {Jeltema}, {Krause}, {Kuropatkin}, {Li}, {Lima}, {MacCrann}, {Maia}, {Marshall}, {Melchior}, {Miquel}, {Ogando}, {Palmese}, {Paz-Chinch{\'o}n}, {Plazas}, {Romer}, {Roodman}, {Sako}, {Sanchez}, {Santiago}, {Scarpine}, {Schubnell}, {Scolnic}, {Serrano}, {Sevilla-Noarbe}, {Smith}, {Soares-Santos}, {Suchyta}, {Tarle}, {Thomas},
  {Walker}, \& {Zhang}}]{2020MNRAS.494.6072S}
{Shajib}, A.~J., {Birrer}, S., {Treu}, T., {et~al.} 2020, \mnras, 494, 6072

\bibitem[{{Shajib} {et~al.}(2023){Shajib}, {Mozumdar}, {Chen}, {Treu}, {Cappellari}, {Knabel}, {Suyu}, {Bennert}, {Frieman}, {Sluse}, {Birrer}, {Courbin}, {Fassnacht}, {Villafa{\~n}a}, \& {Williams}}]{2023A&A...673A...9S}
{Shajib}, A.~J., {Mozumdar}, P., {Chen}, G. C.~F., {et~al.} 2023, \aap, 673, A9

\bibitem[{{Shingles} {et~al.}(2021){Shingles}, {Smith}, {Young}, {Smartt}, {Tonry}, {Denneau}, {Heinze}, {Weiland}, {Flewelling}, {Stalder}, {Clocchiatti}, {F{\"o}rster}, {Pignata}, {Rest}, {Anderson}, {Stubbs}, \& {Erasmus}}]{ATLASForcedPhot}
{Shingles}, L., {Smith}, K.~W., {Young}, D.~R., {et~al.} 2021, Transient Name Server AstroNote, 7, 1

\bibitem[{{Smartt} {et~al.}(2015){Smartt}, {Valenti}, {Fraser}, {Inserra}, {Young}, {Sullivan}, {Pastorello}, {Benetti}, {Gal-Yam}, {Knapic}, {Molinaro}, {Smareglia}, {Smith}, {Taubenberger}, {Yaron}, {Anderson}, {Ashall}, {Balland}, {Baltay}, {Barbarino}, {Bauer}, {Baumont}, {Bersier}, {Blagorodnova}, {Bongard}, {Botticella}, {Bufano}, {Bulla}, {Cappellaro}, {Campbell}, {Cellier-Holzem}, {Chen}, {Childress}, {Clocchiatti}, {Contreras}, {Dall'Ora}, {Danziger}, {de Jaeger}, {De Cia}, {Della Valle}, {Dennefeld}, {Elias-Rosa}, {Elman}, {Feindt}, {Fleury}, {Gall}, {Gonzalez-Gaitan}, {Galbany}, {Morales Garoffolo}, {Greggio}, {Guillou}, {Hachinger}, {Hadjiyska}, {Hage}, {Hillebrandt}, {Hodgkin}, {Hsiao}, {James}, {Jerkstrand}, {Kangas}, {Kankare}, {Kotak}, {Kromer}, {Kuncarayakti}, {Leloudas}, {Lundqvist}, {Lyman}, {Hook}, {Maguire}, {Manulis}, {Margheim}, {Mattila}, {Maund}, {Mazzali}, {McCrum}, {McKinnon}, {Moreno-Raya}, {Nicholl}, {Nugent}, {Pain}, {Pignata}, {Phillips}, {Polshaw}, {Pumo}, {Rabinowitz},
  {Reilly}, {Romero-Ca{\~n}izales}, {Scalzo}, {Schmidt}, {Schulze}, {Sim}, {Sollerman}, {Taddia}, {Tartaglia}, {Terreran}, {Tomasella}, {Turatto}, {Walker}, {Walton}, {Wyrzykowski}, {Yuan}, \& {Zampieri}}]{2015A&A...579A..40S}
{Smartt}, S.~J., {Valenti}, S., {Fraser}, M., {et~al.} 2015, \aap, 579, A40

\bibitem[{{Smith} {et~al.}(2002){Smith}, {Tucker}, {Kent}, {Richmond}, {Fukugita}, {Ichikawa}, {Ichikawa}, {Jorgensen}, {Uomoto}, {Gunn}, {Hamabe}, {Watanabe}, {Tolea}, {Henden}, {Annis}, {Pier}, {McKay}, {Brinkmann}, {Chen}, {Holtzman}, {Shimasaku}, \& {York}}]{2002AJ....123.2121S}
{Smith}, J.~A., {Tucker}, D.~L., {Kent}, S., {et~al.} 2002, \aj, 123, 2121

\bibitem[{{Smith} {et~al.}(2020{\natexlab{a}}){Smith}, {Smartt}, {Young}, {Tonry}, {Denneau}, {Flewelling}, {Heinze}, {Weiland}, {Stalder}, {Rest}, {Stubbs}, {Anderson}, {Chen}, {Clark}, {Do}, {F{\"o}rster}, {Fulton}, {Gillanders}, {McBrien}, {O'Neill}, {Srivastav}, \& {Wright}}]{Smith2020}
{Smith}, K.~W., {Smartt}, S.~J., {Young}, D.~R., {et~al.} 2020{\natexlab{a}}, \pasp, 132, 085002

\bibitem[{{Smith} {et~al.}(2020{\natexlab{b}}){Smith}, {Poulin}, \& {Amin}}]{2020PhRvD.101f3523S}
{Smith}, T.~L., {Poulin}, V., \& {Amin}, M.~A. 2020{\natexlab{b}}, \prd, 101, 063523

\bibitem[{{Stanishev}(2007)}]{Stanishev2007}
{Stanishev}, V. 2007, Astronomische Nachrichten, 328, 948

\bibitem[{{Suyu} {et~al.}(2017){Suyu}, {Bonvin}, {Courbin}, {Fassnacht}, {Rusu}, {Sluse}, {Treu}, {Wong}, {Auger}, {Ding}, {Hilbert}, {Marshall}, {Rumbaugh}, {Sonnenfeld}, {Tewes}, {Tihhonova}, {Agnello}, {Blandford}, {Chen}, {Collett}, {Koopmans}, {Liao}, {Meylan}, \& {Spiniello}}]{2017MNRAS.468.2590S}
{Suyu}, S.~H., {Bonvin}, V., {Courbin}, F., {et~al.} 2017, \mnras, 468, 2590

\bibitem[{{Suyu} {et~al.}(2020){Suyu}, {Huber}, {Ca{\~n}ameras}, {Kromer}, {Schuldt}, {Taubenberger}, {Y{\i}ld{\i}r{\i}m}, {Bonvin}, {Chan}, {Courbin}, {N{\"o}bauer}, {Sim}, \& {Sluse}}]{HOLISMOKES_I}
{Suyu}, S.~H., {Huber}, S., {Ca{\~n}ameras}, R., {et~al.} 2020, \aap, 644, A162

\bibitem[{{Taubenberger} {et~al.}(2019){Taubenberger}, {Suyu}, {Komatsu}, {Jee}, {Birrer}, {Bonvin}, {Courbin}, {Rusu}, {Shajib}, \& {Wong}}]{2019A&A...628L...7T}
{Taubenberger}, S., {Suyu}, S.~H., {Komatsu}, E., {et~al.} 2019, \aap, 628, L7

\bibitem[{{Taylor} {et~al.}(2014){Taylor}, {Cinabro}, {Dilday}, {Galbany}, {Gupta}, {Kessler}, {Marriner}, {Nichol}, {Richmond}, {Schneider}, \& {Sollerman}}]{2014ApJ...792..135T}
{Taylor}, M., {Cinabro}, D., {Dilday}, B., {et~al.} 2014, \apj, 792, 135

\bibitem[{{Teja} {et~al.}(2023){Teja}, {Singh}, {Basu}, {Anupama}, {Sahu}, {Dutta}, {Swain}, {Nakaoka}, {Pathak}, {Bhalerao}, {Barway}, {Kumar}, {A.~J.}, {Imazawa}, {Kumar}, \& {Kawabata}}]{2023ApJ...954L..12T}
{Teja}, R.~S., {Singh}, A., {Basu}, J., {et~al.} 2023, \apjl, 954, L12

\bibitem[{{Thorp} {et~al.}(2024){Thorp}, {Mandel}, {Jones}, {Kirshner}, \& {Challis}}]{2024MNRAS.530.4016T}
{Thorp}, S., {Mandel}, K.~S., {Jones}, D.~O., {Kirshner}, R.~P., \& {Challis}, P.~M. 2024, \mnras, 530, 4016

\bibitem[{{Tody}(1986)}]{1986SPIE..627..733T}
{Tody}, D. 1986, in Society of Photo-Optical Instrumentation Engineers (SPIE) Conference Series, Vol. 627, Instrumentation in astronomy VI, ed. D.~L. {Crawford}, 733

\bibitem[{{Tody}(1993)}]{1993ASPC...52..173T}
{Tody}, D. 1993, in Astronomical Society of the Pacific Conference Series, Vol.~52, Astronomical Data Analysis Software and Systems II, ed. R.~J. {Hanisch}, R.~J.~V. {Brissenden}, \& J.~{Barnes}, 173

\bibitem[{{Tonry} {et~al.}(2018){Tonry}, {Denneau}, {Heinze}, {Stalder}, {Smith}, {Smartt}, {Stubbs}, {Weiland}, \& {Rest}}]{Tonry2018}
{Tonry}, J.~L., {Denneau}, L., {Heinze}, A.~N., {et~al.} 2018, \pasp, 130, 064505

\bibitem[{{Tully}(2015)}]{Tully2015}
{Tully}, R.~B. 2015, \aj, 149, 171

\bibitem[{{Tully} {et~al.}(2023){Tully}, {Kourkchi}, {Courtois}, {Anand}, {Blakeslee}, {Brout}, {Jaeger}, {Dupuy}, {Guinet}, {Howlett}, {Jensen}, {Pomar{\`e}de}, {Rizzi}, {Rubin}, {Said}, {Scolnic}, \& {Stahl}}]{2023ApJ...944...94T}
{Tully}, R.~B., {Kourkchi}, E., {Courtois}, H.~M., {et~al.} 2023, \apj, 944, 94

\bibitem[{{Uddin} {et~al.}(2023){Uddin}, {Burns}, {Phillips}, {Suntzeff}, {Freedman}, {Brown}, {Morrell}, {Hamuy}, {Krisciunas}, {Wang}, {Hsiao}, {Goobar}, {Perlmutter}, {Lu}, {Stritzinger}, {Anderson}, {Ashall}, {Hoeflich}, {Shappee}, {Persson}, {Piro}, {Baron}, {Contreras}, {Galbany}, {Kumar}, {Shahbandeh}, {Davis}, {Anais}, {Busta}, {Campillay}, {Castell{\'o}n}, {Corco}, {Diamond}, {Gall}, {Gonzalez}, {Holmbo}, {Roth}, {Ser{\'o}n}, {Taddia}, {Torres}, {Baltay}, {Folatelli}, {Hadjiyska}, {Kasliwal}, {Nugent}, {Rabinowitz}, \& {Ryder}}]{2023arXiv230801875U}
{Uddin}, S.~A., {Burns}, C.~R., {Phillips}, M.~M., {et~al.} 2023, arXiv e-prints, arXiv:2308.01875

\bibitem[{{Valenti} {et~al.}(2016){Valenti}, {Howell}, {Stritzinger}, {Graham}, {Hosseinzadeh}, {Arcavi}, {Bildsten}, {Jerkstrand}, {McCully}, {Pastorello}, {Piro}, {Sand}, {Smartt}, {Terreran}, {Baltay}, {Benetti}, {Brown}, {Filippenko}, {Fraser}, {Rabinowitz}, {Sullivan}, \& {Yuan}}]{Valenti2016}
{Valenti}, S., {Howell}, D.~A., {Stritzinger}, M.~D., {et~al.} 2016, \mnras, 459, 3939

\bibitem[{{van Dokkum}(2001)}]{2001PASP..113.1420V}
{van Dokkum}, P.~G. 2001, \pasp, 113, 1420

\bibitem[{{Vasylyev} {et~al.}(2022){Vasylyev}, {Filippenko}, {Vogl}, {Brink}, {Brown}, {de Jaeger}, {Matheson}, {Gal-Yam}, {Mazzali}, {Modjaz}, {Patra}, {Rowe}, {Smith}, {Van Dyk}, {Williamson}, {Yang}, {Zheng}, {deGraw}, {Fox}, {Gates}, {Jennings}, \& {Rich}}]{Vasylyev2022}
{Vasylyev}, S.~S., {Filippenko}, A.~V., {Vogl}, C., {et~al.} 2022, \apj, 934, 134

\bibitem[{{Vasylyev} {et~al.}(2023){Vasylyev}, {Vogl}, {Yang}, {Filippenko}, {Brink}, {Brown}, {Matheson}, {Modjaz}, {Gal-Yam}, {Mazzali}, {de Jaeger}, {Patra}, \& {Stewart}}]{Vasylyev2023}
{Vasylyev}, S.~S., {Vogl}, C., {Yang}, Y., {et~al.} 2023, \apjl, 959, L26

\bibitem[{{Vincenzi} {et~al.}(2023){Vincenzi}, {Sullivan}, {M{\"o}ller}, {Armstrong}, {Bassett}, {Brout}, {Carollo}, {Carr}, {Davis}, {Frohmaier}, {Galbany}, {Glazebrook}, {Graur}, {Kelsey}, {Kessler}, {Kovacs}, {Lewis}, {Lidman}, {Malik}, {Nichol}, {Popovic}, {Sako}, {Scolnic}, {Smith}, {Taylor}, {Tucker}, {Wiseman}, {Aguena}, {Allam}, {Annis}, {Asorey}, {Bacon}, {Bertin}, {Brooks}, {Burke}, {Carnero Rosell}, {Carretero}, {Castander}, {Costanzi}, {da Costa}, {Pereira}, {De Vicente}, {Desai}, {Diehl}, {Doel}, {Everett}, {Ferrero}, {Flaugher}, {Fosalba}, {Frieman}, {Garc{\'\i}a-Bellido}, {Gerdes}, {Gruen}, {Gutierrez}, {Hinton}, {Hollowood}, {Honscheid}, {James}, {Kuehn}, {Kuropatkin}, {Lahav}, {Li}, {Lima}, {Maia}, {Marshall}, {Miquel}, {Morgan}, {Ogando}, {Palmese}, {Paz-Chinch{\'o}n}, {Pieres}, {Plazas Malag{\'o}n}, {Reil}, {Roodman}, {Sanchez}, {Schubnell}, {Serrano}, {Sevilla-Noarbe}, {Suchyta}, {Tarle}, {To}, {Varga}, {Weller}, {Wilkinson}, \& {DES Collaboration}}]{Vincenzi23}
{Vincenzi}, M., {Sullivan}, M., {M{\"o}ller}, A., {et~al.} 2023, \mnras, 518, 1106

\bibitem[{{Vogl} {et~al.}(2020){Vogl}, {Kerzendorf}, {Sim}, {Noebauer}, {Lietzau}, \& {Hillebrandt}}]{Vogl2020}
{Vogl}, C., {Kerzendorf}, W.~E., {Sim}, S.~A., {et~al.} 2020, \aap, 633, A88

\bibitem[{{Vogl} {et~al.}(2019){Vogl}, {Sim}, {Noebauer}, {Kerzendorf}, \& {Hillebrandt}}]{Vogl2019}
{Vogl}, C., {Sim}, S.~A., {Noebauer}, U.~M., {Kerzendorf}, W.~E., \& {Hillebrandt}, W. 2019, \aap, 621, A29

\bibitem[{{Wagoner}(1981)}]{1981ApJ...250L..65W}
{Wagoner}, R.~V. 1981, \apjl, 250, L65

\bibitem[{{Wang} {et~al.}(2016){Wang}, {Abdalla}, {Atrio-Barandela}, \& {Pav{\'o}n}}]{2016RPPh...79i6901W}
{Wang}, B., {Abdalla}, E., {Atrio-Barandela}, F., \& {Pav{\'o}n}, D. 2016, Reports on Progress in Physics, 79, 096901

\bibitem[{{Wojtak} {et~al.}(2023){Wojtak}, {Hjorth}, \& {Hjortlund}}]{2023MNRAS.525.5187W}
{Wojtak}, R., {Hjorth}, J., \& {Hjortlund}, J.~O. 2023, \mnras, 525, 5187

\bibitem[{{Wong} {et~al.}(2024){Wong}, {Dux}, {Shajib}, {Suyu}, {Millon}, {Mozumdar}, {Wells}, {Agnello}, {Birrer}, {Buckley-Geer}, {Courbin}, {Fassnacht}, {Frieman}, {Galan}, {Lin}, {Marshall}, {Poh}, {Schuldt}, {Sluse}, \& {Treu}}]{2024A&A...689A.168W}
{Wong}, K.~C., {Dux}, F., {Shajib}, A.~J., {et~al.} 2024, \aap, 689, A168

\bibitem[{{Wong} {et~al.}(2020){Wong}, {Suyu}, {Chen}, {Rusu}, {Millon}, {Sluse}, {Bonvin}, {Fassnacht}, {Taubenberger}, {Auger}, {Birrer}, {Chan}, {Courbin}, {Hilbert}, {Tihhonova}, {Treu}, {Agnello}, {Ding}, {Jee}, {Komatsu}, {Shajib}, {Sonnenfeld}, {Blandford}, {Koopmans}, {Marshall}, \& {Meylan}}]{Wong2020}
{Wong}, K.~C., {Suyu}, S.~H., {Chen}, G. C.~F., {et~al.} 2020, \mnras, 498, 1420

\bibitem[{{Wood-Vasey} {et~al.}(2003){Wood-Vasey}, {Aldering}, {Nugent}, {Mulchaey}, \& {Phillips}}]{Wood-Vasey2003}
{Wood-Vasey}, W.~M., {Aldering}, G., {Nugent}, P., {Mulchaey}, J., \& {Phillips}, M. 2003, \iaucirc, 8088, 2

\bibitem[{{Yao} {et~al.}(2020){Yao}, {De}, {Kasliwal}, {Ho}, {Schulze}, {Li}, {Kulkarni}, {Fruchter}, {Rubin}, {Perley}, {Fuller}, {Piro}, {Fremling}, {Bellm}, {Burruss}, {Duev}, {Feeney}, {Gal-Yam}, {Golkhou}, {Graham}, {Helou}, {Kupfer}, {Laher}, {Masci}, {Miller}, {Rusholme}, {Shupe}, {Smith}, {Sollerman}, {Soumagnac}, \& {Zolkower}}]{Yao2020}
{Yao}, Y., {De}, K., {Kasliwal}, M.~M., {et~al.} 2020, \apj, 900, 46

\bibitem[{{Yaron} \& {Gal-Yam}(2012)}]{Yaron2012}
{Yaron}, O. \& {Gal-Yam}, A. 2012, \pasp, 124, 668

\bibitem[{{Yaron} {et~al.}(2017){Yaron}, {Perley}, {Gal-Yam}, {Groh}, {Horesh}, {Ofek}, {Kulkarni}, {Sollerman}, {Fransson}, {Rubin}, {Szabo}, {Sapir}, {Taddia}, {Cenko}, {Valenti}, {Arcavi}, {Howell}, {Kasliwal}, {Vreeswijk}, {Khazov}, {Fox}, {Cao}, {Gnat}, {Kelly}, {Nugent}, {Filippenko}, {Laher}, {Wozniak}, {Lee}, {Rebbapragada}, {Maguire}, {Sullivan}, \& {Soumagnac}}]{Yaron2017}
{Yaron}, O., {Perley}, D.~A., {Gal-Yam}, A., {et~al.} 2017, Nature Physics, 13, 510

\bibitem[{{Y{\i}ld{\i}r{\i}m} {et~al.}(2023){Y{\i}ld{\i}r{\i}m}, {Suyu}, {Chen}, \& {Komatsu}}]{2023A&A...675A..21Y}
{Y{\i}ld{\i}r{\i}m}, A., {Suyu}, S.~H., {Chen}, G.~C.~F., \& {Komatsu}, E. 2023, \aap, 675, A21

\bibitem[{{Zehavi} {et~al.}(1998){Zehavi}, {Riess}, {Kirshner}, \& {Dekel}}]{1998ApJ...503..483Z}
{Zehavi}, I., {Riess}, A.~G., {Kirshner}, R.~P., \& {Dekel}, A. 1998, \apj, 503, 483

\end{thebibliography}


\begin{appendix}
    \balance
        \section{Time of explosion} \label{app:t0}
	\begin{figure}[H]
		\centering
		\includegraphics[scale=0.95]{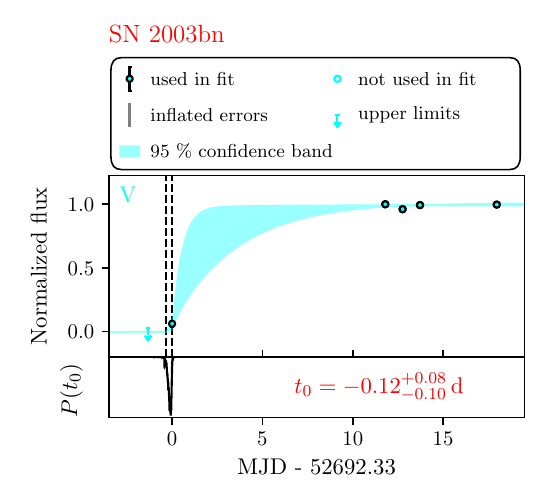}
		\includegraphics[scale=0.95]{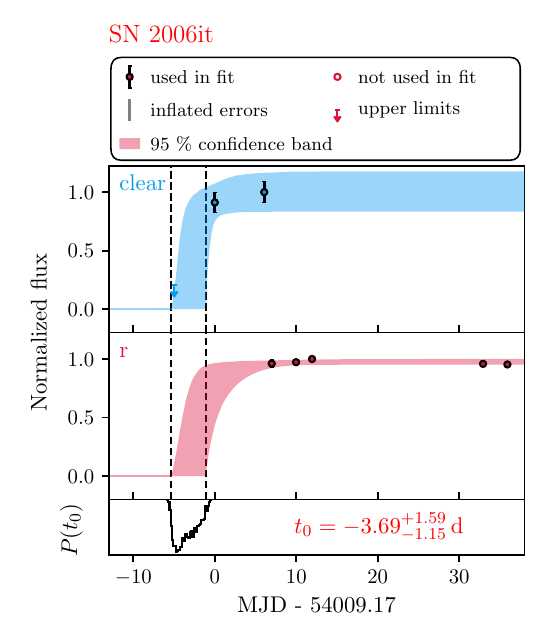}
		\caption{Time-of-explosion fits for SN~2003bn and SN~2006it. See \cref{fig:lc_fit_t0} for an explanation of the plot.}
	\end{figure}

	\begin{figure}[H]
             \vspace{10.5mm}
		\centering
            \includegraphics[scale=0.95]{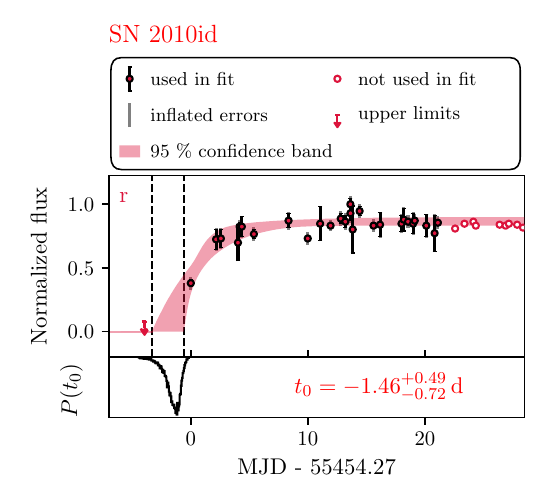}
		\includegraphics[scale=0.95]{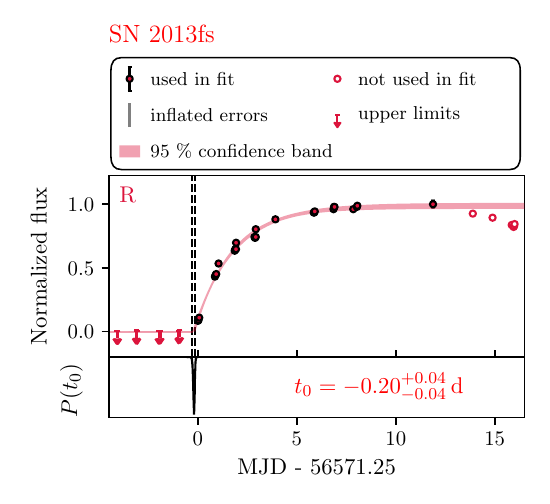}
		\caption{Time-of-explosion fits for SN~2010id and SN~2013fs. See \cref{fig:lc_fit_t0} for an explanation of the plot.}
	\end{figure}

	\begin{figure}[H]
		\centering
  		\includegraphics[scale=0.95]{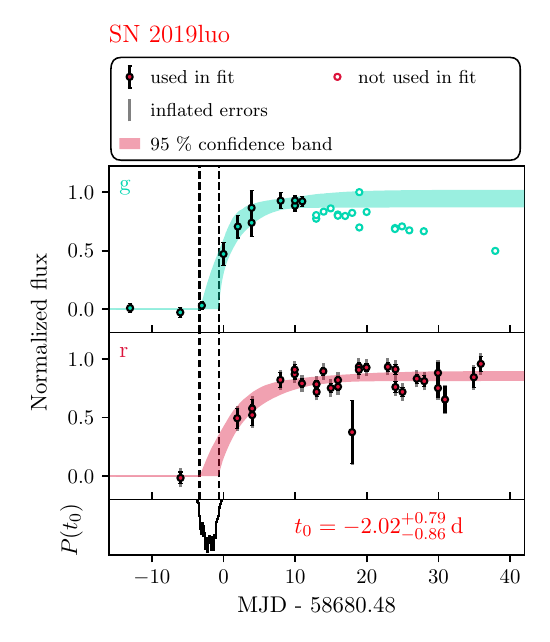}
		\includegraphics[scale=0.95]{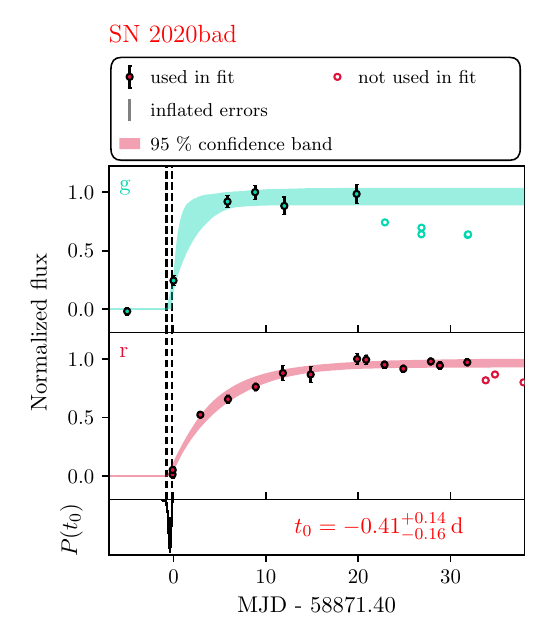}
		\caption{Time-of-explosion fits for SN~2019luo and SN~2020bad. See \cref{fig:lc_fit_t0} for an explanation of the plot.}
	\end{figure}

	\begin{figure}[H]
		\centering
  		\includegraphics[scale=0.95]{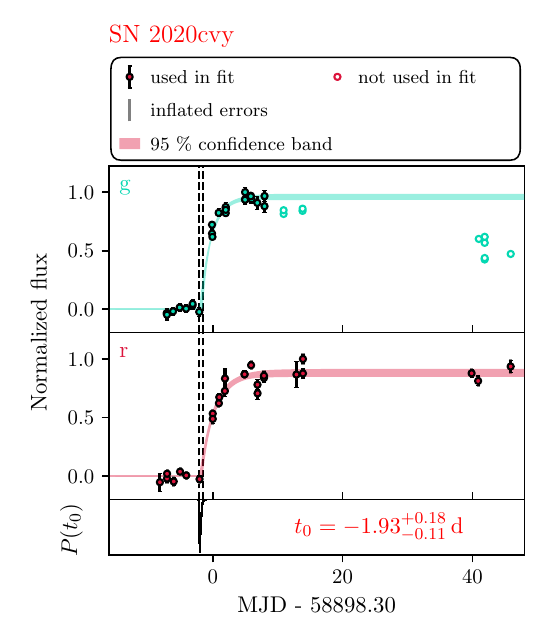}
		\includegraphics[scale=0.95]{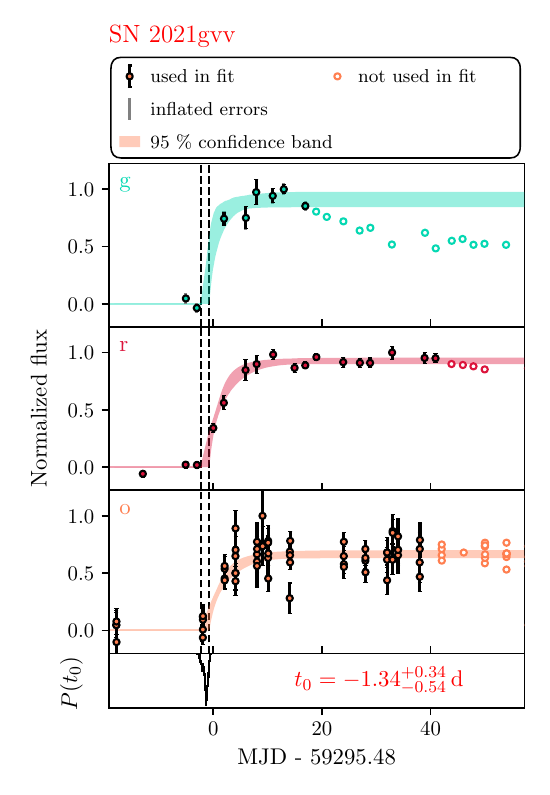}
		\caption{Time-of-explosion fits for SN~2020cvy and SN~2021gvv. See \cref{fig:lc_fit_t0} for an explanation of the plot.}
	\end{figure}

	\begin{figure}[H]
		\centering
  		\includegraphics[scale=0.95]{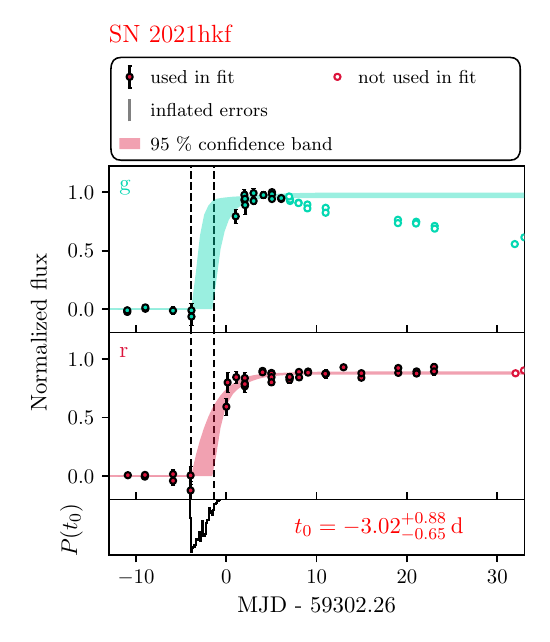}
		\includegraphics[scale=0.95]{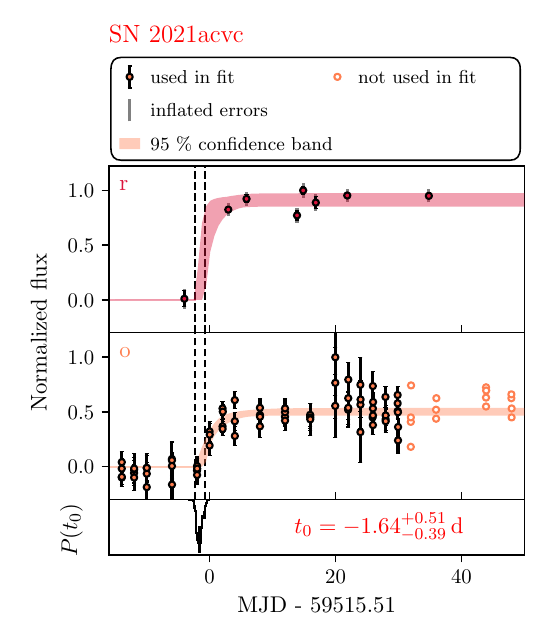}
		\caption{Time-of-explosion fit for SN~2021hkf and SN~2021acvc. See \cref{fig:lc_fit_t0} for an explanation of the plot.}
	\end{figure}

	\onecolumn
	\section{Light curve interpolation} \label{app:lc_interp}
	\begin{figure}[H]
		\begin{center}
			\includegraphics[]{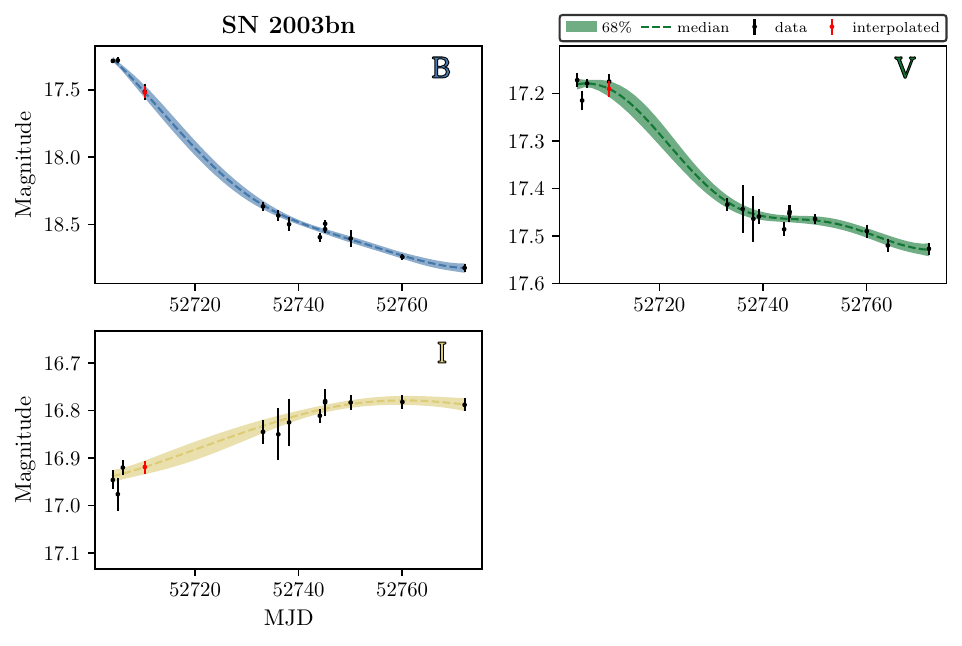}
			\includegraphics[]{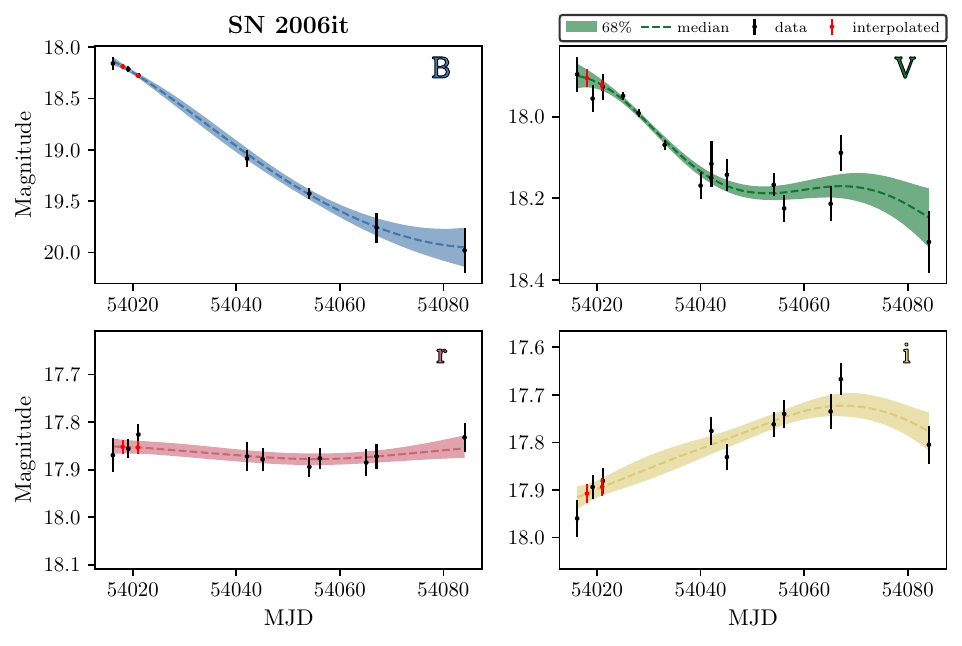}
		\end{center}
		\caption{GP interpolation of the photometry of SN~2003bn and SN~2006it. See \cref{fig:lc_interpolation} for an explanation of the plot.}
	\end{figure}

	\begin{figure*}
		\begin{center}
			\includegraphics[]{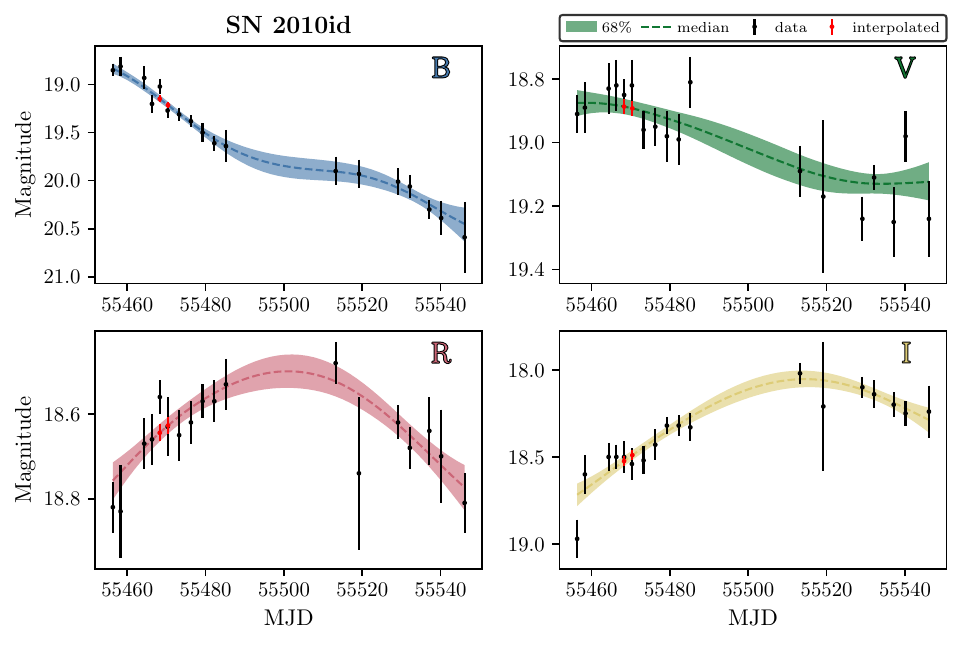}
			\includegraphics[]{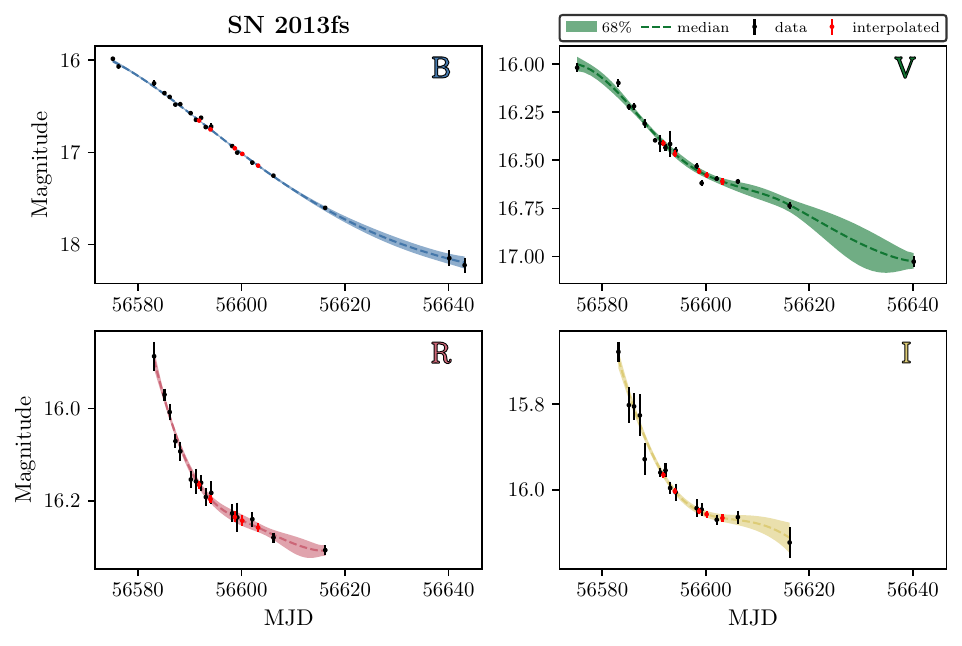}
		\end{center}
		\caption{GP interpolation of the photometry of SN~2010id and SN~2013fs (Bessell). See \cref{fig:lc_interpolation} for an explanation of the plot.}
	\end{figure*}

	\begin{figure*}
		\begin{center}
			\includegraphics[]{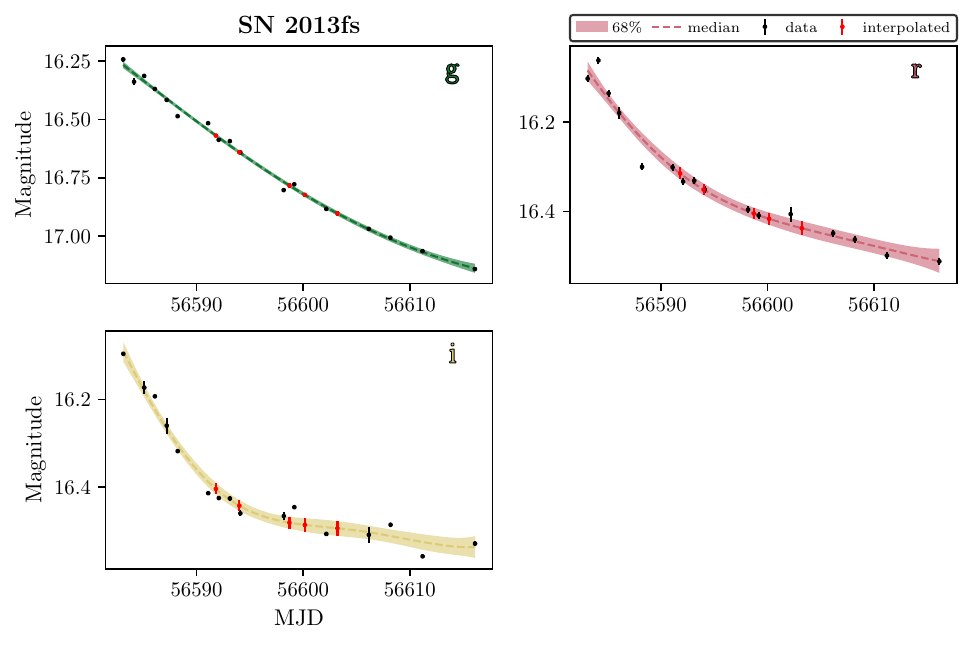}
			\includegraphics[]{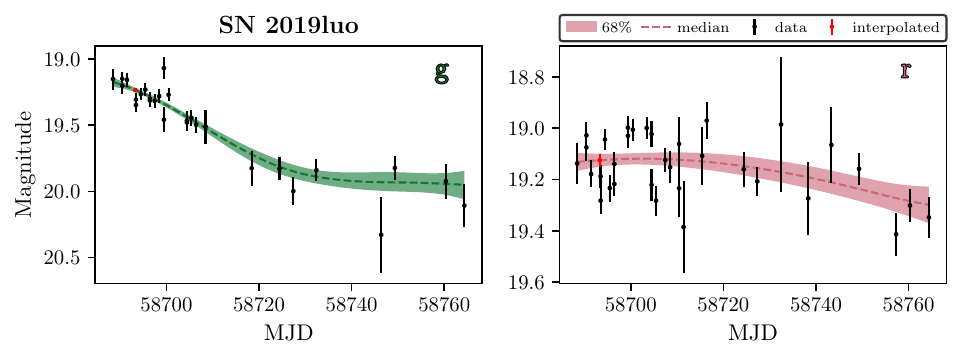}
			\includegraphics[]{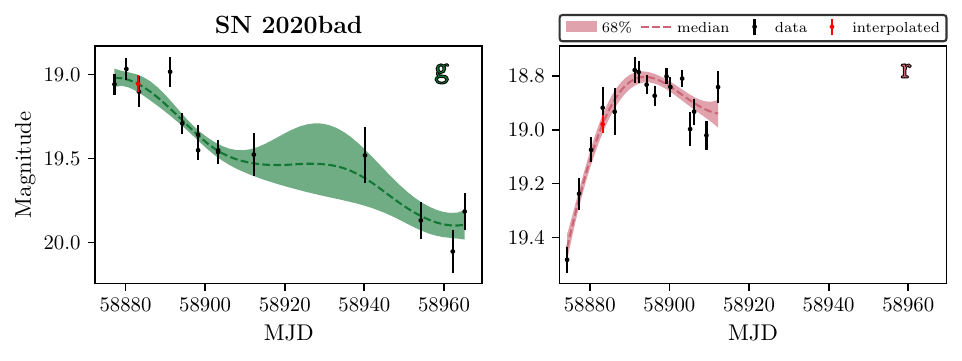}
		\end{center}
		\caption{GP interpolation of the photometry of SN~2013fs (Sloan), SN~2019luo, and SN~2020bad. See \cref{fig:lc_interpolation} for an explanation of the plot.}
	\end{figure*}

	\begin{figure*}
		\begin{center}
			\includegraphics[]{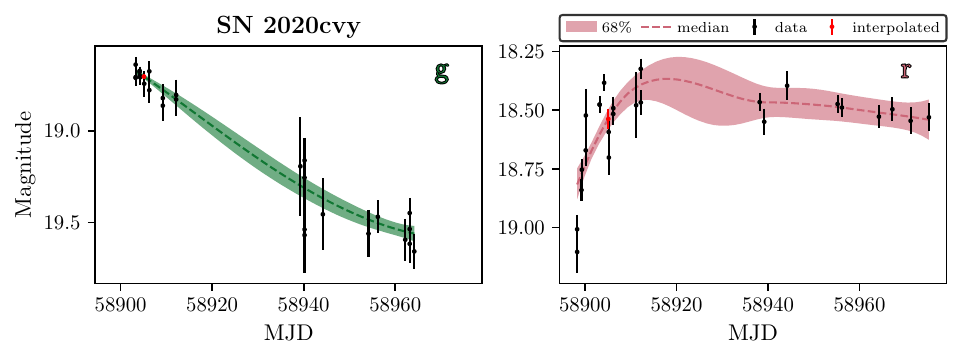}
			\includegraphics[]{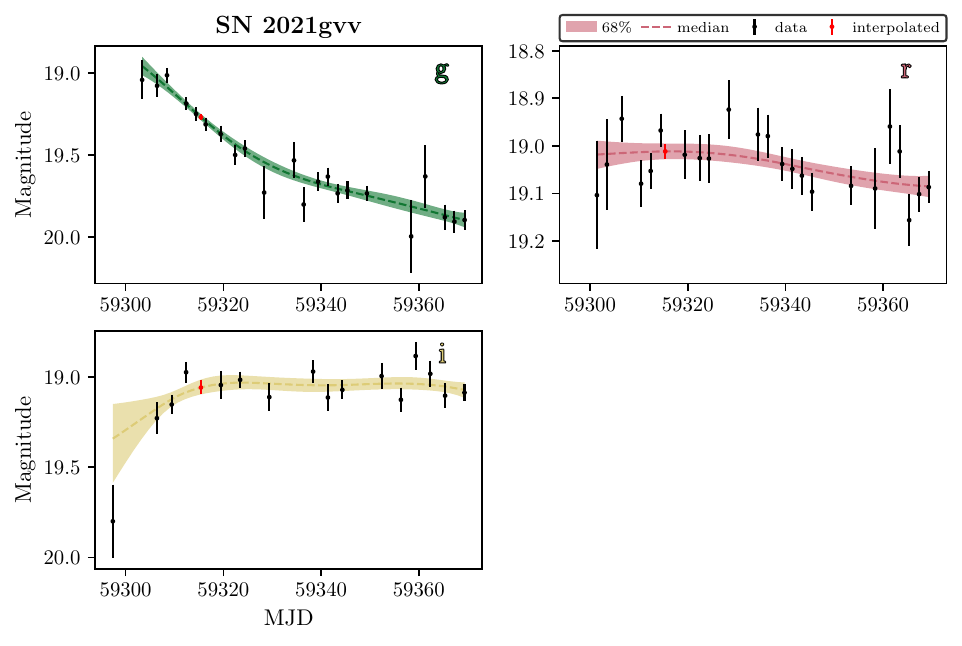}
		\end{center}
		\caption{GP interpolation of the photometry of SN~2020cvy and SN~2021gvv. See \cref{fig:lc_interpolation} for an explanation of the plot.}
	\end{figure*}

	\begin{figure*}
		\begin{center}
			\includegraphics[]{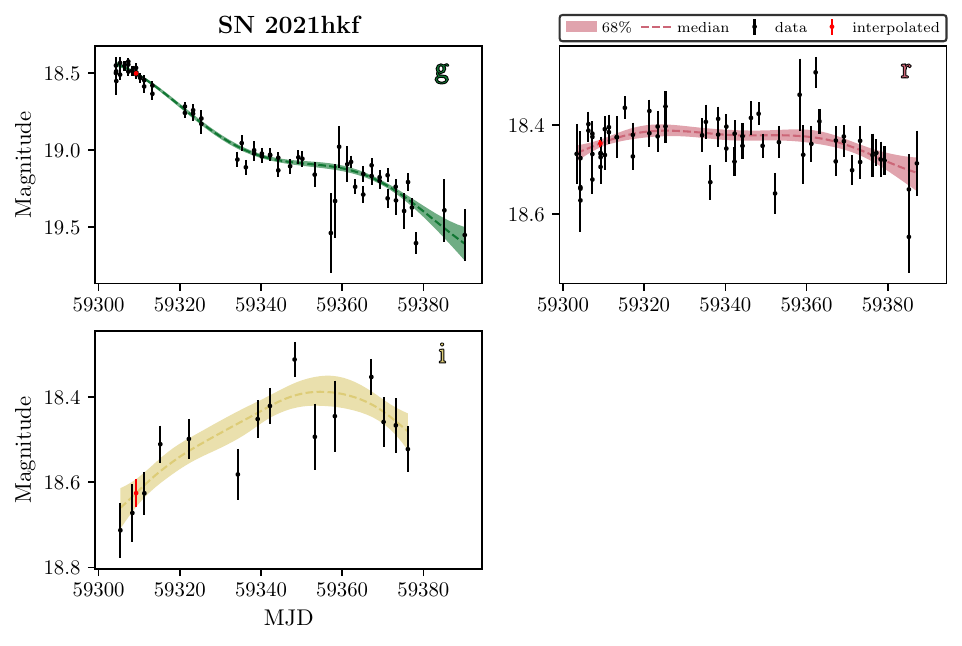}
			\includegraphics[]{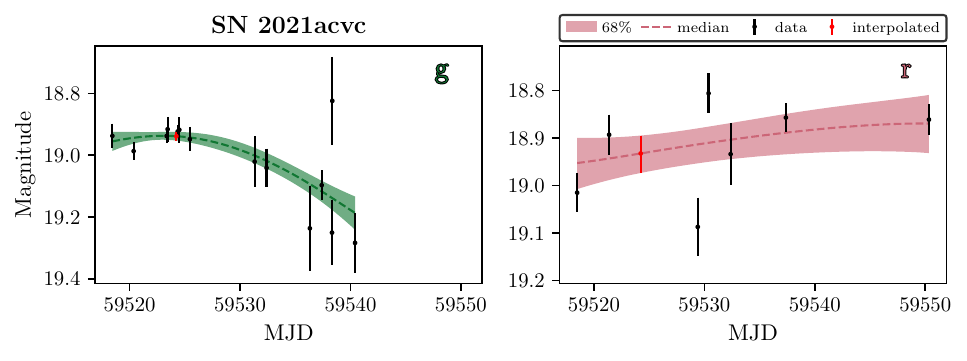}
		\end{center}
		\caption{GP interpolation of the photometry of SN~2021hkf and SN~2021acvc. See \cref{fig:lc_interpolation} for an explanation of the plot.}
	\end{figure*}


	\twocolumn

	\clearpage
	\onecolumn
	\section{Flux calibration} \label{app:flux_cal}
	\begin{figure}[H]
		\centering
		\includegraphics[scale=0.95]{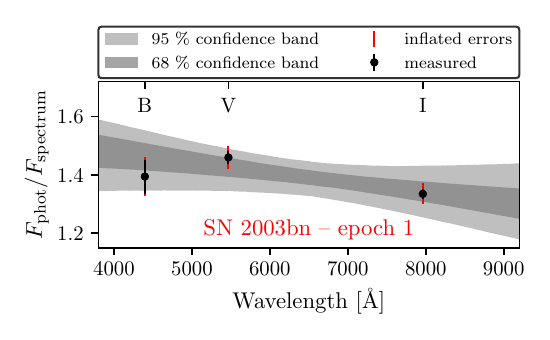}
		\includegraphics[scale=0.95]{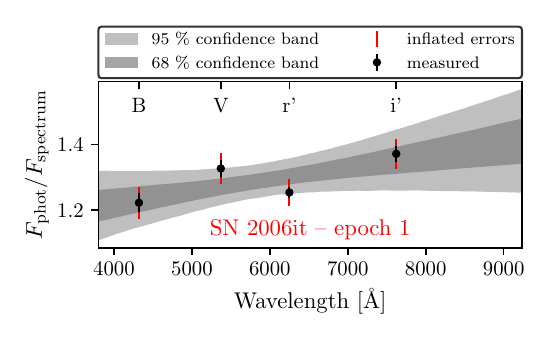}
  		\includegraphics[scale=0.95]{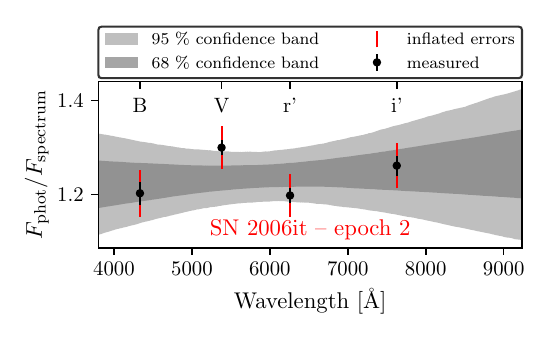}
		\includegraphics[scale=0.95]{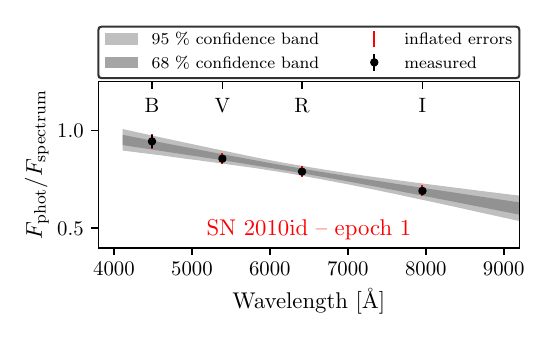}
		\includegraphics[scale=0.95]{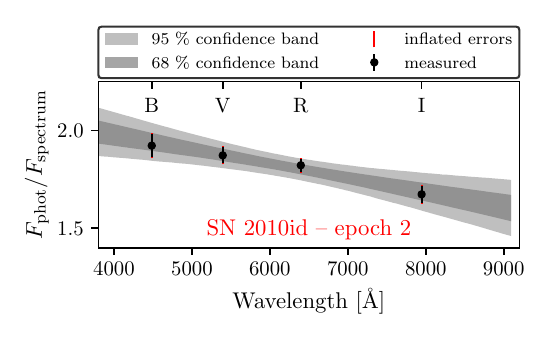}
            \includegraphics[scale=0.95]{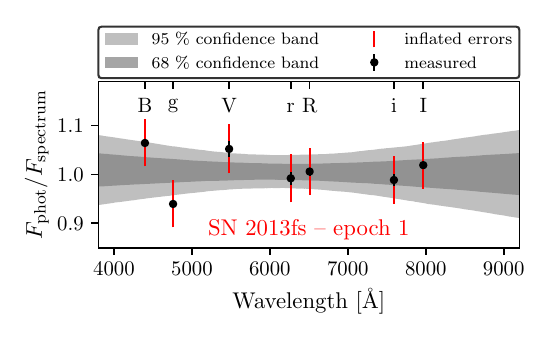}
            \includegraphics[scale=0.95]{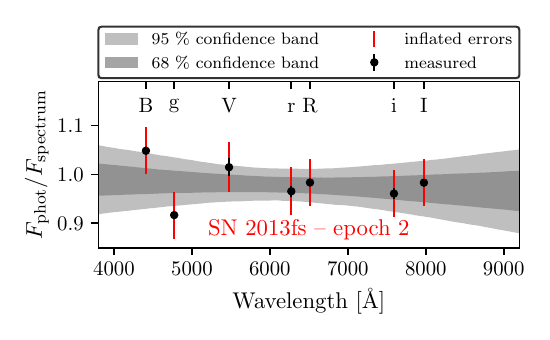}
            \includegraphics[scale=0.95]{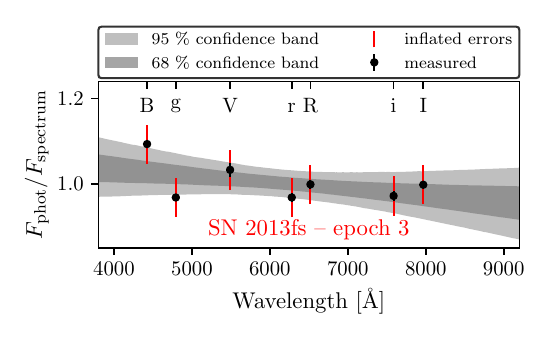}
		\caption{Linear flux calibration procedure for SN~2003bn, SN~2006it, SN~2010id, and SN~2013fs. See \cref{fig:flux_calibration_example} for an explanation of the plot.}
	\end{figure}

	\begin{figure*}
		\centering
            \includegraphics[scale=0.95]{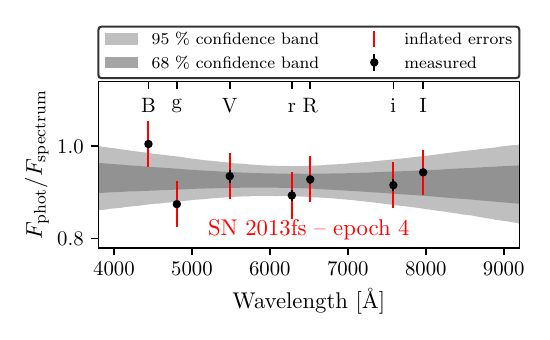}
            \includegraphics[scale=0.95]{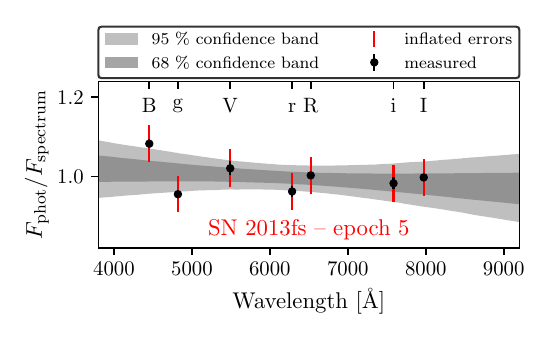}
  		\includegraphics[scale=0.95]{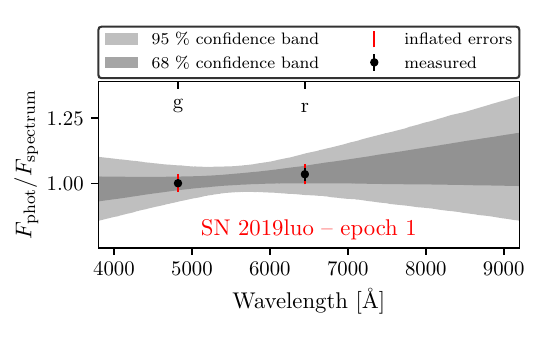}
		\includegraphics[scale=0.95]{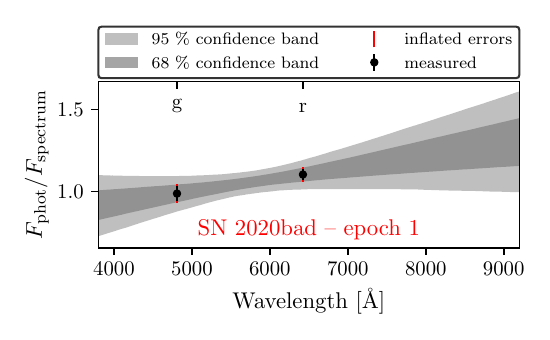}
  		\includegraphics[scale=0.95]{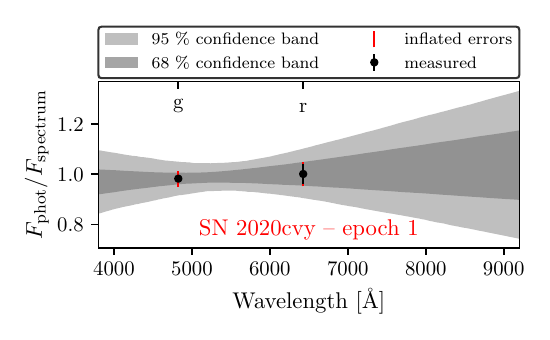}
            \includegraphics[scale=0.95]{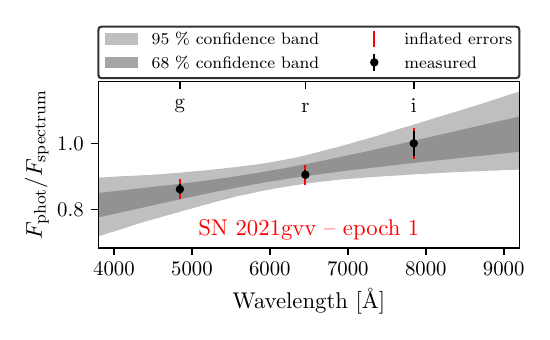}
            \includegraphics[scale=0.95]{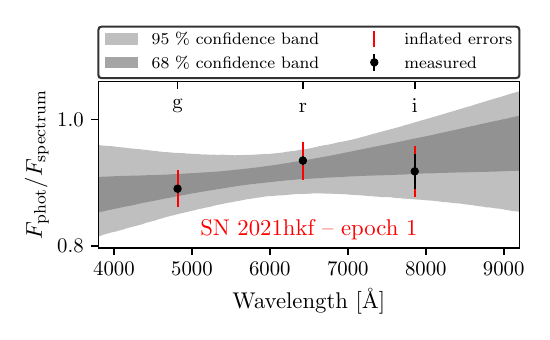}
		\includegraphics[]{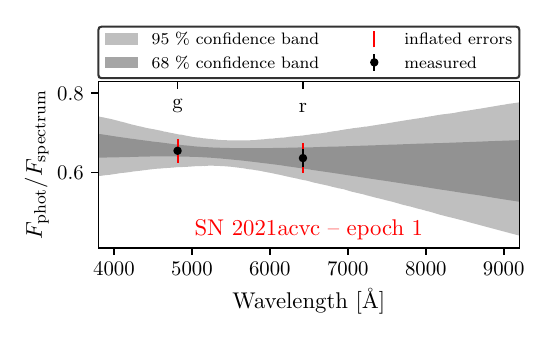}
		\caption{Linear flux calibration procedure for SN~2013fs, SN~2019luo, SN~2020bad, SN~2020cvy, SN~2021gvv, SN~2021hkf, and SN~2021acvc. See \cref{fig:flux_calibration_example} for an explanation of the plot.}
	\end{figure*}

	\clearpage
	\section{Spectral fits} \label{app:spec_fits}
	\begin{figure}[h]
		\setlength{\abovecaptionskip}{2pt}
		\setlength{\belowcaptionskip}{0pt}
		\centering
		\includegraphics[scale=0.90]{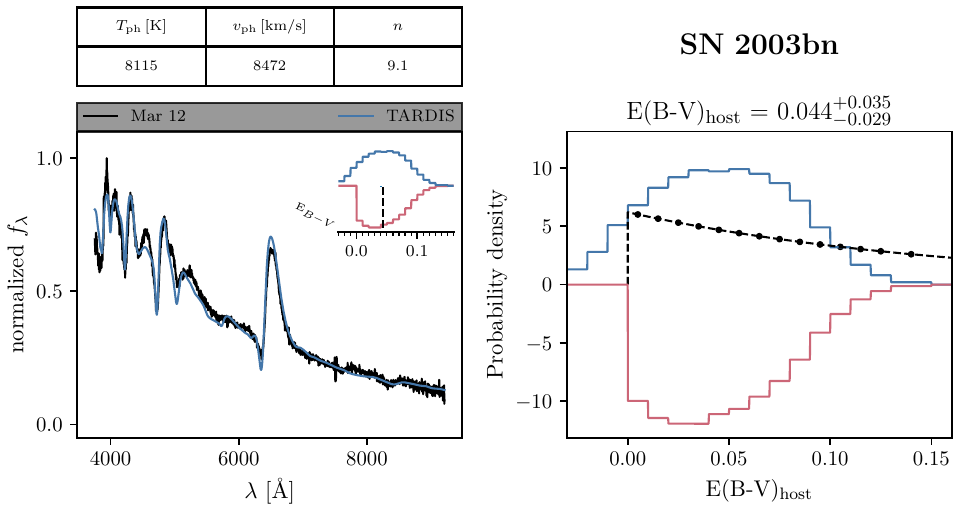}
		\caption{Spectral fits of SN~2003bn. See \cref{fig:spectral_fit} for an explanation of the plot.}
	\end{figure}
	\begin{figure}[h]
		\setlength{\abovecaptionskip}{2pt}
		\setlength{\belowcaptionskip}{0pt}
		\centering
		\includegraphics[scale=0.9]{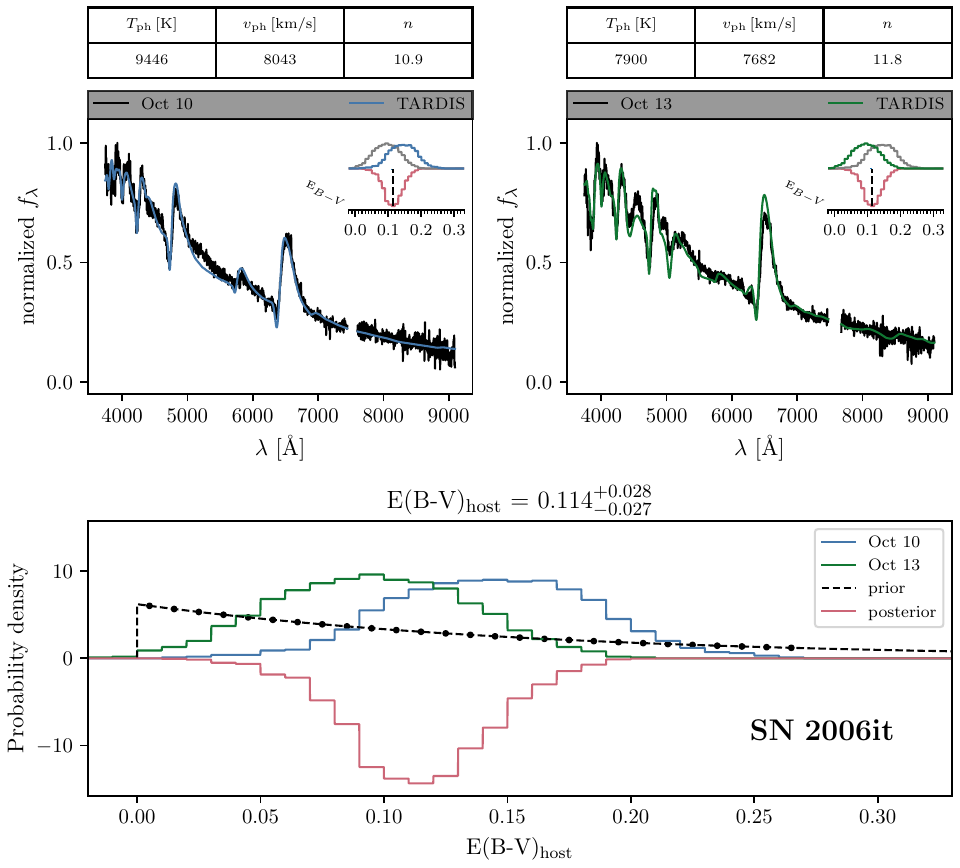}
		\caption{Spectral fits of SN~2006it. See \cref{fig:spectral_fit} for an explanation of the plot.}
	\end{figure}

	\begin{figure}[h]
		\setlength{\abovecaptionskip}{2pt}
		\setlength{\belowcaptionskip}{0pt}
		\centering
		\includegraphics[scale=0.9]{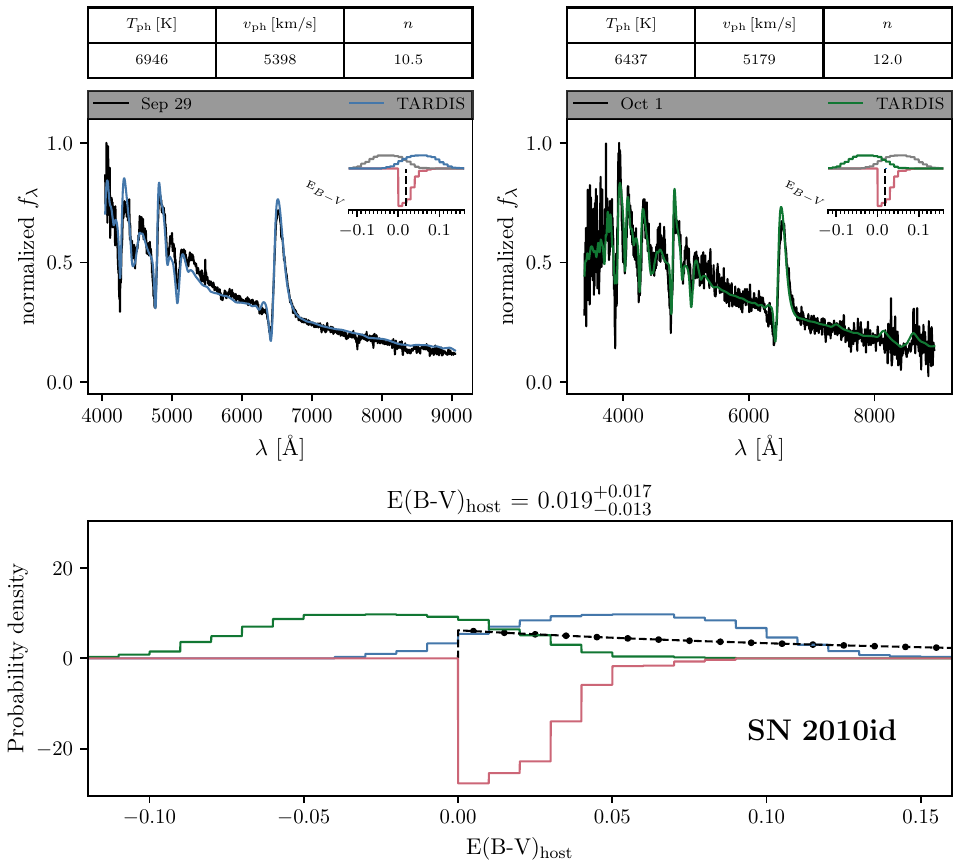}
		\caption{Spectral fits of SN~2010id. See \cref{fig:spectral_fit} for an explanation of the plot.}
	\end{figure}

	\begin{figure*}
		\centering
		\includegraphics[]{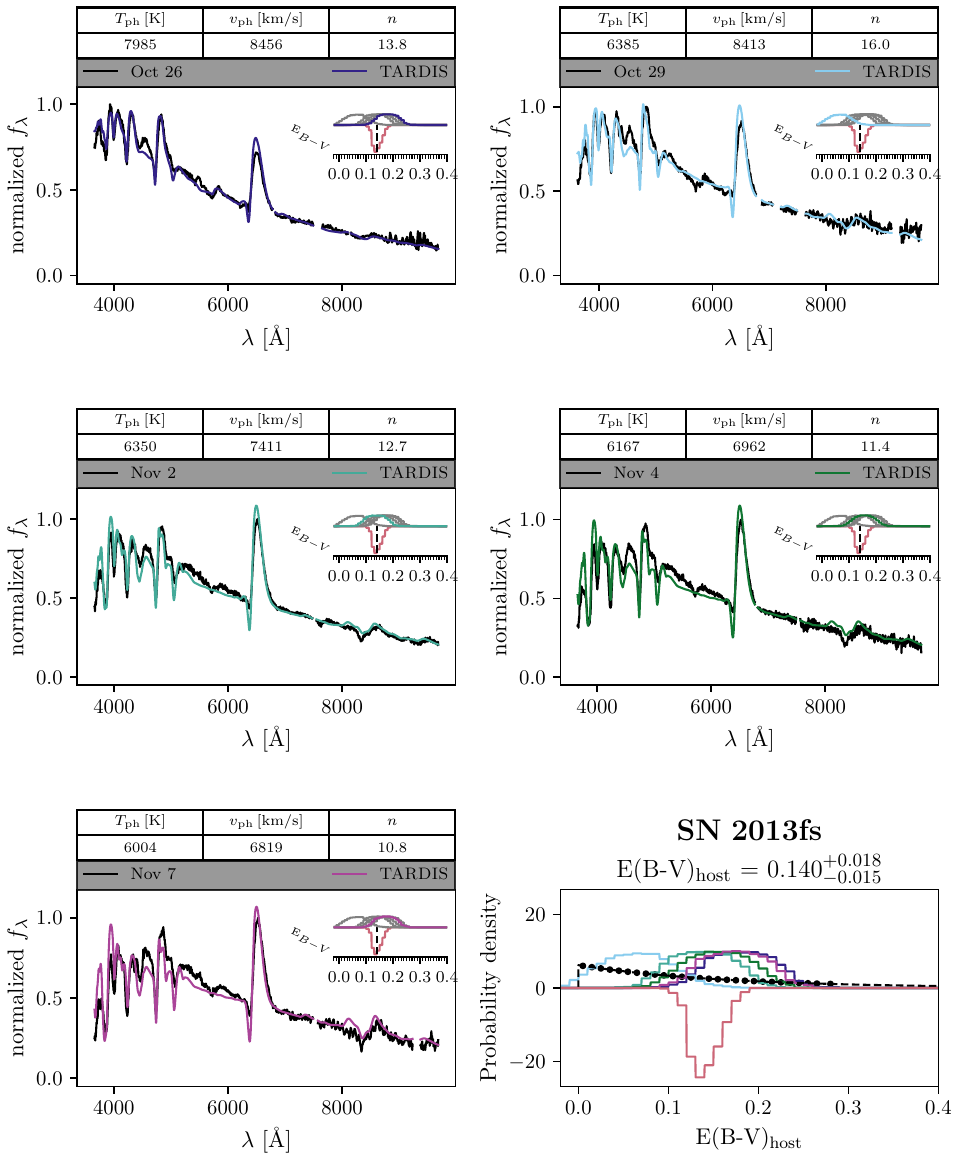}
		\caption{Spectral fit of SN~2013fs. See \cref{fig:spectral_fit} for an explanation of the plot.}
	\end{figure*}
	\begin{figure*}
		\centering
		\includegraphics[]{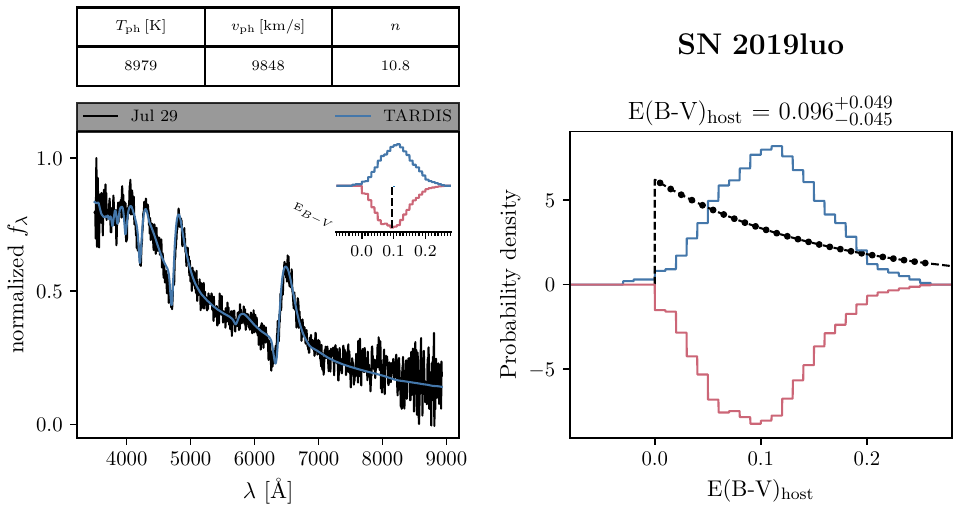}
		\caption{Spectral fits of SN~2019luo. See \cref{fig:spectral_fit} for an explanation of the plot.}
	\end{figure*}

	\begin{figure*}
		\centering
		\includegraphics[]{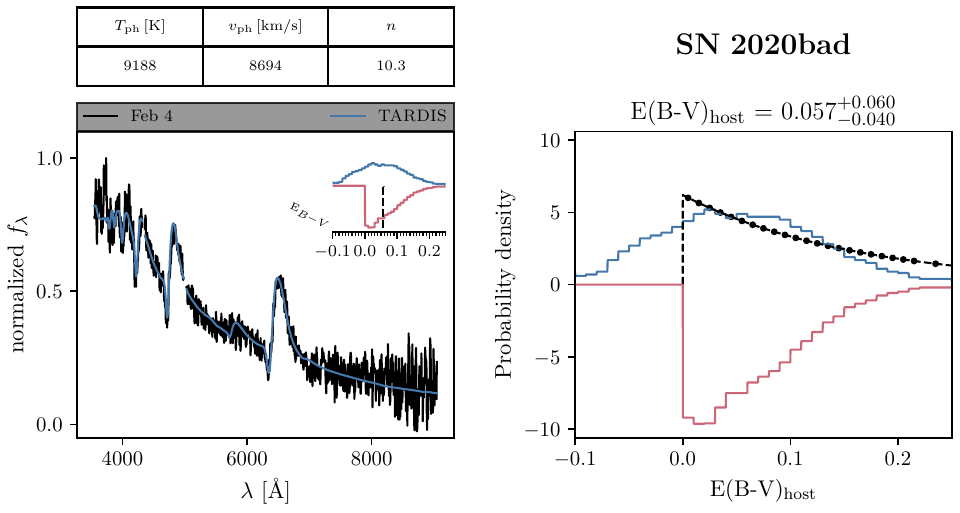}
		\caption{Spectral fits of SN~2020bad. See \cref{fig:spectral_fit} for an explanation of the plot.}
	\end{figure*}
	\begin{figure*}
		\centering
		\includegraphics[]{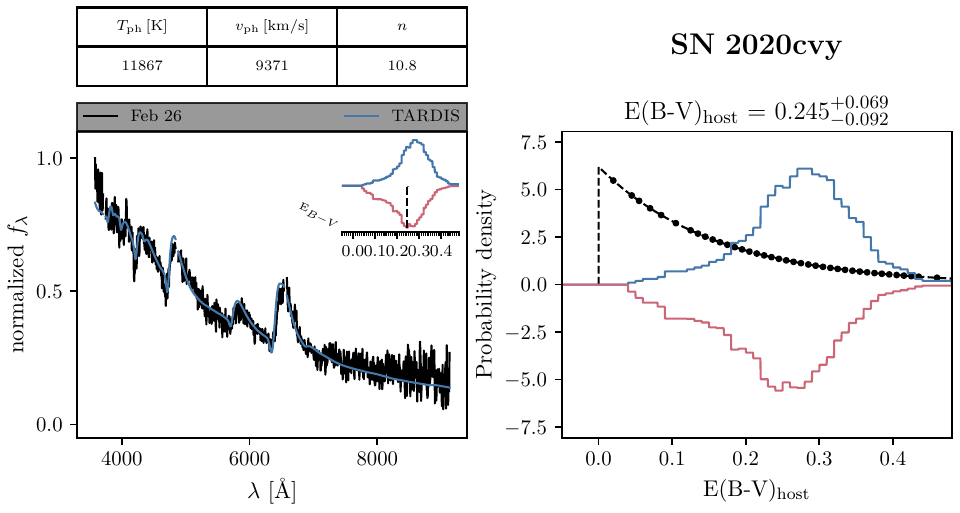}
		\caption{Spectral fits of SN~2020cvy. See \cref{fig:spectral_fit} for an explanation of the plot.}
	\end{figure*}
	\begin{figure*}
		\centering
		\includegraphics[]{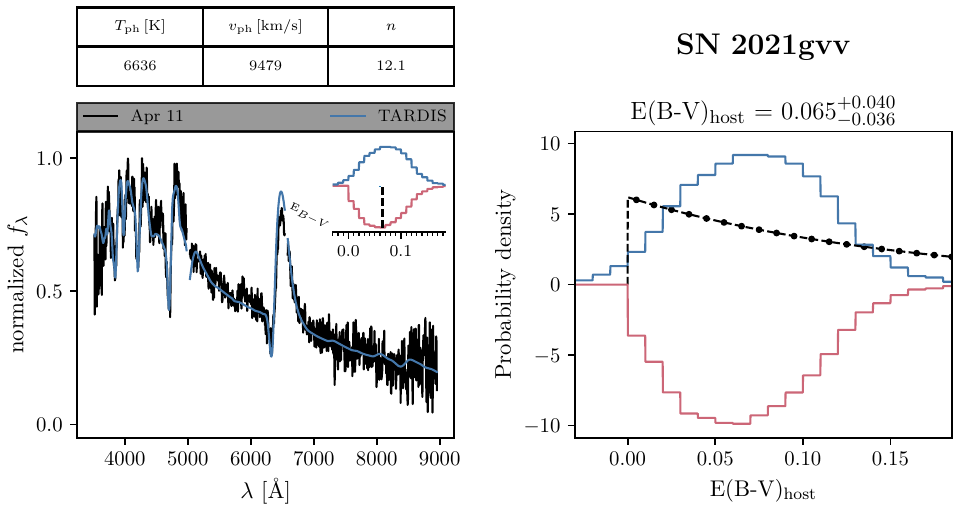}
		\caption{Spectral fits of SN~2021gvv. See \cref{fig:spectral_fit} for an explanation of the plot.}
	\end{figure*}

	\begin{figure*}
		\centering
		\includegraphics[]{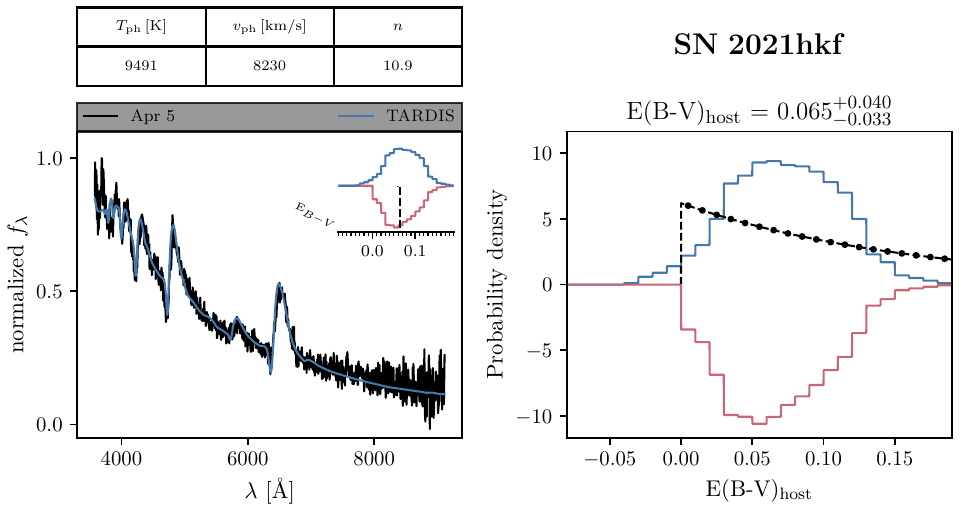}
		\caption{Spectral fits of SN~2021hkf. See \cref{fig:spectral_fit} for an explanation of the plot.}
	\end{figure*}

	\begin{figure*}
		\centering
		\includegraphics[]{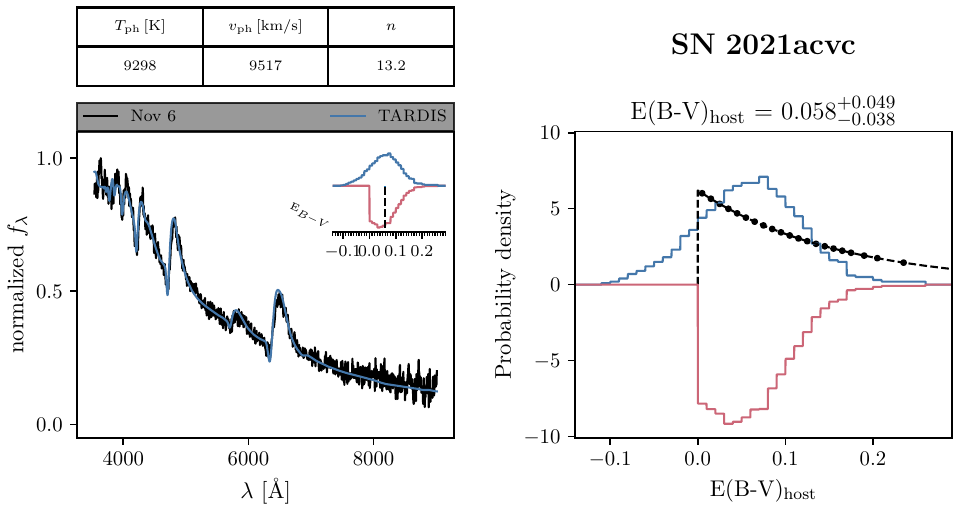}
		\caption{Spectral fits of SN~2021acvc. See \cref{fig:spectral_fit} for an explanation of the plot.}
	\end{figure*}

\end{appendix}

\end{document}